\documentclass[10pt,aps,prd,showpacs,twocolumn,longbibliography,nofootinbib,eqsecnum]{revtex4-2}
\usepackage[english]{babel}
\usepackage{amsfonts}
\usepackage{amssymb}
\usepackage{amsmath}
\usepackage{amsthm}
\usepackage{mathrsfs}
\usepackage{graphicx}
\usepackage{bbm}
\usepackage[unicode=true,bookmarksopen=true]{hyperref}
\usepackage{upgreek}
\usepackage{euscript}
\usepackage{color,soul}
\usepackage{ulem}

\DeclareMathAlphabet{\mathpzc}{OT1}{pzc}{m}{it}

\usepackage{etoolbox}
\apptocmd{\sloppy}{\hbadness 10000\relax}{}{}

\usepackage{mathtools}

\let\originalleft\left
\let\originalright\right
\renewcommand{\left}{\mathopen{}\mathclose\bgroup\originalleft}
\renewcommand{\right}{\aftergroup\egroup\originalright}

\makeatletter

\def\@spliteq#1{\begin{equation}\begin{split}#1\end{split}\end{equation}}
\def\splitequation{\collect@body\@spliteq}

\makeatother

\date{\today}

\let\oldre\Re
\let\oldim\Im
\renewcommand{\Re}{\oldre\mathfrak{e}\,}
\renewcommand{\Im}{\oldim\mathfrak{m}\,}

\newcommand{\eqend}[1]{\,#1}
\newcommand{\tr}{\textrm{tr}}

\newcommand{\vt}[1]{\textbf{\textrm{#1}}}
\renewcommand{\emph}[1]{{\it #1}}

\begin{document}

\title{Hartle-Hawking state in the real-time formalism}

\author{Atsushi Higuchi}
\email{atsushi.higuchi@york.ac.uk}
\affiliation{Department of Mathematics, University of York, Heslington, York YO10 5DD, United Kingdom}

\author{William C.\ C.\ Lima}
\email{william.correadelima@york.ac.uk}
\affiliation{Department of Mathematics, University of York, Heslington, York YO10 5DD, United Kingdom}


\begin{abstract}
We study self-interacting massive scalar field theory in static spacetimes with a bifurcate Killing horizon and a wedge reflection. In this theory the Hartle-Hawking state is defined to have the $N$-point correlation functions obtained by analytically continuing those in the Euclidean theory, whereas the double Kubo-Martin-Schwinger (KMS) state is the pure state invariant under the Killing flow and the wedge reflection which is regular on the bifurcate Killing horizon and reduces to the thermal state at the Hawking temperature in each of the two static regions. We demonstrate in the Schwinger-Keldysh operator formalism of perturbation theory the equivalence between the Hartle-Hawking state and the double KMS state with the Hawking temperature, which was shown before by Jacobson in the path-integral framework. 
\end{abstract}

\maketitle

%%%%%%%%%%%%%%%%%%%%%%%%%%%%%%%%%%%%%%%%%%%%%%%%%%%%%%%%%%%%%%%%%%%%%%%%%%%%%%%%%%%%%%%%%%%%%%%%%%%%%%%%%%%%%%%%%%%%%%%%%%%%%%%%%%%%%%%%%%%%%%%%%%%%%
\section{Introduction}                                                                                                                              %
\label{sec:introduction}                                                                                                                            %
%%%%%%%%%%%%%%%%%%%%%%%%%%%%%%%%%%%%%%%%%%%%%%%%%%%%%%%%%%%%%%%%%%%%%%%%%%%%%%%%%%%%%%%%%%%%%%%%%%%%%%%%%%%%%%%%%%%%%%%%%%%%%%%%%%%%%%%%%%%%%%%%%%%%%

In static spacetimes with a bifurcate Killing horizon and a wedge reflection, the natural state for quantum fields in the static region is a thermal equilibrium state at the Hawking temperature. Important examples of these spacetimes are Schwarzschild (i.e., eternal black hole), de~Sitter, and Minkowski spacetimes. This state was first described by Hartle and Hawking~\cite{hartle_hawking_prd_1976} by computing the Feynman propagator of a free scalar field in the Euclidean section of Schwarzschild spacetime, and then analytically continuing the result to the black-hole exterior region. The thermal property of the Hartle-Hawking (HH) state follows from the assumption that in the Euclidean section the imaginary Killing time is periodic. The periodicity corresponding to the Hawking temperature prevents the appearance of a conical singularity on the Euclideanized  bifurcate horizon. In Minkowski spacetime, the HH state corresponds to the Minkowski vacuum~\cite{dowker_prd_1978}, and its thermal property with respect to a boost Killing vector is directly related to the Unruh effect~\cite{unruh_prd_1976}. The HH state in de~Sitter spacetime is known as the Euclidean (or Bunch-Davies) vacuum~\cite{chernikov_tagirov_aihp_1968, schomblond_spindel_aihp_1976, bunch_davies_prsl_1978}, and its thermal property with respect to a de~Sitter boost is manifested in the Gibbons-Hawking effect~\cite{gibbons_hawking_prd_1977}. The HH state in this spacetime was shown to be the same as the vacuum state in the Poincar\'e patch~\cite{higuchi_marolf_morrison_prd_2011,korai_tanaka_2013} for interacting scalar field theories, and its infrared behavior was investigated in Refs.~\cite{marolf_morrison_prd_2010, marolf_morrison_prd_2011, hollands_cmp_2013,rajaraman_prd_2010}. 

In the free-field case, Kay~\cite{kay_cmp_1985} proved that this state exists on a certain algebra of observables localized on the double-wedged region of the Kruskal-Szekeres extension~\cite{kruskal_pr_1960, szekeres_grg_2002} of Schwarzschild spacetime. As noted earlier by Israel~\cite{israel_pla_1976}, when seen as a state on that region, the HH state is actually a pure state due to correlations between the two wedges. Moreover, as shown by Kay and Wald~\cite{kay_wald_pr_1991}, this state can be extended across the horizon and defines a pure state on the entire spacetime. More precisely, they proved that the HH state, if it exists, is the unique state to be both invariant under the action of the Killing field generating the horizon and regular (i.e., to have the Hadamard form) on and across the horizon. 

For interacting fields, Gibbons and Perry~\cite{gibbons_perry_prsl_1978} have employed perturbation theory to extend the original argument of Hartle and Hawking, and pointed out that the interacting Euclidean theory also defines a thermal field theory on the black-hole exterior. Concrete realizations of this claim were later provided via the path-integral formalism by Unruh and Weiss~\cite{unruh_weiss_prd_1984} in Minkowski spacetime and by Barvinsky, Frolov and Zelnikov~\cite{barvinsky_frolov_zelnikov_prd_1994} in the case of a black hole. Motivated by these discussions, and the regularity results of Kay and Wald, Jacobson~\cite{jacobson_prd_1994} proposed that the HH state should define a good state even across the bifurcate horizon, notwithstanding the Killing time coordinate being not well defined there. He then showed that the HH state is the double KMS state using the path-integral formalism. Jacobson's proposal has been proved to work in a more rigorous framework by Sanders~\cite{sanders_lmp_2015} in the case of free fields only recently.

Perturbative analysis of the double KMS state leads naturally to the Schwinger-Keldysh formalism~\cite{schwinger_jmp_1961, keldysh_zetf_1964}, often used in more general spacetimes~\cite{calzetta_hu_book}. In this paper we demonstrate that the $N$-point correlation functions in the double KMS state in static spacetime with a bifurcate Killing horizon agrees with those in the HH state, i.e., that they agree with those obtained by analytic continuation from the Euclidean theory. We note that the double KMS state in this spacetime has also been studied in axiomatic field theory~\cite{bisognano_wichmann_jmp_1975, bisognano_wichmann_jmp_1976,sewell_ap_1982}.

The remainder of the paper is organized as follows. In Sec.~\ref{sec:preliminaries} we briefly discuss the geometry of the class of spacetimes with a static bifurcate Killing horizon and a wedge reflection considered in this paper.  We also discuss their complexification.  We then discuss the properties of double KMS states in general. In Sec.~\ref{sec:double_kms_states} we consider a massive, self-interacting scalar field theory in these spacetimes and show that the double (i.e., purified) KMS state at the Hawking temperature with respect to the Killing vector field generating the horizon is the HH state.  We first review the equivalence between the double KMS state and HH state for the noninteracting case. Then, we demonstrate this equivalence for the interacting scalar field with nonderivative self-interaction. To do so, we first explain how the $N$-point correlation functions are obtained in the HH state by analytic continuation of the Euclidean theory. Then we show that the $N$-point correlation functions for the double KMS state in the Schwinger-Keldysh operator formalism are the same as those in the HH state.  In Sec.~\ref{sec:example_spacetimes} we briefly explain how the $N$-point functions are given perturbatively in the HH state for Minkowski and de~Sitter spacetimes. We conclude in Sec.~\ref{sec:discussion} with a summary and a discussion of our results. We present some technical details in the Appendices. In Appendix~\ref{appendix:future-right} we discuss the free-theory two-point function in the region in the future of the bifurcate Killing horizon. In Appendix~\ref{appendix:detail-analytic} we present some details of the analytic continuation for the HH state. In Appendix~\ref{appendix:int_with_t_dep_hamiltonian} we discuss the interaction picture with a time-dependent Hamiltonian. Throughout this paper we employ units such that $k_\textrm{B} = \hbar = c = G= 1$ and adopt the signature $-++\dots +$ for the metric.

%%%%%%%%%%%%%%%%%%%%%%%%%%%%%%%%%%%%%%%%%%%%%%%%%%%%%%%%%%%%%%%%%%%%%%%%%%%%%%%%%%%%%%%%%%%%%%%%%%%%%%%%%%%%%%%%%%%%%%%%%%%%%%%%%%%%%%%%%%%%%%%%%%%%%
\section{Preliminaries}                                                                                                                             %
\label{sec:preliminaries}                                                                                                                           %
%%%%%%%%%%%%%%%%%%%%%%%%%%%%%%%%%%%%%%%%%%%%%%%%%%%%%%%%%%%%%%%%%%%%%%%%%%%%%%%%%%%%%%%%%%%%%%%%%%%%%%%%%%%%%%%%%%%%%%%%%%%%%%%%%%%%%%%%%%%%%%%%%%%%%

\subsection{Static spacetime with a bifurcate Killing horizon}
%%%%%%%%%%%%%%%%%%%%%%%%%%%%%%%%%%%%%%%%%%%%%%%%%%%%%%%%%%%%%%%%%%%%%%%%%%%%%%%%%%%%%%%%%%%%%%%%%%%%%%%%%%%%%%%%%%%%%%%%%%%%%%%%%%%%%%%%%%%%%%%%%%%%%

Let us recall that the Kruskal-Szekeres extension of $4$-dimensional Schwarzschild spacetime has the following metric (see, e.g.\ Ref.~\cite{wald_gr_book}):
\begin{splitequation}\label{eq:schwarzschild-metric_2}
g_{ab}^\textrm{Sch}
& = \frac{32M^3}{r}e^{-r/(2M)}\left[ - (dT)_a(dT)_b + (dX)_a (dX)_b\right]\\ 
&\phantom{=} + r^2\omega_{ab}\eqend{,}
\end{splitequation}
where $M$ is the black-hole mass and $\omega_{ab}$ is the metric on the unit $2$-sphere, $S^2$. The variable $r$ is implicitly defined in terms of $X$ and $T$ as
\begin{equation}\label{eq:schwarzshild-rho-r}
    \frac{r-2M}{2M}e^{r/(2M)} = X^2 - T^2\eqend{.}
\end{equation}
Note that the function of $r$ on the left-hand side is monotonically increasing, which makes $r$ a well-defined function of $X^2-T^2$.
It is more common to use the variables $U=T-X$ and $V=T+X$ instead of $T$ and $X$.   

Motivated by this metric, we consider in this paper the $n$-dimensional, globally hyperbolic spacetime $(\mathcal{M},g_{ab})$, with the metric tensor
\begin{equation}
 g_{ab} = f(\rho^2,\boldsymbol{\theta})\left[-(dT)_a(dT)_b + (dX)_a(dX)_b\right] + s_{ab}(\rho^2,\boldsymbol{\theta})\eqend{,} 
 \label{eq:metric_cartesian_coordinates_2}
\end{equation}
where $\rho^2 = X^2 - T^2$ takes values in a real interval containing an open neighborhood of $0$ and $f(\rho^2,\boldsymbol{\theta})$ is a positive function.  We assume that the hypersurfaces with constant $T$ are Cauchy surfaces. [For Schwarzschild spacetime, we have $\rho^2 \in (-1,\infty)$.]  Here, $\boldsymbol{\theta}$ represents all coordinate variables other than $T$ and $X$, and $s_{ab}$ is a linear combination of $(d\theta^i)_a(d\theta^j)_b$.  

\begin{center}
 \begin{figure}[ht]
   \includegraphics[scale=1]{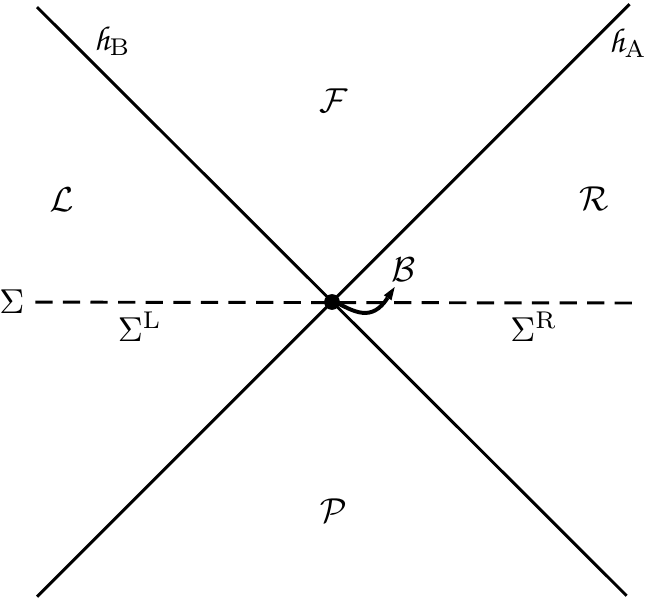}
   \caption{Spacetime with a bifurcate Killing horizon.}
   \label{fig:bKh}
 \end{figure}
\end{center}

The vector
\begin{equation}\label{eq:def_Killing_XT}
    \xi^a = \kappa\left[ X (\partial_T)^a + T (\partial_X)^a\right]\eqend{,}
\end{equation}
with $\kappa > 0$, is a Killing vector because $T$ and $X$ appear in the metric only through $\rho^2 = X^2 - T^2$ and because $\xi^a\nabla_a \rho^2 = 0$. The constant $\kappa$ is chosen suitably for each spacetime.\footnote{For Schwarzschild spacetime it is chosen so that $\xi^a\xi_a \to -1$ at spacelike infinity.  For de~Sitter spacetime we choose $\xi^a\xi_a=-1$ for a particular timelike geodesic.}  An orbit of $\xi^a$ has $\boldsymbol{\theta}$ and $X^2 - T^2$ constant.
The hypersurfaces $X-T=0$ (denoted by $\mathpzc{h}_\mathrm{A}$ in Fig.~\ref{fig:bKh}) and $X+T=0$ (denoted by $\mathpzc{h}_\mathrm{B}$ in Fig.~\ref{fig:bKh}) are null hypersurfaces with $\xi^a$ as the normal vector. (Recall that a vector normal to a null hypersurface is also tangent to it.)  Thus, these hypersurfaces are Killing horizons for $\xi^a$.  These two Killing horizons constitute a bifurcate Killing horizon~\cite{boyer_prsla_1969}.

The Killing vector $\xi^a$ vanishes on the $(n-2)$-dimensional surface given by $X = T = 0$.  This surface is called the bifurcation surface and denoted by $\mathcal{B}$ in Fig.~\ref{fig:bKh}.  The bifurcate Killing horizon divides the spacetime $\mathcal{M}$ into four regions
as follows (see Fig.~\ref{fig:bKh}):
\begin{equation}
\begin{array}{ll}
  \textrm{the right wedge } (\mathcal{R}) &: X>0,\, -X < T < X\eqend{;}\\
  \textrm{the left wedge } (\mathcal{L}) &: X<0,\, X < T < -X\eqend{;}\\
  \textrm{the future wedge } (\mathcal{F}) &: T>0,\, -T < X < T\eqend{;}\\
  \textrm{the past wedge } (\mathcal{P}) &: T<0,\, T < X < -T\eqend{.}
\end{array}
\end{equation}
The right and left wedges, $\mathcal{R}$ and $\mathcal{L}$, are globally hyperbolic on their own right, and the hypersurfaces at $T=0$ with $X>0$ and $X<0$ are the Cauchy surfaces for $\mathcal{R}$ and $\mathcal{L}$, respectively, and denoted here by $\Sigma^\mathrm{R}$ and $\Sigma^\mathrm{L}$.

The constant of proportionality $\kappa$ in Eq.~\eqref{eq:def_Killing_XT} can be shown to be the surface gravity, which is defined on a Killing horizon $\mathpzc{h}$ as follows:  
\begin{equation}
\xi^b \nabla_b \xi^a|_{\mathpzc{h}} = \pm\kappa \xi^a|_{\mathpzc{h}}
\eqend{,}
\end{equation}
with $\kappa > 0$. This equation can also be written as
\begin{equation}\label{eq:surface_gravity_2}
\mp \frac{1}{2}\nabla^a \left(\xi^b\xi_b\right)|_{\mathpzc{h}} = \kappa\xi^a|_{\mathpzc{h}}\eqend{,}    
\end{equation}
by using the Killing equation. This equation can be used to verify that the constant $\kappa$ in Eq.~\eqref{eq:def_Killing_XT} is indeed the surface gravity of the bifurcate Killing horizon.

A form of the metric useful for the right and left wedges, $\mathcal{R}\cup \mathcal{L}$, can be given after the coordinate change 
\begin{splitequation}\label{eq:TX-left-right}
 & (T,X)\\
 & = 
   \begin{cases}
    \big(\rho\sinh(\kappa t),\rho\cosh(\kappa t)\big),\ \rho>0,& \textrm{if}\ (T,X)\in\mathcal{R}, \\
    \big(-\rho\sinh(\kappa t),\rho\cosh(\kappa t)\big),\ \rho < 0,& \textrm{if}\ (T,X)\in \mathcal{L}\eqend{.}
	 \end{cases}
\end{splitequation}
Thus, we find
\begin{splitequation}\label{eq:metric_restricted_2}
  & g_{ab}|_{\mathcal{R}\cup \mathcal{L}}\\ 
	& = f(\rho^2,\boldsymbol{\theta})\left[-\kappa^2\rho^2 (dt)_a(dt)_b + (d\rho)_a(d\rho)_b\right] + s_{ab}(\rho^2,\boldsymbol{\theta})\eqend{.}
\end{splitequation}
In these coordinates the Killing vector $\xi^a$ becomes simply
\begin{equation}
\xi^a = \begin{cases} (\partial_t)^a & \textrm{in}\ \mathcal{R}\eqend{,}\\
- (\partial_t)^a & \textrm{in}\ \mathcal{L}\eqend{.} \end{cases}
\end{equation}
The time variable $t$ increases toward the future both in  the right and left wedges, and the Killing vector is timelike in either wedge, future-directed in the right wedge and past-directed in the left wedge. For $4$-dimensional Schwarzschild spacetime, this coordinate transformation with $\kappa = 1/(4M)$ and Eq.~\eqref{eq:schwarzshild-rho-r}  gives the standard Schwarzschild metric:
\begin{splitequation}
 & g_{ab}^{\textrm{Sch}}|_{\mathcal{R}\cup\mathcal{L}}\\ 
 & = -\left(1 - \frac{2M}{r}\right)(dt)_a(dt)_b + \frac{(dr)_a(dr)_b}{1 - 2M/r} + r^2\,\omega_{ab}\eqend{.} 
\end{splitequation}

The wedge reflection $I$, which is an isometry of $(\mathcal{M},g_{ab})$, is defined by
\begin{equation}
    I: (T,X) \mapsto (-T,-X)\eqend{,}
\end{equation}
or equivalently,
\begin{equation}
    I: (t,\rho)|_\mathcal{R} \mapsto (-t,-\rho)|_\mathcal{L}\eqend{.}
\end{equation}
[We recall that $T$ is given differently in terms of $\rho$ and $t$ on the right and left wedges in Eq.~\eqref{eq:TX-left-right}.]
It maps a point in $\mathcal{R}$ to a point in $\mathcal{L}$ and vice versa.  Note that the time ordering is reversed under the map $I$.

The Euclidean section $\mathcal{M}_\mathrm{E}$ of the manifold $\mathcal{M}$ is obtained by letting $T=iT_\mathrm{E}$ with $T_\mathrm{E}\in \mathbb{R}$.  The metric of $\mathcal{M}_\mathrm{E}$ can be found from Eq.~\eqref{eq:metric_cartesian_coordinates_2} as
\begin{equation}\label{eq:euclidean_2}
 g_{ab}^{\mathrm{E}} =  f(\rho^2,\boldsymbol{\theta})\left[(dT_\mathrm{E})_a(dT_\mathrm{E})_b + (dX)_a(dX)_b\right] + s_{ab}(\rho^2,\boldsymbol{\theta})\eqend{,}
\end{equation}
where $\rho^2 \geq 0$. This metric can also be given in terms of $t_\mathrm{E}$, where $t=it_\mathrm{E}$, and $\rho\geq 0$ as
\begin{splitequation}\label{eq:euclidean-polar-2}
 & g_{ab}^{\mathrm{E}}\\ 
 & = f(\rho^2,\boldsymbol{\theta})\left[\kappa^2\rho^2(dt_\mathrm{E})_a(dt_\mathrm{E})_b + (d\rho)_a(d\rho)_b\right] + s_{ab}(\rho^2,\boldsymbol{\theta})\eqend{.} 
\end{splitequation}
The coordinates $(T_\mathrm{E},X)$ and $(t_\mathrm{E},\rho)$ cover the whole of $\mathcal{M}_\mathrm{E}$, and are related by
\begin{equation}
    (T_\mathrm{E},X) = \big(\rho \sin(\kappa t_\mathrm{E}),\rho\cos(\kappa t_\mathrm{E})\big)\eqend{.}
\end{equation}
Thus, we identify $t_\mathrm{E}$ with period $2\pi/\kappa$:
\begin{equation}\label{eq:periodic-identification}
    t_\mathrm{E} \sim t_\mathrm{E} + 2n\pi/\kappa\eqend{,}\ n \in\mathbb{Z}\eqend{.}
\end{equation}
Note that the hypersurface $T_\mathrm{E}=0$ of $\mathcal{M}_\mathrm{E}$ can be identified with $\Sigma^\mathrm{R}\cup \mathcal{B}\cup\Sigma^\mathrm{L}$, which is the Cauchy surface $T=0$ of the Lorentzian manifold $\mathcal{M}$.  Moreover, in this identification the hypersurface with $t_\mathrm{E}=0$ is identified with the Cauchy surface $\Sigma^\mathrm{R}$ of the right wedge $\mathcal{R}$ whereas that with $t_\mathrm{E}=-\pi/\kappa$ is identified with the Cauchy surface $\Sigma^\mathrm{L}$ of the left wedge $\mathcal{L}$.

\subsection{Double KMS states}
%%%%%%%%%%%%%%%%%%%%%%%%%%%%%%%%%%%%%%%%%%%%%%%%%%%%%%%%%%%%%%%%%%%%%%%%%%%%%%%%%%%%%%%%%%%%%%%%%%%%%%%%%%%%%%%%%%%%%%%%%%%%%%%%%%%%%%%%%%%%%%%%%%%%%

In quantum statistical mechanics there is a standard procedure to reproduce the expectation values computed in a mixed state in terms of those of a pure state~\cite{takahashi_umezawa_ijmpb_1996}. The idea is to double the original system and, in the doubled system, construct an entangled pure state such that the expectation values of operators restricted to the original system reproduce the statistical predictions in the original mixed state. For example, consider the standard density matrix of a thermal state for the Hamiltonian $H$ with inverse temperature $\beta$. This state is given by
\begin{equation}
\varrho(\beta) = \frac{e^{-\beta H}}{Z(\beta)}
\end{equation}
in a finite dimensional space of states $\mathscr{H}$, where $Z(\beta) \equiv \tr\, e^{-\beta H}$ is the partition function. Suppose that $\psi_i$ form a complete set of eigenstates of the Hamiltonian $H$, with $H\psi_i = E_i\psi_i$. Then, the thermal average of an operator $A$ given by $\langle A \rangle_\beta \equiv \tr[\varrho(\beta) A]$ is reproduced in the doubled space of states $\mathscr{H} \otimes \mathscr{H}$ by taking the expectation value of the operator $\mathbbm{1}\otimes A$ in the pure state
\begin{equation}\label{eq:double_kms_state}
\Omega_\beta = \frac{1}{\sqrt{Z(\beta)}}\sum_i e^{-\frac{1}{2}\beta E_i}\psi_i\otimes \psi_i\eqend{.}
\end{equation}

In this doubled system one also defines the Hamiltonian operator
\begin{equation}\label{eq:H_tilde}
 \tilde{H} \equiv \mathbbm{1}\otimes H - H\otimes \mathbbm{1}
\end{equation}
and an antiunitary operator\footnote{We recall that $J$ is said to be antilinear if for any two state vectors $\Psi_1$ and $\Psi_2$ and constants $a,b\in\mathbb{C}$ we have $J(a\Psi_1 + b\Psi_2) = a^* J\Psi_1 + b^* J\Psi_2$, while $J$ is said to be antiunitary if it is antilinear and satisfies $\langle J\Psi_1| J\Psi_2\rangle = \langle \Psi_2| \Psi_1\rangle$.} $J$ such that $J(\psi_i\otimes \psi_j) \equiv \psi_j\otimes \psi_i$. Note that from this definition it follows that
\begin{equation}
J^2 = \mathbbm{1}\eqend{.} 
\end{equation}
It can readily be verified that 
\begin{equation}
[e^{-i\tilde{H}t},\, J] = 0\eqend{,} 
\end{equation}
by letting the left-hand side act on the basis of the doubled space of states.

It is easy to see that the state $\Omega_\beta$ defined by Eq.~(\ref{eq:double_kms_state}) is annihilated by the Hamiltonian $\tilde{H}$ and is left invariant by $J$, i.e.,
\begin{equation}\label{eq:invariance_kms_state_H}
e^{-i\tilde{H}t}\Omega_\beta = \Omega_\beta
\end{equation}
and
\begin{equation}\label{eq:invariance_kms_state_J}
J\Omega_\beta = \Omega_\beta\eqend{.}
\end{equation}
One can also show that the state $O\Omega_\beta$ obtained by letting any operator of the form $O = \mathbbm{1}\otimes A$ act on $\Omega_\beta$ satisfies
\begin{equation}\label{eq:kms_condition}
 e^{-\frac{1}{2}\beta\tilde{H}}O\Omega_\beta = JO^\dagger \Omega_\beta\eqend{.}
\end{equation}
This relation is called the KMS condition~\cite{kubo_jpsj_1957, martin_schwinger_pr_1959, haag_hugenholtz_Winnink_cmp_1967}. Since the state $\Omega_\beta$ is the extension of the thermal state $\varrho(\beta)$ to the enlarged system, we shall refer to it as the double KMS state, following Kay~\cite{kay_hpa_1985}. The converse is also true: a state $\Omega_\beta \in \mathscr{H}\otimes \mathscr{H}$ satisfying the KMS condition~(\ref{eq:kms_condition}) is given by Eq.~(\ref{eq:double_kms_state}). The advantage of defining double KMS states by Eqs.~(\ref{eq:invariance_kms_state_H})-(\ref{eq:kms_condition}) is that they can be used to characterize thermal equilibrium states for systems with an infinite number of degrees of freedom as well~\cite{haag_hugenholtz_Winnink_cmp_1967}.

Although this doubling of the space of states may appear artificial in the case of ordinary thermal systems such as the quantum-mechanical example above, double KMS states appear quite naturally in quantum field theory in spacetimes with a bifurcate Killing horizon and a wedge-reflection symmetry, as first noticed by Israel~\cite{israel_pla_1976}. Thus, let us consider a double KMS state over the left and right wedges, $\mathcal{R}\cup \mathcal{L}$, for a massive real scalar field $\Phi(x)$ in a spacetime described in the previous section. The Hamiltonian operator $H$ then corresponds to the $t$-translation generator in either wedge.  Since these wedges are static with respect to the $t$-translation, this Hamiltonian is conserved.  We assume that there is a unique state with the lowest energy, i.e., the lowest eigenvalue of $H$, in either wedge, which we call the vacuum state.

The Hilbert space $\mathscr{H}$ of (pure) states in the right or left wedge is constructed by applying the (smeared) field operators on the vacuum state with support in the respective wedge.  Then the Hilbert space of states in $\mathcal{L}\cup \mathcal{R}$ is the tensor product $\mathscr{H}\otimes \mathscr{H}$. We define the operator $\tilde{H}$ acting on $\mathscr{H}\otimes \mathscr{H}$ as in Eq.~\eqref{eq:H_tilde}. Moreover, the wedge reflection $I$ is an isometry of $\mathcal{M}\supset \mathcal{L}\cup \mathcal{R}$, and defines in the quantum theory an antiunitary operator $J$ according to
\begin{equation}\label{eq:J_2}
 J\Phi(x)J \equiv \Phi(I(x))\eqend{,}
\end{equation}
for all $x\in\mathcal{M}$. The operator $J$ is antiunitary because it reverses the direction of time.  The double KMS state $\Omega_\beta$ for the scalar field $\Phi$ is then defined by requiring that it satisfy Eqs.~(\ref{eq:invariance_kms_state_H})-(\ref{eq:kms_condition}).

%%%%%%%%%%%%%%%%%%%%%%%%%%%%%%%%%%%%%%%%%%%%%%%%%%%%%%%%%%%%%%%%%%%%%%%%%%%%%%%%%%%%%%%%%%%%%%%%%%%%%%%%%%%%%%%%%%%%%%%%%%%%%%%%%%%%%%%%%%%%%%%%%%%%%
\section{Hartle-Hawking state as a double KMS state}                                                                                                %
\label{sec:double_kms_states}                                                                                                                       %
%%%%%%%%%%%%%%%%%%%%%%%%%%%%%%%%%%%%%%%%%%%%%%%%%%%%%%%%%%%%%%%%%%%%%%%%%%%%%%%%%%%%%%%%%%%%%%%%%%%%%%%%%%%%%%%%%%%%%%%%%%%%%%%%%%%%%%%%%%%%%%%%%%%%%

In this section we consider a massive real scalar field theory with nonderivative self-interaction in a spacetime with metric given by Eq.~\eqref{eq:metric_cartesian_coordinates_2}.  The HH state for this theory is defined as a state such that its $N$-point functions are the analytic continuation from that state in the corresponding Euclidean field theory defined on the Euclidean section with metric~\eqref{eq:euclidean_2}.  The HH state in the noninteracting case is known to be a double KMS state~\cite{kay_wald_pr_1991}, which is a pure state with correlations between the left and right wedges. These correlations give rise to a thermal state at the temperature 
\begin{equation}\label{eq:the-right-temperature}
\beta_\textrm{H}^{-1} \equiv \frac{\kappa}{2\pi}\eqend{,}
\end{equation}
the Hawking temperature, when the HH state is restricted to the right or left wedge. A formal path-integral argument~\cite{jacobson_prd_1994} shows that this is also the case for the interacting case.  In this section we show this fact in perturbation theory using the Schwinger-Keldysh operator formalism.

\subsection{The noninteracting case}
\label{subsection:non-interacting_case}
%%%%%%%%%%%%%%%%%%%%%%%%%%%%%%%%%%%%%%%%%%%%%%%%%%%%%%%%%%%%%%%%%%%%%%%%%%%%%%%%%%%%%%%%%%%%%%%%%%%%%%%%%%%%%%%%%%%%%%%%%%%%%%%%%%%%%%%%%%%%%%%%%%%%%

Before presenting the interacting case, we review the equivalence of the HH and a double KMS state at the Hawking temperature for free massive scalar fields. Most results in this subsection can be found, e.g.\ in Ref.~\cite{fulling_ruijsenaars_pr_1987}. Thus, we let the quantum scalar field $\Phi_\mathrm{I}(x)$ satisfy the Klein-Gordon equation,
\begin{equation}\label{eq:KG-equation_2}
   \left[ - \nabla_a\nabla^a + m^2\right]\Phi_\mathrm{I}(x)=0\eqend{,}
\end{equation}
in the spacetime $(\mathcal{M},g_{ab})$. We put the subscript ``I'' on the field $\Phi$ in anticipation of the application of the results of this subsection to the field in the interaction picture, which is a noninteracting field. First we describe the HH state $\Omega_{\textrm{HH}}^{(0)}$ for this theory. Let $x_\mathrm{E} = (it_\mathrm{E},\vt{x})$, $\vt{x}=(\rho,\boldsymbol{\theta})$, $\rho\geq 0$, with $t_\mathrm{E}$ periodically identified as in Eq.~\eqref{eq:periodic-identification}.  Thus, $x_\mathrm{E}$ can be identified with a point in the Euclidean section $\mathcal{M}_\mathrm{E}$ defined in the previous section. Define the Green's function $G^{(0)}(x_\mathrm{E};x_\mathrm{E}')$ as the function satisfying
\begin{equation}\label{eq:greens-differential-eq}
    \left[ - \nabla_a^{(\mathrm{E})}\nabla^{(\mathrm{E})a} + m^2\right]G^{(0)}(x_\mathrm{E};x_\mathrm{E}') = \frac{1}{\sqrt{g^{\mathrm{E}}}}\delta(x_\mathrm{E};x_\mathrm{E}')\eqend{,}
\end{equation}
where $\nabla_a^{(\mathrm{E})}$ is the covariant derivative compatible with the metric $g^\mathrm{E}_{ab}$.\footnote{We may need to impose boundary conditions at the upper end of the interval for $\rho$ in order to make this Green's function unique.} Dirac's delta-function $\delta(x_\mathrm{E};x_\mathrm{E}')$ is defined by 
\begin{equation}\label{eq:dirac-delta-definition}
    \int_{\mathcal{M}_\mathrm{E}} dt_\mathrm{E}'\,d^{n-1}\mathbf{x}'\,F(x_{\mathrm{E}}')\delta(x_\mathrm{E};x_\mathrm{E}') = F(x_\mathrm{E})\eqend{,}
\end{equation} 
for any smooth and compactly supported function $F$ on $\mathcal{M}_\mathrm{E}$.  The function $G^{(0)}(x_\mathrm{E};x_\mathrm{E}')$ is known to be symmetric under the interchange of the two arguments, $x_\mathrm{E}$ and $x_\mathrm{E}'$.
We define the function $G^{(0)}(x;x')$ for $x=(t,\mathbf{x})$ and $x'=(t',\mathbf{x}')$ with $t$ and $t'$ not purely imaginary by analytic continuation in $t$ and $t'$, where the time variable is periodically identified in the imaginary direction as
\begin{equation}
    t \sim t + 2n\pi i/\kappa,\ \ n\in\mathbb{Z}\eqend{.}
\end{equation}
This is possible in general if $\textrm{Im}(t - t') \neq 0$ because singularities occur only for $\textrm{Im}(t - t')=0$~\cite{fulling_ruijsenaars_pr_1987}. Thus, we analytically continue $G^{(0)}(x,x')$ from $(t,t') = (it_\mathrm{E},it_\mathrm{E}')$ to $(t_\mathrm{R}+it_\mathrm{E},t_\mathrm{R}' + it_\mathrm{E}')$, with $t_\mathrm{R}$ and $t_\mathrm{R}'$ nonzero while keeping $t_\mathrm{E}$ and $t_\mathrm{E}'$ fixed.

We introduce the notation
\begin{equation}
\Phi_\mathrm{I}(x) = 
\begin{cases}
\Phi_\mathrm{I}^{(\mathrm{R})}(x)\eqend{,} & \textrm{if } x\in\mathcal{R}\eqend{,}\\
\Phi_\mathrm{I}^{(\mathrm{L})}(x)\eqend{,} & \textrm{if } x\in\mathcal{L}\eqend{.}\\
\end{cases}
\end{equation}
The Wightman two-point function in the right wedge for the HH state $\Omega_\textrm{HH}^{(0)}$ is defined by
\begin{splitequation}\label{eq:right-wedge-two-point}
 \langle \Omega_{\mathrm{HH}}^{(0)}|\Phi_\mathrm{I}^{(\textrm{R})}(t,\mathbf{x})\Phi_\mathrm{I}^{(\textrm{R})}(t',\mathbf{x}')\Omega_{\textrm{HH}}^{(0)}\rangle\\
 \equiv \lim_{\epsilon\to 0^+}G^{(0)}(t - i\epsilon,\mathbf{x};t',\mathbf{x}')\eqend{,}
\end{splitequation}
with $t,t'\in\mathbb{R}$. A point $(t,\mathbf{y}) \in \mathcal{L}$ with $\mathbf{y}=(\rho,\boldsymbol{\theta})$, $\rho<0$, can be identified with $\big(-t - \tfrac{i\beta_\textrm{H}}{2},\iota(\mathbf{y})\big)$, where 
\begin{equation}\label{eq:definition-of-iota2}
\iota(\mathbf{y}) \equiv (|\rho|,\boldsymbol{\theta})\eqend{,}   
\end{equation}
since 
\begin{splitequation}\label{eq:relation-btw-left-and-right}
 & \left(|\rho|\sinh\left[\kappa(-t - \tfrac{i\beta_\textrm{H}}{2})\right],|\rho|\cosh\left[\kappa(-t - \tfrac{i\beta_\textrm{H}}{2})\right]\right)\\
 &\;\;\; = \big(-\rho\sinh(\kappa t),\rho\cosh(\kappa t)\big)
\end{splitequation}
[see Eq.~\eqref{eq:TX-left-right} for the definition of $t$ and $\rho$ in the left wedge].  Thus, in the case one of the points is in the left wedge, we can define
\begin{splitequation}\label{eq:left-right-two-point}
 \langle \Omega_{\mathrm{HH}}^{(0)}|\Phi_\mathrm{I}^{(\textrm{L})}(t,\mathbf{y})\Phi_\mathrm{I}^{(\textrm{R})}(t',\mathbf{x}')\Omega_{\textrm{HH}}^{(0)}\rangle\\
 \equiv G^{(0)}(-t - \tfrac{i\beta_\textrm{H}}{2},\iota(\mathbf{y});t',\mathbf{x}')\eqend{,}
\end{splitequation}
where $\iota(\vt{y})$ is defined by Eq.~\eqref{eq:definition-of-iota2}.
If both points are in the left wedge, then by the symmetry of the spacetime we have
\begin{splitequation}\label{eq:L-L-case-with-HH}
 & \langle \Omega_{\mathrm{HH}}^{(0)}|\Phi_\mathrm{I}^{(\textrm{L})}(t,\mathbf{y})\Phi_\mathrm{I}^{(\textrm{L})}(t',\mathbf{y}')\Omega_{\textrm{HH}}^{(0)}\rangle\\ 
 &\;\;\; \equiv \lim_{\epsilon\to 0^+}G^{(0)}(t - i\epsilon,\iota(\mathbf{y});t',\iota(\mathbf{y}'))\eqend{.}
\end{splitequation}
Since the two-point function depends on $t$ and $t'$ only through $t-t'$, we can write
\begin{splitequation}\label{eq:left-left-two-point}
 & \langle \Omega_{\mathrm{HH}}^{(0)}|\Phi_\mathrm{I}^{(\textrm{L})}(t,\mathbf{y})\Phi_\mathrm{I}^{(\textrm{L})}(t',\mathbf{y}')\Omega_{\textrm{HH}}^{(0)}\rangle \\
 & \equiv \lim_{\epsilon \to 0^+} G^{(0)}(-t' - \tfrac{i\beta_\textrm{H}}{2}-i\epsilon,\iota(\mathbf{y}');-t - \tfrac{i\beta_\textrm{H}}{2},\iota(\mathbf{y}))\eqend{,}
\end{splitequation}
which is more suggestive because of Eq.~\eqref{eq:relation-btw-left-and-right}. The function $\langle\Omega_{\textrm{HH}}^{(0)}|\Phi_\mathrm{I}(x)\Phi_\mathrm{I}(x')\Omega_{\textrm{HH}}^{(0)}\rangle$ for $x$ or $x'$ in the future region $\mathcal{F}$ or past region $\mathcal{P}$ can be determined by the Cauchy evolution from $G^{(0)}(x,x')$ with $x,x'\in\mathcal{R}\cup \mathcal{L}$.  

The state $\Omega_\textrm{HH}^{(0)}$ is defined to be quasifree with vanishing one-point function: $\langle \Omega_\textrm{HH}^{(0)}|\Phi_\mathrm{I}(x)\Omega_\textrm{HH}^{(0)}\rangle = 0$.  That is, the $N$-point function for $N$ odd vanishes and that for $N$ even is defined as if it obeyed Wick's theorem. Thus, if $N$ is even, it is defined as a sum of products of the two-point functions as follows.  Let $S$ be the set of permutations $\sigma$ of $\{1,2,\ldots,N\}$ such that $\sigma(2i-1) < \sigma(2i)$ and that $\sigma(2i-1) < \sigma(2j-1)$ if $i < j$ for $i,j \in \mathbb{N}$. Then we define 
\begin{splitequation}\label{eq:wicks-theorem}
& \langle \Omega_\textrm{HH}^{(0)}|\Phi_\mathrm{I}(x_{1})\Phi_\mathrm{I}(x_{2})\cdots \Phi_\mathrm{I}(x_{N})\Omega_\textrm{HH}^{(0)}\rangle \\
& = \sum_{\sigma\in S}G^{(0)}(x_{\sigma(1)};x_{\sigma(2)}) \\
& \;\;\;\;\;\;\;\;\;\;\times G^{(0)}(x_{\sigma(3)};x_{\sigma(4)})\cdots G^{(0)}(x_{\sigma(N-1)};x_{\sigma(N)})\eqend{,}
\end{splitequation}
where we have written
\begin{equation}
  \langle\Omega_\textrm{HH}^{(0)}|\Phi_\mathrm{I}(x)\Phi_\mathrm{I}(x')\Omega_\textrm{HH}^{(0)}\rangle \equiv G^{(0)}(x;x')\eqend{,}
\end{equation}
for simplicity, with the understanding that the time coordinates $t$ and $t'$ of $x$ and $x'$, respectively, have infinitesimal imaginary parts satisfying $\textrm{Im}(t-t') < 0$. For example, the four-point function reads
\begin{splitequation}
& \langle\Omega_\textrm{HH}^{(0)}|\Phi_\mathrm{I}(x_1)\Phi_\mathrm{I}(x_2)\Phi_\mathrm{I}(x_3)\Phi_\mathrm{I}(x_4)\Omega_\textrm{HH}^{(0)}\rangle \\
& = G^{(0)}(x_1;x_2)G^{(0)}(x_3;x_4) + G^{(0)}(x_1;x_3)G^{(0)}(x_2;x_4)\\
&\phantom{=} + G^{(0)}(x_1;x_4)G^{(0)}(x_2;x_3)\eqend{.}
\end{splitequation}

Now we discuss the double KMS state $\Omega_\beta^{(0)}$, which later will be compared to the HH state $\Omega_\textrm{HH}^{(0)}$. Thus, consider a complete set of positive-frequency solutions to the Klein-Gordon equation~\eqref{eq:KG-equation_2}, given by $\phi_{\omega\sigma}(\vt{x})e^{-i\omega t}$, $\omega > 0$, and choose $\phi_{\omega\sigma}(\mathbf{x})$ to be real.  Here $\sigma$ represents all the labels of the solutions other than $\omega$.  The differential equation satisfied by $\phi_{\omega\sigma}(\vt{x})$ can be found from the Klein-Gordon equation~\eqref{eq:KG-equation_2} as
\begin{equation}
 \Big[\rho fD_a \rho f D^a - m^2(\rho f)^2 \Big]\phi_{\omega\sigma}(\mathbf{x}) = -\frac{\omega^2}{\kappa^2} \phi_{\omega\sigma}(\mathbf{x})\eqend{,}
\end{equation}
where $D_a$ is the covariant derivative compatible with the spatial metric $h_{ab} \equiv f^2(d\rho)_a(d\rho)_b + s_{ab}$. One can show that these functions with different values of $\omega$ are orthogonal to each other with respect the measure $\sqrt{s}/\rho$, noting that the determinant of the spatial metric is $\sqrt{s}f$, where $s = \textrm{det}(s_{ab})$. 

We normalize the functions $\phi_{\omega\sigma}(\vt{x})$ by requiring 
\begin{equation}\label{eq:normalization-chosen2}
    \frac{1}{\kappa}\int \frac{d\rho\, d^{n-2}\boldsymbol{\theta}\sqrt{s}}{\rho}\phi_{\omega\sigma}(\mathbf{x})\phi_{\omega'\sigma'}(\mathbf{x}) = \delta_{\sigma\sigma'}\delta(\omega-\omega')\eqend{.}
\end{equation}
Here we assume that the labels $\sigma$ are discrete, but the generalization to the continuous case is straightforward.
The completeness of $\phi_{\omega\sigma}(\mathbf{x})$ reads
\begin{equation}\label{eq:completeness-of-phi}
    \int_0^\infty d\omega\,\sum_\sigma\phi_{\omega\sigma}(\mathbf{x})\phi_{\omega\sigma}(\mathbf{x}') = \frac{\kappa\rho}{\sqrt{s}}\delta(\mathbf{x};\mathbf{x}')\eqend{,}
\end{equation}
where Dirac's delta-function $\delta(\mathbf{x};\mathbf{x}')$ is defined in the same way as $\delta(x_\mathrm{E};x_\mathrm{E}')$ in Eq.~\eqref{eq:dirac-delta-definition}. The field operator $\Phi_\mathrm{I}^{(\mathrm{R})}(x)$ can be expanded as
\begin{splitequation}
    \Phi^{(\mathrm{R})}_\mathrm{I}(t,\vt{x})  
		& = \int_0^\infty \frac{d\omega}{\sqrt{2\omega}} \sum_\sigma \left\{ \phi_{\omega\sigma}(\vt{x})e^{-i\omega t}a^{(\mathrm{R})}_\sigma(\omega) \right.\\
    &\phantom{=} \qquad\qquad\left.+ \phi_{\omega\sigma}(\vt{x}) e^{i\omega t}a_\sigma^{(\mathrm{R})\dagger}(\omega)\right\}\eqend{.}
\end{splitequation}
We have chosen the normalization of the functions $\phi_{\omega\sigma}(\mathbf{x})$ in Eq.~\eqref{eq:normalization-chosen2} so that 
\begin{equation}
    \left[a^{(\mathrm{R})}_\sigma(\omega),a^{(\mathrm{R})\dagger}_{\sigma'}(\omega')\right] = \delta_{\sigma\sigma'}\delta(\omega-\omega')\eqend{,}
\end{equation} 
with all other commutators vanishing. The vacuum state in the right wedge $\Psi_\mathrm{R}^{(0)}$ is defined by $a_{\omega\sigma}^{(\mathrm{R})}\Psi_\mathrm{R}^{(0)} = 0$ for all $\omega$ and $\sigma$.

For $x=(t,\mathbf{x}) \in\mathcal{L}$ the field can be expanded as
\begin{splitequation}
    \Phi^{(\mathrm{L})}_\mathrm{I}(t,\vt{x})  
		& =  \int_0^\infty \frac{d\omega}{\sqrt{2\omega}} \sum_\sigma \left\{ \phi_{\omega\sigma}(\iota(\mathbf{x}))e^{-i\omega t}a^{(\mathrm{L})}_\sigma(\omega) \right.\\
    &\phantom{=} \qquad\qquad\left. + \phi_{\omega\sigma}(\iota(\mathbf{x})) e^{i\omega t}a_\sigma^{(\mathrm{L})\dagger}(\omega)\right\}\eqend{,}
\end{splitequation}
where the annihilation and creation operators, $a^{(\mathrm{L})}_\sigma(\omega)$ and $a^{(\mathrm{L})\dagger}_\sigma(\omega)$, satisfy 
\begin{equation}
    \left[a^{(\mathrm{L})}_\sigma(\omega),a^{(\mathrm{L})\dagger}_{\sigma'}(\omega')\right] = \delta_{\sigma\sigma'}\delta(\omega-\omega')\eqend{,}
\end{equation} 
with all other commutators vanishing. The vacuum state $\Psi_\mathrm{L}^{(0)}$ in the left wedge is defined by requiring $a_\sigma^{(\mathrm{L})}(\omega)\Psi_\mathrm{L}^{(0)} = 0$ for all $\omega$ and $\sigma$.  We define the static vacuum state $\Psi^{(0)} \in \mathscr{H}\otimes \mathscr{H}$ by
\begin{equation}
    \Psi^{(0)} \equiv \Psi_\mathrm{L}^{(0)}\otimes\Psi_\mathrm{R}^{(0)}\eqend{.}
\end{equation}
From the definition of the antiunitary operator $J$, Eq.~\eqref{eq:J_2}, we conclude that its action on annihilation and creation operators yields
\begin{splitequation}\label{eq:J-a-J}
  Ja^{(\mathrm{R})}_\sigma(\omega)J & = a^{(\mathrm{L})}_\sigma(\omega)\eqend{,}\\
	Ja^{(\mathrm{R})\dagger}_\sigma(\omega)J & = a^{(\mathrm{L})\dagger}_\sigma(\omega)\eqend{,}
\end{splitequation}
respectively.

The double KMS state $\Omega_\beta^{(0)}$ for the field $\Phi_\mathrm{I}(x)$ with inverse temperature $\beta$ is a thermal state if restricted to the right wedge.  That is,
\begin{splitequation}
\langle \Omega_\beta^{(0)}|a^{(\mathrm{R})\dagger}_\sigma(\omega)a_{\sigma'}^{(\mathrm{R})}(\omega')\Omega_\beta^{(0)}\rangle 
& = \frac{1}{e^{\omega\beta} - 1}\delta_{\sigma\sigma'}\delta(\omega-\omega')\eqend{,} \\
\langle \Omega_\beta^{(0)}|a_\sigma^{(\mathrm{R})}(\omega)a^{(\mathrm{R})\dagger}_{\sigma'}(\omega')\Omega_\beta^{(0)}\rangle 
& = \frac{1}{1 - e^{-\omega\beta}}\delta_{\sigma\sigma'}\delta(\omega-\omega')\eqend{,}
\end{splitequation}
with 
\begin{equation}
    \langle\Omega^{(0)}_\beta|a^{(\mathrm{R})}_\sigma(\omega)a^{(\mathrm{R})}_{\sigma'}(\omega')\Omega_\beta^{(0)}\rangle = 0\eqend{.} 
\end{equation}
Using these formulas one finds
\begin{splitequation}\label{eq:two-point-function_2}
 & \langle\Omega_\beta^{(0)}|\Phi_\mathrm{I}^{(\mathrm{R})}(x)\Phi_\mathrm{I}^{(\mathrm{R})}(x')\Omega_\beta^{(0)}\rangle \\
 & = \int_0^\infty \frac{d\omega}{2\omega}\sum_\sigma  \phi_{\omega\sigma}(\mathbf{x})\phi_{\omega\sigma}(\mathbf{x}')
\left[ \frac{e^{-i\omega(t-t')}}{1-e^{-\omega\beta}} + \frac{e^{i\omega(t-t')}}{e^{\omega\beta} - 1}\right]\eqend{,}\\
\end{splitequation}
where $x=(t,\mathbf{x})$ and $x'=(t',\mathbf{x}')$.
This two-point function can be extended to complex values of $t$ and $t'$ with $- \beta < \textrm{Im}(t-t') < 0$ since the $\omega$-integral here is convergent if this condition is satisfied. For these values of $t$ and $t'$ one finds the KMS condition for the free field:
\begin{splitequation}\label{eq:KMS-free-field2}
& \langle \Omega_\beta^{(0)}|\Phi_\mathrm{I}^{(\mathrm{R})}(t'-i\beta,\vt{x}')\Phi_\mathrm{I}^{(\mathrm{R})}(t,\vt{x})\Omega_\beta^{(0)}\rangle\\
&\;\;\; = \langle\Omega_\beta^{(0)}|\Phi_\mathrm{I}^{(\mathrm{R})}(t,\vt{x})\Phi_\mathrm{I}^{(\mathrm{R})}(t',\vt{x}')\Omega_\beta^{(0)}\rangle\eqend{.}
\end{splitequation}

Now, define for $-\beta < \textrm{Im}(t - t') < \beta$ 
\begin{splitequation}\label{eq:definition-ordering}
&\Delta_\beta^{(0)}(t,\mathbf{x};t',\mathbf{x}')\\
& \;\;\; \equiv 
\begin{cases} 
\langle\Omega_\beta^{(0)}|\Phi_\mathrm{I}^{(\mathrm{R})}(x)\Phi_\mathrm{I}^{(\mathrm{R})}(x')\Omega_\beta^{(0)}\rangle & \textrm{if}\ \textrm{Im}(t-t') < 0\eqend{,} \\
\langle\Omega_\beta^{(0)}|\Phi_\mathrm{I}^{(\mathrm{R})}(x')\Phi_\mathrm{I}^{(\mathrm{R})}(x)\Omega_\beta^{(0)}\rangle & \textrm{if}\ \textrm{Im}(t'-t) < 0\eqend{.} 
\end{cases}
\end{splitequation}
 If $\vt{x}\neq \vt{x}'$, this function is in fact an analytic function in the strip $-\beta < \textrm{Im}(t - t') < \beta$ with branch cuts on the real line $\textrm{Im}(t-t')=0$~\cite{fulling_ruijsenaars_pr_1987}. [It is analytic for $\textrm{Im}(t - t')=0$ as well, if $(t_\mathrm{R},\vt{x})$ and $(t_\mathrm{R}',\vt{x}')$, where $t_\mathrm{R}$ and $t_\mathrm{R}'$ are the real parts of $t$ and $t'$, respectively, are spacelike separated.] On these branch cuts, we have from Eq.~\eqref{eq:definition-ordering} that
\begin{equation}\label{eq:time-ordered_2}
\lim_{\epsilon\to 0^+} \Delta^{(0)}_\beta(t-i\epsilon,\mathbf{x};t',\mathbf{x}')
= \langle\Omega_\beta^{(0)}|\Phi^{(\mathrm{R})}_\mathrm{I}(x)\Phi^{(\mathrm{R})}_\mathrm{I}(x')\Omega_\beta^{(0)}\rangle
\end{equation}
and
\begin{equation}\label{eq:anti-time-ordered_2}
\lim_{\epsilon\to 0^+} \Delta^{(0)}_\beta(t+i\epsilon,\mathbf{x};t',\mathbf{x}')
= \langle\Omega_\beta^{(0)}|\Phi^{(\mathrm{R})}_I(x')\Phi^{(\mathrm{R})}_\mathrm{I}(x)\Omega_\beta^{(0)}\rangle\eqend{.}
\end{equation}

Equation~\eqref{eq:KMS-free-field2} implies that
\begin{equation}
    \Delta^{(0)}_\beta(t,\mathbf{x};t'-i\beta,\mathbf{x}') = \Delta^{(0)}_\beta(t,\mathbf{x};t',\mathbf{x}')\eqend{,}
\end{equation}
for $-\beta < \textrm{Im}(t-t') < 0$ since $-\beta < \textrm{Im}[(t' - i\beta) - t] < 0$.  Therefore, this two-point function is periodic in $t - t'$, with period $i\beta$ in the strip $-\beta < \textrm{Im}(t - t') < \beta$. Hence, it is an analytic function of $t - t'$ on the whole complex plane with period $i\beta$, with branch cuts on the lines $\textrm{Im}(t - t') = n\beta$, where $n \in \mathbb{Z}$. For this reason, we shall regard the two-point function $\Delta^{(0)}_\beta(x,x')$ as a function on $\mathcal{C}_\beta\times \mathcal{C}_\beta$, where $\mathcal{C}_\beta$ is the complex plane quotiented by the equivalence relation $\sim$ defined by
\begin{equation}
    t \sim t + i\beta\eqend{.}
\end{equation}
That is,
\begin{equation}\label{eq:definition-of-C_beta}
    \mathcal{C}_\beta \equiv \mathbb{C}/\sim\eqend{.}
\end{equation}

Now, let us assume that our double KMS state is at the Hawking temperature, i.e., let us take $\beta = \beta_\textrm{H}$. Assuming the Euclidean times $t_\mathrm{E}$ and $t_\mathrm{E}'$ to satisfy $0 < |t_\mathrm{E}-t_\mathrm{E}'| < \beta$, we use Eq.~\eqref{eq:two-point-function_2} to show that
\begin{splitequation}\label{eq:greens-function-expanded2}
& \Delta^{(0)}_{\beta_\textrm{H}}(it_\mathrm{E},\mathbf{x};it'_\mathrm{E},\mathbf{x}') \\
& = \int_0^\infty \frac{d\omega}{2\omega}\sum_\sigma  \phi_{\omega\sigma}(\mathbf{x})\phi_{\omega\sigma}(\mathbf{x}')
\left[ \frac{e^{-\omega|t_\mathrm{E}-t'_\mathrm{E}|}}{1-e^{-\omega\beta_\textrm{H}}} + \frac{e^{\omega|t_\mathrm{E}-t'_\mathrm{E}|}}{e^{\omega\beta_\textrm{H}} - 1}\right]\eqend{.}
\end{splitequation}
This two-point function is defined on the Euclidean section $\mathcal{M}_\mathrm{E}$ with the metric \eqref{eq:euclidean-polar-2}.
One can readily verify that this function satisfies the equation for the Green's function on $\mathcal{M}_\mathrm{E}$, Eq.~\eqref{eq:greens-differential-eq}, by directly differentiating Eq.~\eqref{eq:greens-function-expanded2} and using Eq.~\eqref{eq:completeness-of-phi}. Since $\Delta_{\beta_\textrm{H}}^{(0)}$ and $G^{(0)}$ both satisfy the field equation~\eqref{eq:greens-differential-eq} with the same boundary conditions in the Euclidean section, it follows that     
\begin{equation}\label{eq:delta-beta-equals-G}
\Delta^{(0)}_{\beta_\mathrm{H}}(it_\mathrm{E},\vt{x};it_\mathrm{E}',\vt{x}') = G^{(0)}(it_\mathrm{E},\vt{x};it_\mathrm{E}',\vt{x}')\eqend{.}
\end{equation}

Finally, we can show that the double KMS state at the Hawking temperature $\Omega_{\beta_\textrm{H}}^{(0)}$ is nothing but the HH state $\Omega_\textrm{HH}^{(0)}$ by checking that these states have the same two-point functions. Indeed, if $-\beta<\textrm{Im}(t - t') < 0$, then Eq.~\eqref{eq:delta-beta-equals-G} yields
\begin{equation}\label{eq:kms-equals-HH-0}
    \langle\Omega_{\beta_\textrm{H}}^{(0)}|\Phi_\mathrm{I}^{(\mathrm{R})}(x)\Phi_\mathrm{I}^{(\mathrm{R})}(x')\Omega_{\beta_\textrm{H}}^{(0)}\rangle = G^{(0)}(x,x')\eqend{.}
\end{equation}
This in turn implies
\begin{splitequation}\label{eq:relation-R-R}
  &  \langle \Omega_{\beta_\textrm{H}}^{(0)}|\Phi_\mathrm{I}^{(\textrm{R})}(t,\mathbf{x})\Phi_\mathrm{I}^{(\textrm{R})}(t',\mathbf{x}')\Omega_{\beta_\textrm{H}}^{(0)}\rangle\\
  &\;\;\; = \lim_{\epsilon\to 0^+}G^{(0)}(t - i\epsilon,\mathbf{x};t',\mathbf{x}') \\
  &\;\;\; = \langle \Omega_\textrm{HH}^{(0)}|\Phi_\mathrm{I}^{(\textrm{R})}(t,\mathbf{x})\Phi_\mathrm{I}^{(\textrm{R})}(t',\mathbf{x}')\Omega_\textrm{HH}^{(0)}\rangle \eqend{,}
\end{splitequation}
where the last equality follows from Eq.~\eqref{eq:right-wedge-two-point}.

Next, we note that if $x\in\mathcal{L}$ and $x'\in\mathcal{R}$, then 
\begin{splitequation}\label{eq:free_two_point_funct_aux}
 &  \langle \Omega_{\beta_\textrm{H}}^{(0)}|\Phi_\mathrm{I}^{(\mathrm{L})}(x)\Phi_\mathrm{I}^{(\mathrm{R})}(x')\Omega_{\beta_\textrm{H}}^{(0)}\rangle  \\
 &\;\;\; = \langle J\Phi_\mathrm{I}^{(\mathrm{R})}(I(x))\Omega_{\beta_\textrm{H}}^{(0)}|\Phi_\mathrm{I}^{(\mathrm{R})}(x')\Omega_{\beta_\textrm{H}}^{(0)}\rangle\eqend{,}
\end{splitequation}
where we have used the definition of the operator $J$, Eq.~\eqref{eq:J_2}. Furthermore, the KMS condition~\eqref{eq:kms_condition} yields
\begin{equation}
J\Phi_\mathrm{I}^{(\mathrm{R})}(I(x))\Omega_\beta^{(0)} = e^{-\frac{1}{2}\beta\tilde{H}_0}\Phi_\mathrm{I}^{(\mathrm{R})}(-t,\iota(\mathbf{x}))\Omega_\beta^{(0)}\eqend{,}
\end{equation}
where $\tilde{H}_0 \equiv \mathbbm{1}\otimes H_0 - H_0 \otimes \mathbbm{1}$, with $H_0$ denoting the free Hamiltonian operator that generates the time translation in the right or left wedge. Going back to Eq.~\eqref{eq:free_two_point_funct_aux}, and recalling that $\Omega_\beta^{(0)}$ is annihilated by $\tilde{H}_0$, we use the above result to show that
\begin{splitequation}
 & \langle \Omega_{\beta_\textrm{H}}^{(0)}|\Phi_\mathrm{I}^{(\mathrm{L})}(x)\Phi_\mathrm{I}^{(\mathrm{R})}(x')\Omega_{\beta_\textrm{H}}^{(0)}\rangle \\
 &\;\;\; = \langle\Omega_{\beta_\textrm{H}}^{(0)}|e^{\frac{\beta_\textrm{H}}{2}\tilde{H}_0}\Phi_\mathrm{I}^{(\mathrm{R})}(-t,\iota(\mathbf{x}))e^{-\frac{\beta_\textrm{H}}{2}\tilde{H}_0}\Phi_\mathrm{I}^{(\mathrm{R})}(x')\Omega_{\beta_\textrm{H}}^{(0)}\rangle \\
 &\;\;\; = \langle\Omega_{\beta_\textrm{H}}^{(0)}|\Phi_\mathrm{I}^{(\mathrm{R})}(-t - \tfrac{i\beta_\textrm{H}}{2},\iota(\mathbf{x}))\Phi_\mathrm{I}^{(\mathrm{R})}(t',\mathbf{x}')\Omega_{\beta_\textrm{H}}^{(0)}\rangle \eqend{.}
   \label{eq:relation-L-R}
\end{splitequation}
Together with Eq.~\eqref{eq:kms-equals-HH-0}, this last result implies that
\begin{splitequation}
 & \langle \Omega_{\beta_\textrm{H}}^{(0)}|\Phi_\mathrm{I}^{(\textrm{L})}(t,\mathbf{x})\Phi_\mathrm{I}^{(\textrm{R})}(t',\mathbf{x}')\Omega_{\beta_\textrm{H}}^{(0)}\rangle \\
 &\;\;\; = G^{(0)}(-t - \tfrac{i\beta_\textrm{H}}{2},\iota(\mathbf{x});t',\mathbf{x}') \\
 &\;\;\; = \langle \Omega_\textrm{HH}^{(0)}|\Phi_\mathrm{I}^{(\textrm{L})}(t,\mathbf{x})\Phi_\mathrm{I}^{(\textrm{R})}(t',\mathbf{x}')\Omega_\textrm{HH}^{(0)}\rangle\eqend{,}\label{eq:to-be-shown-LR}
\end{splitequation}
where the last equality follows from Eq.~\eqref{eq:left-right-two-point}. By using Eq.~\eqref{eq:two-point-function_2}, we can write this two-point function explicitly as a mode sum, i.e.,
\begin{splitequation}
& \langle \Omega_{\beta_\textrm{H}}^{(0)}|\Phi_\mathrm{I}^{(\textrm{L})}(t,\mathbf{x})\Phi_\mathrm{I}^{(\textrm{R})}(t',\mathbf{x}')\Omega_{\beta_\textrm{H}}^{(0)}\rangle \\
& = \int_0^\infty \frac{d\omega}{2\omega}\sum_\sigma \phi_{\omega\sigma}(\iota(\mathbf{x}))\phi_{\omega\sigma}(\mathbf{x}') \\
&\phantom{=} \;\;\;\;\;\;\;\times \frac{1}{2\sinh\frac{\omega\beta_\textrm{H}}{2}}\left[e^{-i\omega(t + t')} + e^{i\omega(t + t')}\right]\eqend{.}
\end{splitequation}

Finally, if $x,x'\in\mathcal{L}$, it immediately follows from the symmetry of $\Omega_\beta^{(0)}$ that
\begin{splitequation}\label{eq:to-be-shown-LL}
&  \langle \Omega_{\beta_\textrm{H}}^{(0)}|\Phi_\mathrm{I}^{(\textrm{L})}(t,\mathbf{x})\Phi_\mathrm{I}^{(\textrm{L})}(t',\mathbf{x}')\Omega_{\beta_\textrm{H}}^{(0)}\rangle \\
&\;\;\;  = \langle \Omega_{\beta_\textrm{H}}^{(0)}|\Phi_\mathrm{I}^{(\textrm{R})}(t,\iota(\mathbf{x}))\Phi_\mathrm{I}^{(\textrm{R})}(t',\iota(\mathbf{x}'))\Omega_{\beta_\textrm{H}}^{(0)}\rangle \\
&\;\;\; = \lim_{\epsilon\to 0^+}G^{(0)}(t - i\epsilon,\iota(\vt{x});t',\iota(\vt{x})) \\
&\;\;\; = \langle \Omega_\textrm{HH}^{(0)}|\Phi_\mathrm{I}^{(\textrm{L})}(t,\mathbf{x})\Phi_\mathrm{I}^{(\textrm{L})}(t',\mathbf{x}')\Omega_\textrm{HH}^{(0)}\rangle\eqend{,}
\end{splitequation}
where the last equality follows from Eq.~\eqref{eq:L-L-case-with-HH}.

Equations~\eqref{eq:relation-R-R}, \eqref{eq:to-be-shown-LR} and \eqref{eq:to-be-shown-LL} show that, if $x,x'\in \mathcal{R}\cup \mathcal{L}$, then
\begin{equation}
    \langle\Omega_{\beta_\mathrm{H}}^{(0)}|\Phi_\mathrm{I}(x)\Phi_\mathrm{I}(x')\Omega_{\beta_\mathrm{H}}^{(0)}\rangle = \langle\Omega_\textrm{HH}^{(0)}|\Phi_\mathrm{I}(x)\Phi_\mathrm{I}(x')\Omega_\textrm{HH}^{(0)}\rangle\eqend{.} 
\end{equation}
The two-point function for any points in the spacetime $(\mathcal{M},g_{ab})$ outside the double-wedged region $\mathcal{R}\cup \mathcal{L}$ can be uniquely determined from this case by the Cauchy evolution.  Since the double KMS state $\Omega_\beta^{(0)}$ is quasifree for any $\beta$~\cite{kay_hpa_1985second,kay_cmp_1985}, all its $N$-point functions are determined by its two-point function.  Hence, we conclude that 
\begin{equation}
 \Omega_{\beta_\textrm{H}}^{(0)} = \Omega_\textrm{HH}^{(0)}
\end{equation} 
for the free-field theory.

If the mode functions $\phi_{\omega\sigma}(\mathbf{x})$ are analytic in $\rho$, then one should be able to find the mode expansion for the field $\Phi_\mathrm{I}(x)$ and its two-point function also for $x\in \mathcal{F}\cup \mathcal{P}$ by analytic continuation of the corresponding expressions in $\mathcal{R}\cup \mathcal{L}$. Assuming that this analytic continuation is analogous to the special cases of Minkowski and de~Sitter spacetimes~\cite{yamamoto_prd_minkowski,yamamoto_prd_de_sitter}, we can write down the two-point function with one or two points in $\mathcal{F}\cup \mathcal{P}$. As an example, let us consider the case when the two-point function has points in $\mathcal{R}\cup \mathcal{F}$. Note that
\begin{equation}
    (T,X) = \big(\rho\cosh(\kappa t),\rho\sinh(\kappa t)\big)\ \ \textrm{for}\ (T,X)\in\mathcal{F}\eqend{,}
\end{equation}
with $\rho>0$. Define $\tilde{\phi}_{\omega\sigma}(\mathbf{x})$ to be the function obtained from $\phi_{\omega\sigma}(\mathbf{x})$ by making the substitution $\rho \to e^{\frac{i\pi}{2}}\rho$.  Then, by defining $\Phi_\mathrm{I}^{(\mathrm{F})}(x) = \Phi_\mathrm{I}(x)$ for $x\in\mathcal{F}$, we find
\begin{splitequation}
& \langle \Omega_{\beta_\mathrm{H}}^{(0)}|\Phi_\mathrm{I}^{(\textrm{F})}(t,\mathbf{x})\Phi_\mathrm{I}^{(\textrm{F})}(t',\mathbf{x}')\Omega_{\beta_\mathrm{H}}^{(0)}\rangle = \\
&\int \frac{d\omega}{2\omega}\sum_\sigma \tilde{\phi}_{\omega\sigma}(\mathbf{x})\overline{\tilde{\phi}_{\omega\sigma}(\mathbf{x}')} \frac{e^{-i\omega(t - t')} + e^{i\omega(t - t')}}{2\sinh\frac{\omega\beta_\mathrm{H}}{2}}\eqend{,} \label{eq:first-eq-for-appB}
\end{splitequation}
and
\begin{splitequation}\label{eq:second-eq-for-appB}
& \langle \Omega_{\beta_\mathrm{H}}^{(0)}|\Phi_\mathrm{I}^{(\textrm{F})}(t,\mathbf{x})\Phi_\mathrm{I}^{(\textrm{R})}(t',\mathbf{x}')\Omega_{\beta_\mathrm{H}}^{(0)}\rangle = \\
& \int \frac{d\omega}{2\omega}\sum_\sigma \tilde{\phi}_{\omega\sigma}(\mathbf{x})\phi_{\omega\sigma}(\mathbf{x}')\frac{e^{\frac{\omega\beta}{4}}e^{-i\omega(t - t')} + e^{-\frac{\omega\beta}{4}}e^{i\omega(t - t')}}{2\sinh\frac{\omega\beta_\mathrm{H}}{2}}\eqend{.}
\end{splitequation}
A derivation of these formulas, together with the assumptions we made, can be found in Appendix~\ref{appendix:future-right}.

\subsection{The interacting case}
\label{subsection:interacting}
%%%%%%%%%%%%%%%%%%%%%%%%%%%%%%%%%%%%%%%%%%%%%%%%%%%%%%%%%%%%%%%%%%%%%%%%%%%%%%%%%%%%%%%%%%%%%%%%%%%%%%%%%%%%%%%%%%%%%%%%%%%%%%%%%%%%%%%%%%%%%%%%%%%%%

We will now consider a massive, real scalar field theory with nonderivative self-interaction in spacetimes with a bifurcate Killing horizon considered in Sec.~\ref{sec:preliminaries}. Our aim is to show that a double KMS state at the Hawking temperature~\eqref{eq:the-right-temperature} is the HH state for this interacting theory.

The interacting HH state will be defined here via the Euclidean perturbation theory~\cite{gibbons_perry_prsl_1978}. We assume that the Hamiltonian $H$ in either right or left wedge can be written as a free part $H_0$ plus an interaction perturbation term, 
\begin{equation}
    H = H_0 + H_\mathrm{I}\eqend{,}
\end{equation}
where
\begin{equation}\label{eq:interaction-hamlitonian-as-integral}
    H_\mathrm{I} \equiv \int_{\Sigma^\mathrm{R}}d\rho d^{n-2}\boldsymbol{\theta}\,\kappa\rho f(\rho^2,\boldsymbol{\theta})
		\sqrt{s(\boldsymbol{\theta})}\,\mathcal{H}_\mathrm{I}(\Phi)
\end{equation}
is the interaction Hamiltonian, with $\mathcal{H}_\mathrm{I}(\Phi)$ as a polynomial in $\Phi$.  The $\Phi^4$-theory, for example, is given by $\mathcal{H}_\mathrm{I}(\Phi) = (\lambda/4!)\Phi^4$, where $\lambda$ is a real constant.  Then, the Euclidean $N$-point function for the Heisenberg operator $\Phi(x)$ is given by
\begin{widetext}
\begin{splitequation}\label{eq:euclidean-N-pt-fun2}
 &  G(x_{\mathrm{E},1},x_{\mathrm{E},2},\dots,x_{\mathrm{E},N})  \\
 & = \Big{\langle}\Phi_\mathrm{I}(x_{\mathrm{E},1})\Phi_\mathrm{I}(x_{\mathrm{E},2})\cdots \Phi_\mathrm{I}(x_{\mathrm{E},N})\exp\left(  \int_{\mathcal{M}_\mathrm{E}}d^n x_\mathrm{E} \sqrt{g^\mathrm{E}}\,\mathcal{H}_\mathrm{I}(\Phi_\mathrm{I})\right)\Big{\rangle}_\mathrm{E}\Big{/}
 \Big{\langle}\exp\left(  \int_{\mathcal{M}_\mathrm{E}}d^n x_\mathrm{E} \sqrt{g^\mathrm{E}}\,\mathcal{H}_\mathrm{I}(\Phi_\mathrm{I})\right)\Big{\rangle}_\mathrm{E} \\
 & = \Big{\langle}\Phi_\mathrm{I}(x_{\mathrm{E},1})\Phi_\mathrm{I}(x_{\mathrm{E},2})\cdots \Phi_\mathrm{I}(x_{\mathrm{E},N})\exp\left(  \int_{\mathcal{M}_\mathrm{E}}d^n x_\mathrm{E} \sqrt{g^\mathrm{E}}\,\mathcal{H}_\mathrm{I}(\Phi_\mathrm{I})\right)\Big{\rangle}_\textrm{E, connected}
\eqend{,}
\end{splitequation}
and we recall that $x_\mathrm{E} = (it_\mathrm{E},\vt{x})$. In the notation of Eq.~\eqref{eq:euclidean-N-pt-fun2}, $\langle\,\dots\,\rangle_\mathrm{E}$ denotes the expectation value of the interaction field operator $\Phi_\mathrm{I}(x_\mathrm{E})$ such that
\begin{splitequation}
 \langle \Phi_\mathrm{I}(x_\mathrm{E})\rangle_\mathrm{E} & = 0\eqend{,}\\
 \langle \Phi_\mathrm{I}(x_{\mathrm{E},1})\Phi_\mathrm{I}(x_{\mathrm{E},2})\rangle_\mathrm{E} & = G^{(0)}(x_{\mathrm{E},1},x_{\mathrm{E},2})
\end{splitequation}
and the expectation value of a higher number of field is obtained with Wick's theorem, while ``connected'' indicates the sum of diagrams with all parts connected to some of the external points $x_{\mathrm{E},1}$, $x_{\mathrm{E},2}$,$\ldots$,$x_{\mathrm{E},N}$ in the diagrammatic expansion.  The integral is parametrized as
\begin{equation}
    \int_{\mathcal{M}_\mathrm{E}} d^n x_\mathrm{E}  =  \int_0^{-\beta_\mathrm{H}}dt_\mathrm{E}\int_{\Sigma^\mathrm{R}} d^{n-1}\mathbf{x} =  - \int_{-\beta_\mathrm{H}}^0 dt_\mathrm{E}\int_{\Sigma^\mathrm{R}} d^{n-1}\mathbf{x}\eqend{.}
\end{equation}

For the HH state $\Omega_{\textrm{HH}}$, the $N$-point functions are given by the analytic continuation of Eq.~\eqref{eq:euclidean-N-pt-fun2} to the Lorentzian section of the spacetime.  Let us now describe how this analytic continuation is carried out.  For this purpose it is useful to write Eq.~\eqref{eq:euclidean-N-pt-fun2} as follows:
\begin{equation}\label{eq:expression-of-N-pt-fun2}
 G(x_{\mathrm{E},1},x_{\mathrm{E},2},\dots,x_{\mathrm{E},N}) = \Big{\langle}\Phi_\mathrm{I}(x_{\mathrm{E},1})\Phi_\mathrm{I}(x_{\mathrm{E},2})\cdots \Phi_\mathrm{I}(x_{\mathrm{E},N})\exp\left(  -i\int_C dt \int_{\Sigma^\mathrm{R}} d^{n-1}\vt{x}\, \sqrt{-g}\,\mathcal{H}_\mathrm{I}(\Phi_\mathrm{I})\right)\Big{\rangle}_\textrm{E, connected}\eqend{,}
\end{equation}
\end{widetext}
where the directed contour $C$ is the straight line segment from $0$ to $-i\beta_\mathrm{H}$, and corresponds to a circle in $\mathcal{C}_{\beta_\mathrm{H}}$, see Eq.~\eqref{eq:definition-of-C_beta}.  We also have used the fact that $g^\mathrm{E} = -g$.  Note that $x_{\mathrm{E},i} \in C\times \Sigma^\mathrm{R}$, with $i=1,2,\dots,N$. 

These $N$-point functions are ultraviolet divergent in general because the two-point function $G^{(0)}(x,x')$ diverges in the coincidence limit $x\to x'$. They need to be regularized and renormalized.  We assume that the regularization is done in such a way that one may first restrict the integration over $\Sigma^\mathrm{R}$ by requiring that for any two internal points $\|\vt{y}_i-\vt{y}_j\| > \epsilon$ and for any external point $\|\vt{x}_i-\vt{y}_j\|>\epsilon$ for all $i$ and $j$ for some $\epsilon >0$, and then take the limit $\epsilon\to 0$ after the counterterms are included to cancel the ultraviolet divergences.   We also assume that the (multiple) vertex integrals over $\Sigma^\mathrm{R}$ can be performed by first cutting them off in the infrared with $\|\mathbf{x}\| < \Lambda$ and then removing the cutoff.  With these assumptions we can show that the $N$-point function in Eq.~\eqref{eq:expression-of-N-pt-fun2} is analytically continued in $t_1, t_2, \ldots, t_N$, by changing the real part of each $t_i$ while keeping its imaginary part the same.

The analytic continuation of the Euclidean $N$-point function $G$ is defined by extending the external points to $x_i=(t_i,\vt{x}_i)$ with $t_i \in \mathcal{C}_\beta$ and satisfying $\mathrm{Im}(t_i - t_j) \neq 0$ for all $i$ and $j$. Hence, in the diagrammatic expansion of the right-hand side of Eq.~\eqref{eq:expression-of-N-pt-fun2}, this analytic continuation amounts to employing the analytically continued Euclidean Green's function $G^{(0)}(x,x')$ for the free field discussed in the previous subsection. In moving the complex times $t_1, t_2,\ldots, t_N$ we have to avoid hitting the branch cuts of the Green's functions $G^{(0)}(x,x')$, with $x$ and $x'$ being various external and internal points forming a given diagram, while performing the vertex integrations, so that the result is the same analytic function of the external points. This entails that we also need to deform the integration contour $C$ appearing in Eq.~\eqref{eq:expression-of-N-pt-fun2}. This contour is deformed in such a way that it contains the external complex-time coordinates $t_1, t_2, \ldots, t_N$ and its imaginary part is monotonically decreasing. We refer the reader to Appendix~\ref{appendix:detail-analytic} for more details on this point.

The $N$-point functions of the HH state $\Omega_\textrm{HH}$ are defined by the analytic continuation of Eq.~\eqref{eq:expression-of-N-pt-fun2} to the right wedge. Thus, taking $x_1,x_2,\ldots, x_N \in \mathcal{R}$ and using Eq.~\eqref{eq:delta-beta-equals-G} we have
\begin{widetext}
\begin{splitequation}\label{eq:general-def-of-Npt-function2}
 \langle \Omega_\textrm{HH}|\Phi(x_1)\Phi(x_2)\cdots\Phi(x_N)\Omega_\textrm{HH}\rangle
 & \equiv G(x_1,x_2,\dots,x_N)\\
 & = \langle\Omega_{\beta_\mathrm{H}}^{(0)}|\mathcal{P}\Big{[}\Phi_\mathrm{I}^{(\mathrm{R})}(x_1)\Phi_\mathrm{I}^{(\mathrm{R})}(x_2)\cdots 
	 \Phi_\mathrm{I}^{(\mathrm{R})}(x_N) \exp\left(-i\int_C dt H_\mathrm{I}^{(\mathrm{R})}(t) \right)\Big{]}\Omega_{\beta_\mathrm{H}}^{(0)}\rangle_\textrm{E, connected}\eqend{,}
\end{splitequation}
where the points $x_1,x_2,\ldots,x_N$ appear in this order on $C$, and $H_\mathrm{I}^{(\mathrm{R})}(t)$ is defined by Eq.~\eqref{eq:interaction-hamlitonian-as-integral} with $\Phi$ replaced by the interaction-picture operator $\Phi_\mathrm{I}^{(\mathrm{R})}$. The path-ordering $\mathcal{P}$ indicates that the operators $\Phi_\mathrm{I}^{(\mathrm{R})}(x)$ are ordered according to the order on $C$, which coincides with the decreasing order in the imaginary part of $t$ in $x=(t,\vt{x})$.
 
To show that the HH state $\Omega_\textrm{HH}$ is a double KMS state at the Hawking temperature, we consider the following $N$-point function:
\begin{equation}
\Delta_\textrm{HH}(y_1,y_2,\ldots,y_L;x_1,x_2,\ldots,x_R) \equiv  \langle \Omega_\textrm{HH}|\Phi(y_1)\Phi(y_2)\cdots \Phi(y_L)\Phi(x_1)\Phi(x_2)\cdots \Phi(x_R)\Omega_\textrm{HH}\rangle\eqend{,}
\end{equation}
\end{widetext}
where $y_1,y_2,\ldots,y_L \in \mathcal{L}$ and $x_1,x_2,\ldots,x_R\in \mathcal{R}$ with $x_i = (t_i,\mathbf{x}_i)$ and $y_i = (\tau_i,\mathbf{y}_i)$, with $R + L = N$. We shall write this $N$-point function in a form that can readily be compared to the $N$-point function of the state $\Omega_{\beta_\mathrm{H}}$. In particular, we write this $N$-point function in terms of the operator $\Phi_\mathrm{I}^{(\mathrm{R})}(x)$.

Assuming that $t_i$ and $\tau_i$ are real, one needs to define this $N$-point function as a limit of the $N$-point function defined by Eq.~\eqref{eq:general-def-of-Npt-function2} in which the imaginary parts of some time variables coincide. For the points in $\mathcal{R}$, Eq.~\eqref{eq:relation-R-R} implies that one should let $t_i \to t_i - i\epsilon_i$ with $\epsilon_i > \epsilon_j$ if $i < j$ in taking the limit $\epsilon_i\to 0^+$.  A point $y=(\tau,\vt{y}) \in \mathcal{L}$ is represented by $\big(-\tau-\frac{i\beta_\mathrm{H}}{2},\iota(\vt{y})\big)$ according to Eqs.~\eqref{eq:left-right-two-point} and \eqref{eq:left-left-two-point}, and the latter equation implies that we should let $\tau_i-\frac{i\beta_\mathrm{H}}{2} \to - \tau_i - \frac{i\beta_\mathrm{H}}{2} - i\varepsilon_i$, where $\varepsilon_i <\varepsilon_j$ if $i < j$ in taking the limit $\varepsilon_i\to 0^+$. (Notice the reversed order here in comparison with the case for the right wedge.) The imaginary parts of the time coordinates dictate the order on $C$, and these points should appear on $C$ in the following order:
\begin{widetext}
\begin{equation}
-\tau_L -\frac{i\beta_\mathrm{H}}{2} \leftarrow -\tau_{L-1}-\frac{i\beta_\mathrm{H}}{2} \leftarrow \cdots \leftarrow -\tau_1 - \frac{i\beta_\mathrm{H}}{2} \leftarrow t_1 \leftarrow t_2 \leftarrow \cdots \leftarrow t_R\eqend{.}
\end{equation}
Thus, we find
\begin{splitequation}\label{eq:hartle-hawking-N-pt-2}
 & \Delta_\textrm{HH}(y_1,y_2,\ldots,y_L;x_1,x_2,\ldots,x_R)\\
 & = \langle\Omega_{\beta_\mathrm{H}}^{(0)}|\mathcal{P}\Bigg{\{}\Phi_\mathrm{I}^{(\mathrm{R})}(-\tau_L-\tfrac{i\beta_\mathrm{H}}{2},\iota(\mathbf{y}_L)\Phi_\mathrm{I}^{(\mathrm{R})}(-\tau_{L-1} - \tfrac{i\beta_\mathrm{H}}{2},\iota(\mathbf{y}_{L - 1}))\cdots\Phi_\mathrm{I}^{(\mathrm{R})}(-\tau_1 - \tfrac{i\beta_\mathrm{H}}{2},\iota(\mathbf{y}_1))\\
 &\phantom{=}\quad \times \Phi_\mathrm{I}^{(\mathrm{R})}(t_1,\mathbf{x}_1)\Phi_\mathrm{I}^{(\mathrm{R})}(t_2,\mathbf{x}_2)\cdots \Phi_\mathrm{I}^{(\mathrm{R})}(t_R,\mathbf{x}_R)
 \exp\left( - i\int_C dt H_\mathrm{I}^{(\mathrm{R})}(t)\right)\Bigg{\}}\Omega_{\beta_\mathrm{H}}^{(0)}\rangle_{\textrm{connected}}\eqend{.}
\end{splitequation}
\end{widetext}

\begin{center}
 \begin{figure}[ht]
   \includegraphics[scale = 1]{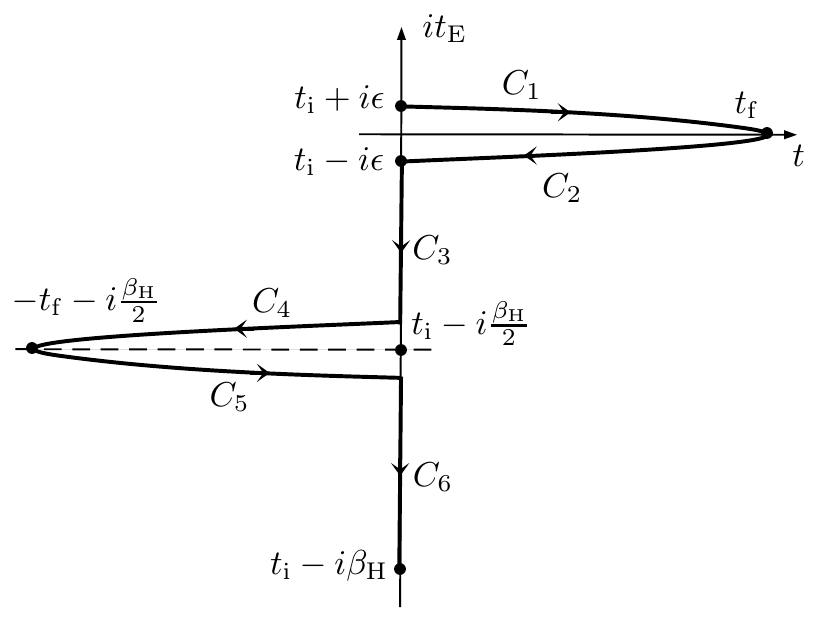}
   \caption{The contour $C$ in the complex Killing-time plane with monotonically decreasing imaginary part. The path $C_1$ runs forward in the real 
	time along the real axis, while the path $C_2$ runs along the real axis but backwards. The path $C_3$ runs from $t_\mathrm{i} - i\epsilon$ down to $t_\mathrm{i} - \frac{i\beta_\mathrm{H}}{2} + i\epsilon$ parallel to the imaginary axis. The path $C_4$ runs backward in the real time along the line $t = -\frac{i\beta_\mathrm{H}}{2}$, while the path $C_5$ runs along the same line but forward. The path $C_6$ runs from $t_\mathrm{i} - \frac{i\beta_\mathrm{H}}{2} - i\epsilon$ to $t_\mathrm{i} - i\beta_\mathrm{H}$ parallel to the imaginary axis. The time $t_\mathrm{i}$ is the initial time and the final time $t_\mathrm{f}$ is assumed to be larger than the time coordinate of any external point on $C$, but otherwise arbitrary. At the end of the computation we let $\epsilon \to 0^+$, so the paths $C_1$, $C_2$, $C_4$, and $C_5$ become parallel to the real axis.}
   \label{fig:sk_path_Killing_time}
 \end{figure}
\end{center}

To illustrate the integration contour for the complex-time coordinate, we consider the special case where the $L$ points $y_1,y_2,\ldots,y_L$ and the $R$ points $x_1,x_2,\ldots,x_R$ have positive time coordinates and are time-ordered, i.e., $\tau_1>\tau_2>\cdots >\tau_L>0$ and $t_1>t_2>\cdots >t_R > 0$. In this case, the contour $C$ can be chosen as shown in Fig.~\ref{fig:sk_path_Killing_time} (with the initial time $t_\mathrm{i}$ set to $0$). The time coordinates $t_i$ (with small positive imaginary parts), $i=1,2,\ldots,R$, satisfying $0 < t_i < t_\mathrm{f}$ are on the path $C_1$ whereas the coordinates $-\tau_j - \frac{i\beta_\mathrm{H}}{2}$, $0 < \tau_j < t_\mathrm{f}$, with further small negative imaginary parts, are on the path $C_5$.

Now, we consider a double KMS state $\Omega_\beta$, defined by Eqs.~(\ref{eq:invariance_kms_state_H})-(\ref{eq:kms_condition}) with respect to the timelike Killing vector field $\xi^a$ and the wedge reflection $I$. For simplicity we consider the $4$-point function of this state with points $y_1$ and $y_2$ in the left wedge and points $x_1$ and $x_2$ in the right wedge. The generalization to the $N$-point function with $N > 4$ is straightforward. Our aim is to express the $4$-point function
\begin{equation} \label{eq:4pt-function-in-KMS}
    \Delta_\beta(y_1,y_2;x_1,x_2) \equiv \langle\Omega_\beta|\Phi(y_1)\Phi(y_2)\Phi(x_1)\Phi(x_2)\Omega_\beta\rangle
\end{equation}
in terms of the interaction-picture field $\Phi_\mathrm{I}^{(\mathrm{R})}(x)$ in the right wedge, so we can compare it to the corresponding HH state $4$-point function for $\beta = \beta_\mathrm{H}$. Hence, we first use Eqs.~\eqref{eq:invariance_kms_state_H}-\eqref{eq:J_2} to write the right-hand side of the $4$-point function above as
\begin{splitequation}\label{eq:interim_4p_function}
& \Delta_\beta(y_1,y_2;x_1,x_2) \\
& = \langle \Phi(y_2)\Phi(y_1)\Omega_\beta|\Phi(x_1)\Phi(x_2)\Omega_\beta\rangle\\
& = \langle J\Phi(I(y_2))\Phi(I(y_1))\Omega_\beta|\Phi(x_1)\Phi(x_2)\Omega_\beta\rangle \\
& = \langle e^{-\frac{1}{2}\beta\tilde{H}}\Phi(I(y_1))\Phi(I(y_2))\Omega_\beta|\Phi(x_1)\Phi(x_2)\Omega_\beta\rangle\\
& = \langle\Omega_\beta|\Phi(I(y_2))\Phi(I(y_1))e^{-\frac{1}{2}\beta\tilde{H}}\Phi(x_1)\Phi(x_2)\Omega_\beta\rangle \\
& = \langle\Omega_\beta|e^{\frac{1}{2}\beta\tilde{H}}\Phi(I(y_2))\Phi(I(y_1))e^{-\frac{1}{2}\beta\tilde{H}}\Phi(x_1)\Phi(x_2)\Omega_\beta\rangle\eqend{.}
\end{splitequation}
To fully express this 4-point function as a correlator of field operators evaluated on the right wedge, we notice that $e^{\pm\frac{1}{2}\beta\tilde{H}} = (e^{\mp\frac{1}{2}\beta H}\otimes \mathbbm{1})(\mathbbm{1}\otimes e^{\pm\frac{1}{2}\beta H})$ and use the fact that the operators $e^{\mp\frac{1}{2}\beta H}\otimes \mathbbm{1}$ commutes with the operators in the right wedge. This allows us to write
\begin{splitequation}\label{eq:4_point_function_trace}
 & \Delta_\beta(y_1,y_2;x_1,x_2)\\
 & = \frac{1}{Z(\beta)}\tr\left\{e^{-\frac{1}{2}\beta H}\Phi^{(\mathrm{R})}(I(y_2))\Phi^{(\mathrm{R})}(I(y_1))\right.\\ 
 & \qquad \qquad \qquad\left. \times e^{-\frac{1}{2}\beta H} \Phi^{(\mathrm{R})}(x_2)\Phi^{(\mathrm{R})}(x_1)\right\}\eqend{,}
\end{splitequation}
where we have (formally) expressed the expectation value in the state $\Omega_\beta$ as the trace over the space of states in the right wedge with the operator $e^{-\beta H}/Z(\beta)$ inserted.

The right-hand side of Eq.~\eqref{eq:4_point_function_trace} is a correlator involving the Heisenberg field operator $\Phi(t,\mathbf{x})$. In the right wedge, this operator evolves in time with the full Hamiltonian $H = H_0 + H_\mathrm{I}$ according to
\begin{equation}\label{eq:def-heisenberg-operator}
    \Phi(t,\mathbf{x}) = e^{iH(t - t_\mathrm{i})}\Phi(t_\mathrm{i},\mathbf{x})e^{-iH(t - t_\mathrm{i})}\eqend{,}
\end{equation}
where $t_\mathrm{i}$ is an arbitrary time coordinate, which will be set to $0$ later. %(We note in passing that the operators $H_0$ and $H_\mathrm{I}$ are both Schr\"odinger operators, and thus time-independent.) 
To write Eq.~\eqref{eq:4_point_function_trace} as a perturbative series, we have to go to the interaction picture. The interaction-picture field operator is defined by
\begin{equation}\label{eq:def-interaction-operator}
    \Phi_\mathrm{I}(t,\mathbf{x}) \equiv e^{iH_0(t - t_\mathrm{i})}\Phi(t_\mathrm{i},\mathbf{x})e^{-iH_0(t - t_\mathrm{i})}\eqend{,}
\end{equation}
so it satisfies the free-field equation because its evolution is governed by the free-theory Hamiltonian $H_0$. From Eqs.~\eqref{eq:def-heisenberg-operator} and \eqref{eq:def-interaction-operator}, the Heisenberg field can be expressed in terms of the interaction-picture field as
\begin{equation}\label{eq:interaction_pic_field}
 \Phi(t,\vt{x}) = U_\mathrm{I}(t_\mathrm{i},t)\Phi_\mathrm{I}(t,\vt{x})U_\mathrm{I}(t,t_\mathrm{i})\eqend{,}
\end{equation}
where
\begin{equation}\label{eq:U_I}
    U_\mathrm{I}(t,t_\mathrm{i}) \equiv e^{iH_0(t-t_\mathrm{i})}e^{-iH(t-t_\mathrm{i})}\eqend{,}
\end{equation}
and $U_\mathrm{I}(t,t_\mathrm{i}) \equiv U_\mathrm{I}^\dagger(t_\mathrm{i},t)$. The operator $U_\textrm{I}(t,t_\textrm{i})$ is unitary, being the product of two unitary operators. We define the operator $U_\textrm{I}(t,t')$ for arbitrary values of $t$ and $t'$ by
\begin{equation}\label{eq:U_I_composition_dagger}
 U_\textrm{I}(t,t') = U_\textrm{I}(t,t_\textrm{i})U_\textrm{I}(t_\textrm{i},t')\eqend{.}
\end{equation}
From this definition and Eq.~\eqref{eq:U_I} we find
\begin{equation}\label{eq:U_I_explicit}
    U_\mathrm{I}(t,t') = e^{i(t-t_\mathrm{i})H_0}e^{-i(t-t')H}e^{-i(t'-t_\mathrm{i})H_0}\eqend{.}
\end{equation}
From this it follows that
\begin{equation}\label{eq:U_I_composition}
 U_\textrm{I}(t'',t')U_\textrm{I}(t',t) =  U_\textrm{I}(t'',t)\eqend{.}
\end{equation}

The operator $U_\textrm{I}(t,t')$ can be shown to satisfy the following differential equation by direct differentiation using Eq.~\eqref{eq:U_I_explicit}:
\begin{equation}
    i\frac{d\ }{dt}U_\mathrm{I}(t,t') = H_\mathrm{I}(t)U_\mathrm{I}(t,t')\eqend{.}
\end{equation}
The unique solution to this equation is given in terms of Dyson's series~\cite{dyson_pr_1949a}:
\begin{equation}\label{eq:dyson_series}
 U_\textrm{I}(t,t') = \mathcal{P}\exp\left(-i\int_{t'}^tdt''H_\textrm{I}(t'')\right)\eqend{,}
\end{equation}
where the path-ordering $\mathcal{P}$ indicates time-ordering if $t > t'$ and anti-time-ordering if $t < t'$, and the interaction Hamiltonian operator $H_\textrm{I}(t)$ is defined by
\begin{equation}\label{eq:H_I}
 H_\textrm{I}(t) \equiv e^{iH_0 (t - t_\textrm{i})}H_\mathrm{I}e^{-iH_0 (t - t_\textrm{i})}\eqend{.}
\end{equation}
Thus, $H_\mathrm{I}(t)$ is the interaction Hamiltonian written in terms of the interaction-picture field $\Phi_\mathrm{I}(x)$ satisfying the free-field equation.

Furthermore, the operator $e^{-a\beta H}$, with $a$ real and positive, can be seen as an evolution operator in the imaginary time $-ia\beta$ and conveniently expressed as~\cite{matsubara_ptp_1955}
\begin{equation}\label{eq:matsubara_trick}
 e^{-a\beta H} = e^{-a\beta H_0}U_\mathrm{I}(t_\mathrm{i} - ia\beta, t_\mathrm{i})\eqend{,}
\end{equation}
which can readily be verified by using Eq.~\eqref{eq:U_I}. We note that, since the right-hand side is expressed entirely in terms of the free field $\Phi_\mathrm{I}(x)$, it can be given explicitly in the mode expansion order by order in perturbation theory.

Now, let us set the initial time $t_\mathrm{i} = 0$ for simplicity. We then use Eqs.~\eqref{eq:interaction_pic_field}, (\ref{eq:U_I_composition}), and (\ref{eq:matsubara_trick}) to write the right-hand side of the $4$-point function~(\ref{eq:4_point_function_trace}) in the interaction picture. The result is
\begin{splitequation}\label{eq:4_point_function_interaction_pic}
 & \Delta_\beta(y_1,y_2;x_1,x_2) \\
 & = \frac{1}{Z(\beta)}\tr\Big{\{}e^{-\frac{1}{2}\beta H_0}U_\mathrm{I}(- \tfrac{i\beta}{2},0)U_\mathrm{I}(0,-\tau_2)\Phi_\mathrm{I}^{(\mathrm{R})}(-\tau_2,\iota(\vt{y}_2))\\
 & \qquad\times U_\mathrm{I}(-\tau_2,-\tau_1)\Phi_\mathrm{I}^{(\mathrm{R})}(-\tau_1,\iota(\vt{y}_1))U_\mathrm{I}(-\tau_1,0)e^{-\frac{1}{2}\beta H_0}\\
 & \qquad\times U_\mathrm{I}(- \tfrac{i\beta}{2},0)U_\mathrm{I}(0,t_1)\Phi^{(\mathrm{R})}_\mathrm{I}(t_1,\vt{x}_1)U_\mathrm{I}(t_1,t_2)\\
 & \qquad\times\Phi_\mathrm{I}^{(\mathrm{R})}(t_2,\vt{x}_2)U_\mathrm{I}(t_2,0)\Big{\}}\eqend{.}
\end{splitequation}
The right-hand side of Eq.~(\ref{eq:4_point_function_interaction_pic}) can be expressed as a thermal average of an operator by moving the operator $e^{-\frac{1}{2}\beta H_0}$ in the middle to the left to combine it with the operator $e^{-\frac{1}{2}\beta H_0}$ on the far left. To do so, we first note that
\begin{equation}\label{eq:U_I_boltzmann_factor}
U_\mathrm{I}(t,t')e^{-a\beta H_0} =  e^{-a\beta H_0}U_\mathrm{I}(t - ia\beta,t' - ia\beta)\eqend{,}
\end{equation}
which can be shown using Eq.~\eqref{eq:U_I_explicit} on both sides. This equation can be used to express the partition function $Z(\beta)$ in terms of the free-theory thermal state $\varrho_0(\beta) \equiv e^{-\beta H_0}/Z_0(\beta)$ as
\begin{equation}\label{eq:def_Z_beta}
 Z(\beta) = Z_0(\beta)\tr\{\varrho_0(\beta)U_\mathrm{I}(t_\mathrm{i} - i\beta,t_\mathrm{i})\}\eqend{.}
\end{equation}
Moreover, since the interaction-picture field evolves with the free-field Hamiltonian operator, we have that
\begin{equation}
 \Phi_\mathrm{I}^{(\mathrm{R})}(t,\vt{x})e^{-a\beta H_0} 
  =  e^{-a\beta H_0}\Phi_\mathrm{I}^{(\mathrm{R})}(t - ia\beta,\vt{x})\eqend{.}\label{eq:O_I_boltzmann_factor}
\end{equation}
Using these three equations we can write Eq.~(\ref{eq:4_point_function_interaction_pic}) as follows:
\begin{widetext}
\begin{splitequation}
 \Delta_\beta(y_1,y_2;x_1,x_2) 
 & = \tr\Big{\{}\varrho_0(\beta)U_\mathrm{I}(- i\beta,-\tfrac{i\beta}{2})U_\mathrm{I}(-\tfrac{i\beta}{2},
 -\tau_2 - \tfrac{i\beta}{2})\Phi_\mathrm{I}^{(\mathrm{R})}(-\tau_2 - \tfrac{i\beta}{2},\iota(\vt{y}_2))U_\mathrm{I}(-\tau_2 - \tfrac{i\beta}{2},-\tau_1 - \tfrac{i\beta}{2}) \\
 & \qquad \times \Phi_\mathrm{I}^{(\mathrm{R})}(-\tau_1 - \tfrac{i\beta}{2},\iota(\vt{y}_1))U_\mathrm{I}(-\tau_1 - \tfrac{i\beta}{2},-\tfrac{i\beta}{2})U_\mathrm{I}(-\tfrac{i\beta}{2},0)U_\mathrm{I}(0,t_2)\Phi_\mathrm{I}^{(\mathrm{R})}(t_2,\vt{x}_2)U_\mathrm{I}(t_2,t_1)\\
 & \qquad \times\Phi_\mathrm{I}^{(\mathrm{R})}(t_1,\vt{x}_1)U_\mathrm{I}(t_1,0)\Big{\}}
 \Big{/}\tr\{\varrho_0(\beta)U_\mathrm{I}(- i\beta,0)\}\eqend{.}\label{eq:4_point_function_thermal_average}
\end{splitequation}
Now, since the operators on the right-hand side of this equation are all in the right wedge, both the numerator and denominator can be expressed as expectation values of operators in the state $\Omega_\beta^{(0)}$, with $\textrm{tr}\{\varrho_0(\beta)\cdots\}$ replaced by $\langle\Omega_\beta^{(0)}|\cdots\Omega_\beta^{(0)}\rangle$. Then, recalling Eq.~\eqref{eq:dyson_series}, we can write this equation more concisely as
\begin{splitequation}
& \Delta_\mathrm{\beta}(y_1,y_2;x_1,x_2) \\
& = \langle\Omega_\beta^{(0)}|\mathcal{P}\Bigg{\{}\Phi_\mathrm{I}^{(\mathrm{R})}(-\tau_2-\tfrac{i\beta}{2},\iota(\mathbf{y}_2))\Phi_\mathrm{I}^{(\mathrm{R})}(-\tau_1-\tfrac{i\beta}{2},\iota(\mathbf{y}_1)) \Phi_\mathrm{I}^{(\mathrm{R})}(t_1,\mathbf{x}_1)\Phi_\mathrm{I}^{(\mathrm{R})}(t_2,\mathbf{x}_2)\exp\left( - i\int_C dt H_\mathrm{I}^{(\mathrm{R})}(t)\right)\Bigg{\}}\Omega_\beta^{(0)}\rangle_{\textrm{connected}}\eqend{,}\label{eq:4-pt-fun-beta}
\end{splitequation}
\end{widetext}
where $C$ is a directed contour from $0$ to $-i\beta$ with decreasing imaginary parts and with $-\tau_2-\frac{i\beta}{2}$, $-\tau_1-\frac{i\beta}{2}$, $t_1$ and $t_2$ on it in this order. For the inverse temperature $\beta = \beta_\mathrm{H}$, the right-hand side of Eq.~\eqref{eq:4-pt-fun-beta} is the case with $R=L=2$ of Eq.~\eqref{eq:hartle-hawking-N-pt-2}.  It is straightforward to generalize this result to the $N$-point case. Hence, the $N$-point functions of the double KMS state at the Hawking temperature $\Omega_{\beta_\mathrm{H}}$ are equal to those of the HH state $\Omega_\textrm{HH}$.  This concludes our proof that $\Omega_{\beta_\mathrm{H}} = \Omega_\textrm{HH}$, and is the main result of this paper.

Next, we shall write down a perturbative expression of the double KMS state $\Omega_\beta$ itself as an operator acting on the free-theory double KMS state $\Omega_\beta^{(0)}$. To do so, we go back to Eq.~\eqref{eq:4_point_function_thermal_average} and express the trace over the free-field states as the expectation value in the state $\Omega_\beta^{(0)}$. It is convenient to introduce the following notation for interaction-picture propagators acting on the left or the right wedge: 
\begin{splitequation}
U_\mathrm{I}^{(\mathrm{L})}(t,t') & \equiv U_\mathrm{I}(t,t')\otimes\mathbbm{1}\eqend{,}\\
U_\mathrm{I}^{(\mathrm{R})}(t,t') & \equiv \mathbbm{1}\otimes U_\mathrm{I}(t,t')\eqend{.} 
\end{splitequation}
Then, the operator $U_\mathrm{I}(-i\beta,-\tfrac{i\beta}{2})$ in Eq.~\eqref{eq:4_point_function_thermal_average} can be rearranged in the following way. For any operator $O$ with support on the right wedge,
\begin{widetext}
\begin{splitequation}
 \langle\Omega_\beta^{(0)}|U_\mathrm{I}^{(\mathrm{R})}(-i\beta,-\tfrac{i\beta}{2})O\Omega_\beta^{(0)}\rangle
  & = \langle\Omega_\beta^{(0)}|U_\mathrm{I}^{(\mathrm{R})}(-i\beta,-ic_\mathrm{L}\beta-\tfrac{i\beta}{2})U_\mathrm{I}^{(\mathrm{R})}(-ic_\mathrm{L}\beta-\tfrac{i\beta}{2},-\tfrac{i\beta}{2})O \Omega_\beta^{(0)}\rangle\\
	& = \langle\Omega_\beta^{(0)}|\mathcal{P}\exp\left(-i\int_{-ic_\mathrm{L}\beta - \tfrac{i\beta}{2}}^{-i\beta}dtH^{(\mathrm{R})}_\mathrm{I}(t)\right)U_\mathrm{I}^{(\mathrm{R})}(-ic_\mathrm{L}\beta - \tfrac{i\beta}{2},-\tfrac{i\beta}{2})O \Omega_\beta^{(0)}\rangle\\
	& = \langle\Omega_\beta^{(0)}|\mathcal{P}\exp\left(-i\int_{i(\tfrac{1}{2} - c_\mathrm{L})\beta}^0dtH_\mathrm{I}^{(\mathrm{R})}(t - i\beta)\right)U_\mathrm{I}^{(\mathrm{R})}(-ic_\mathrm{L}\beta - \tfrac{i\beta}{2},-\tfrac{i\beta}{2})O \Omega_\beta^{(0)}\rangle\\
	& = \langle\Omega_\beta^{(0)}|e^{\beta\tilde{H}_0}U_\mathrm{I}^{(\mathrm{R})}(0,ic_\mathrm{R}\beta)e^{-\beta\tilde{H}_0}U_\mathrm{I}^{(\mathrm{R})}(-ic_\mathrm{L}\beta - \tfrac{i\beta}{2},-\tfrac{i\beta}{2})O \Omega_\beta^{(0)}\rangle\\
   & =  \langle\Omega_\beta^{(0)}|U_\mathrm{I}^{(\mathrm{R})}(-ic_\mathrm{L}\beta-\tfrac{i\beta}{2},-\tfrac{i\beta}{2})O U_\mathrm{I}^{(\mathrm{R})}(0,ic_\mathrm{R}\beta)\Omega_\beta^{(0)}\rangle\eqend{,}
\end{splitequation}
where in the last equality we have used the invariance of $\Omega_\beta^{(0)}$ under the free Hamiltonian and the KMS condition, Eqs.~\eqref{eq:invariance_kms_state_H} and~\eqref{eq:kms_condition}, respectively. Here, the non-negative numbers $c_\mathrm{R}$ and $c_\mathrm{L}$ satisfy
\begin{equation}\label{eq:c_R_plus_c_L}
    c_\mathrm{L} + c_\mathrm{R} = \tfrac{1}{2}\eqend{.}
\end{equation}
Thus, Eq.~\eqref{eq:4_point_function_thermal_average} can be cast in the form
\begin{splitequation}\label{eq:interim_for_state}
 &  \Delta_\beta(y_1,y_2;x_1,x_2)  \\
 & = \langle \Omega^{(0)}_\beta|U_\mathrm{I}^{(\mathrm{R})}(-ic_\mathrm{L}\beta - \tfrac{i\beta}{2}, - \tfrac{i\beta}{2})U_\mathrm{I}^{(\mathrm{R})}(- \tfrac{i\beta}{2},-\tau_2 - \tfrac{i\beta}{2})\Phi_\mathrm{I}^{(\mathrm{R})}(-\tau_2 - \tfrac{i\beta}{2},\iota(\vt{y}_2))U_\mathrm{I}^{(\mathrm{R})}(-\tau_2 - \tfrac{i\beta}{2},-\tau_1 - \tfrac{i\beta}{2})\\
 &\phantom{=}\quad \times \Phi_\mathrm{I}^{(\mathrm{R})}(-\tau_1 - \tfrac{i\beta}{2},\iota(\vt{y}_1))U_\mathrm{I}^{(\mathrm{R})}(-\tau_1 - \tfrac{i\beta}{2}, - \tfrac{i\beta}{2}) U_\mathrm{I}^{(\mathrm{R})}(- \tfrac{i\beta}{2}, ic_\mathrm{L}\beta - \tfrac{i\beta}{2}) U_\mathrm{I}^{(\mathrm{R})}( -ic_\mathrm{R}\beta,0)U_\mathrm{I}^{(\mathrm{R})}(0,t_2) \\
 &\phantom{=}\quad \times \Phi_\mathrm{I}^{(\mathrm{R})}(t_2,\vt{x}_2)U_\mathrm{I}^{(\mathrm{R})}(t_2,t_1)\Phi_\mathrm{I}^{(\mathrm{R})}(t_1,\vt{x}_1)U_\mathrm{I}^{(\mathrm{R})}(t_1,0)U_\mathrm{I}^{(\mathrm{R})}(0,ic_\mathrm{R}\beta)\Omega_\beta^{(0)}\rangle\Big{/}\langle \Omega_\beta^{(0)}|U_\mathrm{I}^{(\mathrm{R})}(-i\beta,0)\Omega_\beta^{(0)}\rangle\eqend{,}
\end{splitequation}
The next step is to express some of the operators appearing in Eq.~\eqref{eq:interim_for_state} as operators acting on the left wedge. This can been done by retracing the steps taken in Eq.~\eqref{eq:interim_4p_function}. In particular, we observe that for $z,z'\in\mathbb{C}$ we can employ the properties defining the free KMS state $\Omega_\beta^{(0)}$, Eqs.~\eqref{eq:invariance_kms_state_H}-\eqref{eq:J_2}, to show that
\begin{equation}
 \langle\Omega_\beta^{(0)}|\Phi^{(\mathrm{R})}_\mathrm{I}(z - \tfrac{i\beta}{2},\iota(\vt{y}))\cdots\Phi^{(\mathrm{R})}_\mathrm{I}(z' - \tfrac{i\beta}{2},\iota(\vt{y}'))O\Omega_\beta^{(0)}\rangle =  
 \langle\Omega_\beta^{(0}|\Phi^{(\mathrm{L})}_\mathrm{I}(-z',\vt{y}')\cdots \Phi^{(\mathrm{L})}_\mathrm{I}(-z,\vt{y})O\Omega_\beta^{(0)}\rangle\eqend{,}
\end{equation}
where again $O$ is any operator with support on $\mathcal{R}$. (Notice the reversing of the operator ordering when the right-wedge operators are converted to left-wedge operators here.) Using this observation, we convert the string of operators $U_\mathrm{I}^{(\mathrm{R})}(ic_\mathrm{L}\beta-\tfrac{i\beta}{2},-\tfrac{i\beta}{2})\cdots U_\mathrm{I}^{(\mathrm{R})}(-\tfrac{i\beta}{2},ic_\mathrm{L}\beta - \tfrac{i\beta}{2})$ in Eq.~\eqref{eq:interim_for_state} into a string of operators on the left wedge. The result is
\begin{splitequation}
 &\Delta_\beta(y_1,y_2;x_1,x_2)\\
 & = \langle \Omega^{(0)}_\beta|U_\mathrm{I}^{(\mathrm{L})}(-ic_\mathrm{L}\beta,0)U_\mathrm{I}^{(\mathrm{L})}(0,\tau_1)\Phi_\mathrm{I}^{(\mathrm{L})}(\tau_1,\vt{y}_1)U_\mathrm{I}^{(\mathrm{L})}(\tau_1,\tau_2)\Phi^{(\mathrm{L})}_\mathrm{I}(\tau_2,\vt{y}_2)U_\mathrm{I}^{(\mathrm{L})}(\tau_2,0)U_\mathrm{I}^{(\mathrm{L})}(0,ic_\mathrm{L}\beta)U_\mathrm{I}^{(\mathrm{R})}(-ic_\mathrm{R}\beta,0)\\
 & \phantom{=} \quad \times U_\mathrm{I}^{(\mathrm{R})}(0,t_1)\Phi_\mathrm{I}^{(\mathrm{R})}(t_1,\vt{x}_1)U_\mathrm{I}^{(\mathrm{R})}(t_1,t_2)\Phi_\mathrm{I}^{(\mathrm{R})}(t_2,\vt{x}_2)U_\mathrm{I}^{(\mathrm{R})}(t_2,0)U^{(\mathrm{R})}_\mathrm{I}(0,ic_\mathrm{R}\beta)\Omega_\beta^{(0)}\rangle\Big{/}\langle \Omega_\beta^{(0)}|U_\mathrm{I}^{(\mathrm{R})}(-i\beta,0)\Omega_\beta^{(0)}\rangle\eqend{.}\\
\end{splitequation}
\end{widetext}
Finally, we use Eq.~\eqref{eq:interaction_pic_field} to write some strings of interaction-picture operators as the Heisenberg operator (with $t_\mathrm{i} = 0$) and use the fact that operators on the right wedge commute with those on the left ones to cast the expression above into the form
\begin{splitequation}
 & \Delta_\beta(y_1,y_2;x_1,x_2) \\
 & = \langle \Omega^{(0)}_\beta|U_\mathrm{I}^{(\mathrm{L})}(-ic_\mathrm{L}\beta,0)U_\mathrm{I}^{(\mathrm{R})}(-ic_\mathrm{R}\beta,0) \\
 &\phantom{=}\quad \times \Phi(y_1)\Phi(y_2)\Phi(x_1)\Phi(x_2)U^{(\mathrm{L})}_\mathrm{I}(0,ic_\mathrm{L}\beta)\\
 &\phantom{=}\quad \times U^{(\mathrm{R})}_\mathrm{I}(0,ic_\mathrm{R}\beta)\Omega_\beta^{(0)}\rangle\Big{/}\langle \Omega_\beta^{(0)}|U^{(\mathrm{R})}_\mathrm{I}(-i\beta,0)\Omega_\beta^{(0)}\rangle\eqend{.}
\end{splitequation}
It is clear from this equation that the interacting double KMS state is related to the free one according to
\begin{equation}\label{eq:interacting_free_Omega}
\Omega_\beta = \frac{U_\mathrm{I}^{(\mathrm{L})}(0,ic_\mathrm{L}\beta) U_\mathrm{I}^{(\mathrm{R})}(0,ic_\mathrm{R}\beta)}{\sqrt{\langle\Omega_\beta^{(0)}|U_\mathrm{I}^{(\mathrm{R})}(-i\beta,0)\Omega_\beta^{(0)}\rangle}}\Omega_\beta^{(0)}\eqend{,}
\end{equation}
and we recall that the constants $c_\mathrm{L}, c_\mathrm{R} > 0$ and satisfy Eq.~\eqref{eq:c_R_plus_c_L}.

In the case of a quantum system with a finite-dimensional Hilbert space, it is possible to give a simple interpretation of Eq.~\eqref{eq:interacting_free_Omega}. The adjoint of Eq.~\eqref{eq:matsubara_trick} is written as
\begin{splitequation}
  & U_\mathrm{I}^{(\mathrm{L})}(0,ic_\mathrm{L}\beta) U_\mathrm{I}^{(\mathrm{R})}(0,ic_\mathrm{R}\beta) \\
  & \quad = e^{-(c_\mathrm{L}H\otimes\mathbbm{1} + c_\mathrm{R}\mathbbm{1}\otimes H)\beta}e^{(c_\mathrm{L}H_0\otimes\mathbbm{1} + c_\mathrm{R}\mathbbm{1}\otimes H_0)\beta}\eqend{,}
\end{splitequation}
The free-theory counterpart of Eq.~\eqref{eq:double_kms_state} is
\begin{equation}
    \Omega_\beta^{(0)} = \frac{1}{\sqrt{Z_0(\beta)}}\sum_i e^{-\frac{1}{2}\beta E^{(0)}_i}\psi^{(0)}_i \otimes \psi^{(0)}_i\eqend{,}
\end{equation}
where the $\psi^{(0)}_i$ are eigenstates of the free Hamiltonian and satisfy $H_0\psi^{(0)}_i = E^{(0)}_i\psi^{(0)}_i$, with $E^{(0)}_i$ the respective energy eigenvalues. We first note that
\begin{equation}
e^{(c_\mathrm{L}H_0\otimes\mathbbm{1} + c_\mathrm{R}\mathbbm{1}\otimes H_0)\beta}\Omega_\beta^{(0)}
= \frac{1}{\sqrt{Z_0(\beta)}}\sum_i \psi^{(0)}_i \otimes \psi^{(0)}_i\eqend{,}
\end{equation}
since $c_\mathrm{R} + c_\mathrm{L} = \frac{1}{2}$. We assume that both free and full Hamiltonians are invariant under an antiunitary transformation $T$ satisfying $T^2 = \mathbbm{1}$, such as the time-reversal transformation. In this case, the orthonormal energy eigenstates in the free and full theories can be chosen to be invariant under $T$. Then, at the initial time the free energy eigenstates can be expanded in terms of the interacting ones as
\begin{equation}
\psi_i^{(0)} = \sum_j A_{ij}\psi_j\eqend{,}
\end{equation}
with $A_{ij}$ as the elements of an orthogonal matrix.  This is because $\langle \psi^{(0)}_i|\psi_j\rangle = \langle T\psi^{(0)}_i|T\psi_j\rangle = \langle \psi_j|\psi^{(0)}_i\rangle$. The fact that this matrix is orthogonal implies that
\begin{equation}
  \sum_i \psi_i^{(0)}\otimes \psi_i^{(0)} = \sum_i \psi_i\otimes \psi_i\eqend{.} 
\end{equation}
Using this equality, we find
\begin{splitequation}
& U_\mathrm{I}^{(\mathrm{L})}(0,ic_\mathrm{L}\beta) U_\mathrm{I}^{(\mathrm{R})}(0,ic_\mathrm{R}\beta)\Omega_\beta^{(0)}\\
& \quad =\frac{1}{\sqrt{Z_0(\beta)}}e^{-(c_\mathrm{L}H\otimes\mathbbm{1}+ c_\mathrm{R}\mathbbm{1}\otimes H)\beta}\sum_i\psi_i\otimes \psi_i \\
& \quad = \frac{1}{\sqrt{Z_0(\beta)}}\sum_i e^{-\frac{1}{2}\beta E_i}\psi_i\otimes \psi_i \\
& \quad =\sqrt{\frac{Z(\beta)}{Z_0(\beta)}}\Omega_\beta\eqend{.} \label{eq:intermediate-eq2}
\end{splitequation}
A similar argument shows that 
\begin{equation}
    \langle\Omega^{(0)}_\beta|U_\mathrm{I}^{(\mathrm{R})}(-i\beta,0)\Omega^{(0)}_\beta\rangle = \frac{Z(\beta)}{Z_0(\beta)}\eqend{.}
\end{equation}
Equation~\eqref{eq:interacting_free_Omega} then follows from this last result and Eq.~\eqref{eq:intermediate-eq2}.

Equation~\eqref{eq:interacting_free_Omega} allows us to evaluate the expectation value of operators which are not necessarily in the right or left wedge.   Thus, with any points $X_1, X_2,\ldots, X_N \in \mathcal{M}$, we have with the choice $c_\mathrm{R} = \frac{1}{2}$ and $c_\mathrm{L} = 0$
\begin{splitequation}\label{eq:Phi-X-N-pt}
 & \langle \Omega_\beta|\Phi(X_1)\Phi(X_2)\ldots\Phi(X_N)\Omega_\beta\rangle\\
 & \quad = \langle\Omega^{(0)}_\beta|U_\mathrm{I}^{(\mathrm{R})}(-\tfrac{i\beta}{2},0) \Phi(X_1)\Phi(X_2)\ldots\Phi(X_N)\\
 & \quad\phantom{=} \times U^{(\mathrm{R})}_\mathrm{I}(0,\tfrac{i\beta}{2})\Omega_\beta^{(0)}\rangle\Big{/}\langle\Omega_\beta^{(0)}|U_\mathrm{I}^{(\mathrm{R})}(-i\beta,0)\Omega_\beta^{(0)}\rangle\eqend{.}
\end{splitequation}
If some of the points $X_k$, $k=1,2,\ldots,N$, are not in either wedge, we need to use the global time $T$ to construct the Heisenberg operator $\Phi(X_k)$ in terms of the interaction-picture operators satisfying the free-field equation.  The Hamiltonian in this construction is time-dependent, in general.  Nevertheless, as shown in Appendix~\ref{appendix:int_with_t_dep_hamiltonian}, the Heisenberg operator can still be given in terms of the interaction-picture operators.  Thus, if we write $X = (T,\vt{X})$, we have
\begin{equation}
    \Phi(T,\vt{X}) = U_\mathrm{I}(0,T)\Phi_\mathrm{I}(T,\mathbf{X})U_\mathrm{I}(T,0)\eqend{,}
\end{equation}
where the operator $U_\mathrm{I}(T,T')$ is defined in Appendix~\ref{appendix:int_with_t_dep_hamiltonian}. By substituting this formula into Eq.~\eqref{eq:Phi-X-N-pt}, assuming that $T_k >0$ for all $k$ for simplicity, we find for time-ordered products,
\begin{splitequation}\label{eq:time-ordered-product-OK}
 & \langle \Omega_\beta|\mathcal{T}\left[\Phi(X_1)\Phi(X_2)\ldots\Phi(X_N)\right]\Omega_\beta\rangle\\
 & \quad = \langle\Omega^{(0)}_\beta|U_\mathrm{I}^{(\mathrm{R})}(-\tfrac{i\beta}{2},0)U_\mathrm{I}(0,T_\mathrm{f})\\
 & \quad\phantom{=}\times \mathcal{T}\left[U_\mathrm{I}(T_\mathrm{f},0)\Phi_\mathrm{I}(X_1)\Phi_\mathrm{I}(X_2)\ldots\Phi_\mathrm{I}(X_N)\right]\\
 & \quad\phantom{=}\times U^{(\mathrm{R})}_\mathrm{I}(0,\tfrac{i\beta}{2})\Omega_\beta^{(0)}\rangle\Big{/}\langle\Omega_\beta^{(0)}|U_\mathrm{I}^{(\mathrm{R})}(-i\beta,0)\Omega_\beta^{(0)}\rangle \eqend{,}
\end{splitequation}
where $T_\mathrm{f} > T_k$ for all $k$.

The right-hand side of Eq.~\eqref{eq:time-ordered-product-OK} is evaluated perturbatively as follows.  Let $\mathcal{M}_{\mathrm{E}}^{(+)}$ and $\mathcal{M}_{\mathrm{E}}^{(-)}$ be the two halves of the Euclidean section with $T_\mathrm{E}>0$ (or $0 < t_\mathrm{E} < \beta_\mathrm{H}/2 = \pi/\kappa$) and $T_\mathrm{E} <0$ (or $-\pi/\kappa = - \beta_\mathrm{H}/2 < t_\mathrm{E} < 0$), respectively.  Let $\mathcal{M}_1$ and $\mathcal{M}_2$ be two copies of the part of the global Lorentzian manifold $\mathcal{M}$ with $0 < T < T_\mathrm{f}$.  Let
\begin{equation}
 \mathcal{M}_\textrm{total} \equiv \mathcal{M}_{\mathrm{E}}^{(-)}\cup \mathcal{M}_1 \cup \mathcal{M}_2 \cup \mathcal{M}_{\mathrm{E}}^{(+)}\eqend{.}
\end{equation}
Then Eq.~\eqref{eq:time-ordered-product-OK} can be expressed as
\begin{splitequation}
&  \langle \Omega_\beta|\mathcal{T}\left[\Phi(X_1)\Phi(X_2)\ldots\Phi(X_N)\right]\Omega_\beta\rangle\\
&  \quad= \langle\Omega_\beta^{(0)}|\Phi_\mathrm{I}(X_1)\Phi_\mathrm{I}(X_2)\cdots\Phi_\mathrm{I}(X_N)\\
&  \quad\phantom{=}\times\exp\left(-i\int_{\mathcal{M}_\textrm{total}}\sqrt{-g}d^n x\, \mathcal{H}_\mathrm{I}(x)\right)\Omega_\beta^{(0)}\rangle_\textrm{connected}\eqend{,}
\end{splitequation}
where $X_1,X_2,\ldots,X_N \in \mathcal{M}_2$. The right-hand side is expanded using Wick's theorem in terms of the free-field two-point function. If the two points $x$ and $x'$ are on $\mathcal{M}_\mathrm{E}^{(+)}\cup \mathcal{M}_\mathrm{E}^{(-)}$, then the two-point function $\Delta^{(0)}_\beta(x,x')$ is the Green's function on the Euclidean section, $G^{(0)}(x,x')$.  If one point is on $\mathcal{M}_\mathrm{E}^{(+)}\cup \mathcal{M}_\mathrm{E}^{(-)}$ and the other is on $\mathcal{M}_1\cup \mathcal{M}_2$, then $\Delta^{(0)}_\beta(x,x')$ is the analytic continuation of this function. For the other cases we have
\begin{splitequation}
& \Delta^{(0)}_\beta(x_1,x_2) \\
& = \begin{cases} \langle \Omega^{(0)}_\beta|\Phi_\mathrm{I}(x_1)\Phi_\mathrm{I}(x_2)\Omega^{(0)}_\beta\rangle & \textrm{if}\ x_i\in \mathcal{M}_i,\ i=1,2\eqend{,}\\
\langle\Omega_\beta^{(0)}|\mathcal{T}\left[\Phi_\mathrm{I}(x_1)\Phi_\mathrm{I}(x_2)\right]\Omega^{(0)}_\beta\rangle & \textrm{if}\ x_1,x_2\in \mathcal{M}_2\eqend{,} \\
\langle\Omega_\beta^{(0)}|\overline{\mathcal{T}}\left[\Phi_\mathrm{I}(x_1)\Phi_\mathrm{I}(x_2)\right]\Omega^{(0)}_\beta\rangle & \textrm{if}\ x_1,x_2\in \mathcal{M}_1\eqend{,}
\end{cases}
\end{splitequation}
where $\overline{\mathcal{T}}$ denotes anti-time-ordering.

%%%%%%%%%%%%%%%%%%%%%%%%%%%%%%%%%%%%%%%%%%%%%%%%%%%%%%%%%%%%%%%%%%%%%%%%%%%%%%%%%%%%%%%%%%%%%%%%%%%%%%%%%%%%%%%%%%%%%%%%%%%%%%%%%%%%%%%%%%%%%%%%%%%%%
\section{Examples}                                                                                                                                  %
\label{sec:example_spacetimes}                                                                                                                      %
%%%%%%%%%%%%%%%%%%%%%%%%%%%%%%%%%%%%%%%%%%%%%%%%%%%%%%%%%%%%%%%%%%%%%%%%%%%%%%%%%%%%%%%%%%%%%%%%%%%%%%%%%%%%%%%%%%%%%%%%%%%%%%%%%%%%%%%%%%%%%%%%%%%%%

In this section we briefly comment on the Schwinger-Keldysh formulation of the HH state in Sec.~\ref{subsection:interacting} for some spacetimes with a static bifurcate Killing horizon and a wedge reflection.  The application of this formulation to Schwarzschild spacetime is straightforward since it was constructed with this spacetime in mind, except that we need to change the upper limit $T_\mathrm{f}$ for the $T$ integration depending on the coordinate $X$ because the $T$ coordinate is bounded by the singularities as $|T|< \sqrt{1+X^2}$.

\subsection{Minkowski spacetime}
%%%%%%%%%%%%%%%%%%%%%%%%%%%%%%%%%%%%%%%%%%%%%%%%%%%%%%%%%%%%%%%%%%%%%%%%%%%%%%%%%%%%%%%%%%%%%%%%%%%%%%%%%%%%%%%%%%%%%%%%%%%%%%%%%%%%%%%%%%%%%%%%%%%%%

The metric of $n$-dimensional Minkowski spacetime can be given as
\begin{equation}
    g_{ab}^\mathrm{M} = - (dT)_a(dT)_b + (dX)_a(dX)_b + \sum_{i=1}^{n-2}(dx^i)_a(dx^i)_b\eqend{,}
\end{equation}
where $T$ is the usual inertial time. With $(T,X)=\big(\rho\sinh(\kappa t),\rho\cosh(\kappa t)\big)$, $\rho > 0$, we obtain the metric on the right Rindler wedge satisfying $-X < T <X$:
\begin{splitequation}
  g_{ab}^\mathrm{M}|_\mathcal{R} 
	& = \left[- \kappa^2\rho^2(dt)_a(dt)_b + (d\rho)_a(d\rho)_b\right]\\ 
	& \phantom{=} + \sum_{i=1}^{n-2}(dx^i)_a(dx^i)_b\eqend{.}
\end{splitequation}
The metric covering the left Rindler wedge is obtained by letting $(T,X) = \big(\rho\sinh(\kappa t), - \rho\cosh(\kappa t)\big)$, $\rho < 0$, and takes the same form.

The Euclidean section is obtained by letting $T=iT_\mathrm{E}$, which has the metric
\begin{equation}
    g_{ab}^{\mathrm{M},\mathrm{E}} = (dT_\mathrm{E})_a(dT_\mathrm{E})_b + (dX)_a(dX)_b + \sum_{i=1}^{n-2}(dx^i)_a(dx^i)_b\eqend{.}
\end{equation}
The manifold $\mathcal{M}^{(+)}_\mathrm{E}$  ($\mathcal{M}^{(-)}_\mathrm{E}$) is the $T_\mathrm{E}>0$ ($T_\mathrm{E}<0$) part of the Euclidean section and $\mathcal{M}_1$ and $\mathcal{M}_2$ are two copies of the portion satisfying $0<T<T_\mathrm{f}$ of the Lorentzian manifold.  Thus, in a double KMS state at the inverse temperature $\beta = \beta_\mathrm{H}$ with respect to the boost Killing vector $\partial/\partial t$, the time-ordered $N$-point function is obtained by analytic continuation of that for the Euclidean theory. As is well known, the $N$-point function in the vacuum state is obtained in the same way.  Thus, the vacuum state is a double KMS state with respect to $\partial/\partial t$. This corresponds to the result of Bisognano and Wichmann~\cite{bisognano_wichmann_jmp_1975} in axiomatic field theory.

If all $N$ operators for the time-ordered $N$-point function are in the right Rindler wedge with positive time coordinates, then it can be given by Eq.~(\ref{eq:hartle-hawking-N-pt-2}) with the contour $C$ given by Fig~\ref{fig:sk_path_Killing_time} with the path $C_4\cup C_5$ missing. It is interesting that in the limit $t_\mathrm{i}\to - \infty$ the paths $C_3\cup C_6$ can be omitted~\cite{higuchi_lima_prd_2020}. This property is called ``factorization'' (see, e.g.\ Ref.~\cite{landsman_van_weert_pr_1987}).

\subsection{De~Sitter spacetime}
%%%%%%%%%%%%%%%%%%%%%%%%%%%%%%%%%%%%%%%%%%%%%%%%%%%%%%%%%%%%%%%%%%%%%%%%%%%%%%%%%%%%%%%%%%%%%%%%%%%%%%%%%%%%%%%%%%%%%%%%%%%%%%%%%%%%%%%%%%%%%%%%%%%%%

The metric of de~Sitter spacetime with the Hubble constant set to $1$ can be given as
\begin{splitequation}
   g_{ab}^\mathrm{dS} 
	 & = \frac{4}{(1+\rho^2)^2}\left[ -(dT)_a(dT)_b + (dX)_a(dX)_b\right] \\
   &\phantom{=} +\left( \frac{1-\rho^2}{1+\rho^2}\right)^2 \omega_{ab}\eqend{,}
\end{splitequation}
with $\rho^2 = X^2 - T^2 \in (-1,1]$, where $\omega_{ab}$ denotes the metric on the unit $(n-2)$-dimensional sphere, $S^{n - 2}$. This metric tensor can be cast into a more familiar form by defining the coordinates $\tau$ and $\chi$ by
\begin{splitequation}\label{eq:X-T-def-dS}
  X & = \frac{\cos\chi}{\cos\tau + \sin\chi}\eqend{,}\\
  T & = \frac{\sin\tau}{\cos\tau + \sin\chi}\eqend{,}
\end{splitequation}
where $-\frac{\pi}{2} < \tau < \frac{\pi}{2}$ and $0 \leq \chi \leq \pi$. The transformation to these new coordinates yields
\begin{equation}\label{eq:dS_metric_global_patch_2}
g_{ab}^\mathrm{dS} = \frac{1}{\cos^2\tau}\left[-(d\tau)_a(d\tau)_b + (d\chi)_a(d\chi)_b + \sin^2\chi\, \omega_{ab}\right]\eqend{.}
\end{equation}
Notice that $(d\chi)_a(d\chi)_b + \sin^2\chi\,\omega_{ab}$ corresponds to the metric of $S^{n - 1}$.

The static metric on the right wedge is found by letting $(T,X) =(\rho\sinh t, \rho\cosh t)$, $\rho>0$, as
\begin{splitequation}\label{eq:dS_metric_Rindler_form_2}
 g_{ab}^\mathrm{dS}|_\mathcal{R}
 & = \frac{4}{(1 + \rho^2)^2}\left[-\rho^2(dt)_a(dt)_b + (d\rho)_a(d\rho)_b\right]\\
 &\phantom{=} + \left(\frac{1 - \rho^2}{1 + \rho^2}\right)^2\omega_{ab}\eqend{.}
\end{splitequation}
This can be cast into a more familiar form by defining the following radial coordinate:
\begin{equation}
  r \equiv \frac{1-\rho^2}{1+\rho^2}\eqend{.}
\end{equation}
In terms of the coordinate $r$, the metric tensor~\eqref{eq:dS_metric_Rindler_form_2} then reads
\begin{equation}\label{eq:dS_metric_static_patch_2}
g_{ab}^\mathrm{dS}|_\mathcal{R} = - (1 - r^2)(dt)_a(dt)_b + \frac{(dr)_a(dr)_b}{1 - r^2} + r^2\omega_{ab}\eqend{.}
\end{equation}
The metric on the left wedge is identical. The state $\Omega_{\beta_\mathrm{H}}$ [now with $\beta_\mathrm{H} = 1/(2\pi)$] is the double KMS state constructed for these wedges.  

The Euclidean section is given by letting $T = iT_\mathrm{E}$.  This can also be achieved by letting $\tau = i \tau_\mathrm{E}$ in Eq.~\eqref{eq:X-T-def-dS}, and the metric~\eqref{eq:dS_metric_global_patch_2} becomes
\begin{splitequation}
  & g_{ab}^{\textrm{dS},\mathrm{E}}\\ 
  & \quad = \frac{1}{\cosh^2\tau_\mathrm{E}}\left[(d\tau_\mathrm{E})_a(d\tau_\mathrm{E})_b + (d\chi)_a(d\chi)_b + \sin^2\chi\, \omega_{ab}\right]\eqend{.}
\end{splitequation}
Finally, the coordinate change $\cosh\tau_\mathrm{E} = \sec\eta$ (with $d\tau_\mathrm{E}/d\eta >0$) leads to the familiar metric on $S^n$:
\begin{equation}
    g_{ab}^{\textrm{dS},\mathrm{E}} = 
    (d\eta)_a(d\eta)_b + \cos^2\eta\left[(d\chi)_a(d\chi)_b + \sin^2\chi\, \omega_{ab}\right]\eqend{.}
\end{equation}
The regions with $T_\mathrm{E} > 0$ and $T_\mathrm{E}< 0$ in the Euclidean section correspond to the hemisphere with $0 < \eta < \frac{\pi}{2}$ and that with $-\frac{\pi}{2} < \eta < 0$, respectively.  The manifold taken in perturbation theory in the previous section in this case is given as follows.  The manifolds $\mathcal{M}_\mathrm{E}^{(+)}$ and $\mathcal{M}_\mathrm{E}^{(-)}$ are the hemispheres of $S^n$ with $0 < \eta < \frac{\pi}{2}$ and $-\frac{\pi}{2} < \eta < 0$, respectively.  The manifolds $\mathcal{M}_1$ and $\mathcal{M}_2$ can both be replaced by the part of the global Lorentzian manifold with metric~\eqref{eq:dS_metric_global_patch_2} with $0 < \tau < \tau_\mathrm{f}$, where $\tau_\mathrm{f}$ is larger than the $\tau$-coordinate of any external point.

%%%%%%%%%%%%%%%%%%%%%%%%%%%%%%%%%%%%%%%%%%%%%%%%%%%%%%%%%%%%%%%%%%%%%%%%%%%%%%%%%%%%%%%%%%%%%%%%%%%%%%%%%%%%%%%%%%%%%%%%%%%%%%%%%%%%%%%%%%%%%%%%%%%%%
\section{Summary and Discussion}                                                                                                                    %
\label{sec:discussion}                                                                                                                              %
%%%%%%%%%%%%%%%%%%%%%%%%%%%%%%%%%%%%%%%%%%%%%%%%%%%%%%%%%%%%%%%%%%%%%%%%%%%%%%%%%%%%%%%%%%%%%%%%%%%%%%%%%%%%%%%%%%%%%%%%%%%%%%%%%%%%%%%%%%%%%%%%%%%%%

In this paper we discussed the relationship between Euclidean and Lorentzian perturbative formalisms for quantum fields in static spacetimes with a bifurcate Killing horizon and a wedge reflection. The natural state in such a spacetime is the HH state, which is a thermal equilibrium state at the Hawking temperature, as measured by static observers in either of the two static wedges. The naturalness of the HH state comes from the fact that it shares the background symmetries and is regular on the bifurcation surface. Since this state has been originally defined via the analytic continuation of the Euclidean theory to real times, the question we have addressed is how it relates to the double KMS state constructed in an intrinsically Lorentz-signature approach.

We first reviewed the equivalence of the HH state and a double KMS state at the Hawking temperature for a noninteracting scalar field theory.  Then, in an interacting scalar field theory with nonderivative interaction, we clarified how the $N$-point correlation functions are analytically continued from the Euclidean theory with imaginary time to real time in the Schwinger-Keldysh perturbation theory for the HH state. Then, we showed that these $N$-point functions are equal to those in a double KMS state at the Hawking temperature if the points are in the union of the right and left wedges. This gives a perturbative demonstration of the equivalence between the HH state and this double KMS state, shown formally by using path-integral by Jacobson~\cite{jacobson_prd_1994}. We also found a perturbative expression of this interacting double KMS state in terms of the noninteracting one and the free-field operators.  We used this result to express the $N$-point functions when the points are not necessarily in the wedges. 

It is interesting to compare Jacobson's path-integral argument with our operator approach. He started by pointing out that the HH state can be characterized as the Schr\"odinger wave functional on the Cauchy surface $\Sigma^\mathrm{L}\cup \mathcal{B}\cup\Sigma^\mathrm{R}$ constructed by the path integral over the ``lower half'' (with $-\pi < t_\mathrm{E} < 0$) of the Euclidean manifold with metric \eqref{eq:euclidean-polar-2} bounded by this Cauchy surface.  To identify this state as a double KMS state he first noted that this path integral can be interpreted as the following imaginary-time evolution operator up to a normalization factor:
\begin{equation}
    e^{-\frac{1}{2}\beta_\mathrm{H}H}: \mathscr{H}^{(\mathrm{S})}_{\Sigma^\mathrm{R}} \to \mathscr{H}^{(\mathrm{S})}_{\Sigma^\mathrm{L}}\eqend{,} 
\end{equation}
where $\mathscr{H}^{(\mathrm{S})}_{\Sigma^\mathrm{R}}$ and $\mathscr{H}^{(\mathrm{S})}_{\Sigma^\mathrm{L}}$ are the spaces of the Schr\"odinger wave functionals of states on $\Sigma^\mathrm{R}$ and $\Sigma^\mathrm{L}$, respectively.  Let $\{\psi^{(i)}_{\Sigma^\mathrm{R}}\}$ and $\{\psi^{(i)}_{\Sigma^\mathrm{L}}\}$ be complete sets of orthonormal wave functionals on $\Sigma^\mathrm{R}$ and $\Sigma^\mathrm{L}$, respectively.  Then, if $\langle\psi|\psi'\rangle$ is the inner product between the wave functionals $\psi,\psi'\in\mathscr{H}^{(\mathrm{S})}_{\Sigma^\mathrm{L}}$, this path integral gives the following Schr\"odinger wave functional on $\Sigma^\mathrm{L}\cup \mathcal{B}\cup\Sigma^\mathrm{R}$ (in a heuristic notation):
\begin{equation}
    \Psi_\mathrm{HH} \propto \sum_{i,j} \langle\psi^{(i)}_{\Sigma^\mathrm{L}}|e^{-\frac{1}{2}\beta_\mathrm{H}H}\psi^{(j)}_{\Sigma^\mathrm{R}}\rangle \psi^{(i)}_{\Sigma^\mathrm{L}}\otimes \psi^{(j)*}_{\Sigma^\mathrm{R}}\eqend{,}
\end{equation}
which is indeed a double KMS state with the Hawking temperature [see Eq.~\eqref{eq:double_kms_state}].

Jacobson's argument corresponds to the demonstration in this paper that the $4$-point function in the double KMS state $\Omega_\beta$ in Eq.~\eqref{eq:4pt-function-in-KMS} is given by a thermal average as in Eq.~\eqref{eq:4_point_function_trace}, which is a $4$-point function in the HH state if $\beta =\beta_\mathrm{H}$. In our operator-formalism derivation we needed to rewrite the $4$-point function involving operators in both wedges as a $4$-point function only with those in the right wedge using the KMS condition. There is no corresponding step in the path-integral derivation.

The operator approach in this paper makes it clear how the $N$-point functions in the HH state are found in perturbation theory. Also our derivation, based on Hamiltonian perturbation theory, can readily be extended to other quantum field theory, e.g.\ perturbative quantum gravity. 

The detailed discussion we presented of the analytic continuation of the Euclidean $N$-point functions both in the free and interacting theories is the main contribution of this paper. In summary, it shows that the Euclidean theory defines a bona fide state in the Lorentzian section in perturbation theory. Although this does not come as a surprise in the scalar field case, some authors have raised doubts about the validity of  Euclidean methods in perturbative quantum gravity around de~Sitter background~\cite{miao_et_al_prd_2014,woodard_ijmpd_2014}. An interesting application we foresee of the results of this paper is in investigating whether the gauge-fixed Euclidean partition function for quantum gravity in de~Sitter defines a good state when analytically continued to the global patch of de~Sitter spacetime. The free Euclidean vacuum for the graviton is known to be well defined, as it does not display IR divergences~\cite{allen_turyn,higuchi_kouris,faizal_higuchi_2011}. Thus, it will be interesting to use the Schwinger-Keldysh contour presented in this paper to define the interacting Euclidean vacuum, i.e., the HH state, in the global de~Sitter spacetime.  

In defining the HH state for perturbative gravity we would need to confront the infrared problem in the Faddeev-Popov ghost sector~\cite{allen_prd_1986, polchinski_plb_1989, taylor_veneziano_npb_1990}. Recently it has been proposed to solve this problem using certain conserved charges in this sector of the theory~\cite{faizal_higuchi_2008,gibbons_higuchi_lima_prd_2021}. Another challenge would be the conformal-mode problem in the Euclidean quantum gravity~\cite{gibbons_hawking_perry_npb_1978, schleich_prd_1987, mazur_mottola_npb_1990}. It would be interesting to see whether this problem could be circumvented by the Schwinger-Keldysh approach, which is intrinsically Lorentzian.

%%%%%%%%%%%%%%%%%%%%%%%%%%%%%%%%%%%%%%%%%%%%%%%%%%%%%%%%%%%%%%%%%%%%%%%%%%%%%%%%%%%%%%%%%%%%%%%%%%%%%%%%%%%%%%%%%%%%%%%%%%%%%%%%%%%%%%%%%%%%%%%%%%%%%
\acknowledgments

We thank Bernard Kay for discussions concerning the double KMS state in free-field theory and Ted Jacobson for useful correspondence.  We also thank Albert Roura and Markus Fr\"ob for discussions on Schwinger-Keldysh contours in de~Sitter spacetime. This work was supported in part by the Grant No.~RPG-2018-400, ``Euclidean and in-in formalisms in static spacetimes with Killing horizons'' from the Leverhulme Trust.
%%%%%%%%%%%%%%%%%%%%%%%%%%%%%%%%%%%%%%%%%%%%%%%%%%%%%%%%%%%%%%%%%%%%%%%%%%%%%%%%%%%%%%%%%%%%%%%%%%%%%%%%%%%%%%%%%%%%%%%%%%%%%%%%%%%%%%%%%%%%%%%%%%%%%

\appendix
%%%%%%%%%%%%%%%%%%%%%%%%%%%%%%%%%%%%%%%%%%%%%%%%%%%%%%%%%%%%%%%%%%%%%%%%%%%%%%%%%%%%%%%%%%%%%%%%%%%%%%%%%%%%%%%%%%%%%%%%%%%%%%%%%%%%%%%%%%%%%%%%%%%%%

%%%%%%%%%%%%%%%%%%%%%%%%%%%%%%%%%%%%%%%%%%%%%%%%%%%%%%%%%%%%%%%%%%%%%%%%%%%%%%%%%%%%%%%%%%%%%%%%%%%%%%%%%%%%%%%%%%%%%%%%%%%%%%%%%%%%%%%%%%%%%%%%%%%%%
\section{FREE FIELD IN THE FUTURE REGION}                                                                                                           %
\label{appendix:future-right}                                                                                                                       %
%%%%%%%%%%%%%%%%%%%%%%%%%%%%%%%%%%%%%%%%%%%%%%%%%%%%%%%%%%%%%%%%%%%%%%%%%%%%%%%%%%%%%%%%%%%%%%%%%%%%%%%%%%%%%%%%%%%%%%%%%%%%%%%%%%%%%%%%%%%%%%%%%%%%%

For the annihilation operator on the right wedge, the KMS condition~\eqref{eq:kms_condition} reads
\begin{splitequation}\label{eq:kms_annihilation_1}
  e^{-\frac{1}{2}\beta_\mathrm{H}\tilde{H}_0}a^{(\mathrm{R})}_\sigma(\omega)\Omega_{\beta_\mathrm{H}}^{(0)} 
	& = Ja_{\sigma}^{(\mathrm{R})\dagger}(\omega)\Omega_{\beta_\mathrm{H}}^{(0)}\\
	& = a_{\sigma}^{(\mathrm{L})\dagger}(\omega)\Omega_{\beta_\mathrm{H}}^{(0)}\eqend{,}
\end{splitequation}
where we have used Eqs.~\eqref{eq:invariance_kms_state_J} and~\eqref{eq:J-a-J} in the second equality. On the other hand, the commutator 
\begin{equation}
\left[\tilde{H}_0,a^{(\mathrm{R})}_\sigma(\omega)\right] = - \omega a^{(\mathrm{R})}_\sigma(\omega)\eqend{,}
\end{equation}
together with Eq.~\eqref{eq:invariance_kms_state_H}, implies that
\begin{equation}\label{eq:kms_annihilation_2}
e^{-\frac{1}{2}\beta_\mathrm{H}\tilde{H}_0}a_\sigma^{(\mathrm{R})}(\omega)\Omega_{\beta_\mathrm{H}}^{(0)} 
 = e^{\frac{\omega\beta_\mathrm{H}}{2}}a_\sigma^{(\mathrm{R})}(\omega)\Omega_{\beta_\mathrm{H}}^{(0)}\eqend{.}
\end{equation}
Then, by subtracting Eq.~\eqref{eq:kms_annihilation_2} from Eq.~\eqref{eq:kms_annihilation_1}, we obtain
\begin{equation}
    A^{(\mathrm{R})}_\sigma(\omega)\Omega_{\beta_\mathrm{H}}^{(0)} = 0\eqend{,}
\end{equation}
where
\begin{equation}
    A^{(\mathrm{R})}_\sigma(\omega) \equiv \frac{1}{\sqrt{1 - e^{-\omega\beta_\mathrm{H}}}}
    \left[ a^{(\mathrm{R})}_\sigma(\omega) - e^{-\frac{\omega\beta_\mathrm{H}}{2}}a^{(\mathrm{L})\dagger}_\sigma(\omega)\right]\eqend{.}
\end{equation}
One finds similarly,
\begin{equation}
    A^{(\mathrm{L})}_\sigma(\omega)\Omega_{\beta_\mathrm{H}}^{(0)} = 0\eqend{,}
\end{equation}
where
\begin{equation}
    A^{(\mathrm{L})}_\sigma(\omega) \equiv \frac{1}{\sqrt{1 - e^{-\omega\beta_\mathrm{H}}}}
    \left[ a^{(\mathrm{L})}_\sigma(\omega) - e^{-\frac{\omega\beta_\mathrm{H}}{2}}a^{(\mathrm{R})\dagger}_\sigma(\omega)\right]\eqend{.}
\end{equation}
The operators $A^{(\mathrm{R})}_\sigma(\omega)$ and $A^{(\mathrm{L})}_\sigma(\omega)$ are normalized so that
\begin{equation}
    \left[ A^{(\mathrm{R})}_\sigma(\omega),A^{(\mathrm{R})\dagger}_{\sigma'}(\omega')\right] = \delta_{\sigma\sigma'}\delta(\omega-\omega')\eqend{,}
\end{equation}
and the other commutators vanish. Similar commutators are found for $A^{(\mathrm{L})}_\sigma(\omega)$. The field operator with support on the right wedge can then be expanded as follows:
\begin{splitequation}\label{eq:state-expansion-R}
 \Phi_\mathrm{I}^{(\mathrm{R})}(t,\mathbf{x}) & 
 =\int_0^\infty \frac{d\omega}{\sqrt{2\omega(1-e^{-\omega\beta_\mathrm{H}})}}\sum_\sigma\\
 &\phantom{=} \times \phi_{\sigma\omega}(\mathbf{x})\left[ A^{(\mathrm{R})}_\sigma(\omega) e^{-i\omega t} + A^{(\mathrm{L})}_\sigma(\omega)
 e^{-\frac{\omega\beta_\mathrm{H}}{2} + i\omega t}\right. \\
 &\phantom{=} \left.+ A^{(\mathrm{R})\dagger}_\sigma (\omega)e^{i\omega t} + A^{(\mathrm{L})\dagger}_\sigma(\omega) e^{-\frac{\omega\beta_\mathrm{H}}{2} - i\omega t}  \right]\eqend{.}
\end{splitequation}

The coefficient functions of annihilation operators $A_\sigma^{(\mathrm{R})}(\omega)$ and $A_\sigma^{(\mathrm{L})}(\omega)$ are analytically continued to other regions as global positive-frequency modes, whereas those of creation operators are analytically continued as global negative-frequency modes. As the point $(T,\vt{X})$ goes across the horizon $X-T=0$ from $\mathcal{R}$ to $\mathcal{F}$, the positive frequency solution must be analytically continued with the following conditions: (i) its $T+X$ dependence should be the same in $\mathcal{R}$ and $\mathcal{F}$; (ii) the singularity $X-T$ must be avoided by letting $X-T \to X-T + i\epsilon$.  The condition (ii) comes from the fact that, since its high-frequency components with respect to the global time $T$ are of the form $e^{-ikT}$ with $k$ large and positive, this solution should be regarded as a distribution obtained by taking the $\epsilon\to 0^+$ limit with $T\to T - i\epsilon$ so that $e^{-ikT}\to 0$ as $k\to +\infty$.

To find out the implications of conditions (i) and (ii) to the static coordinates, we note that the combinations $X \pm T$ are expressed in terms of $t$ and $\rho$ as  
\begin{equation}
    X + T = \rho e^{\kappa t}\ \ \textrm{for}\ (T,\vt{X}) \in \mathcal{R} \cup \mathcal{F}
\end{equation}
and
\begin{equation}
    X - T = \begin{cases} \rho e^{-\kappa t} & \textrm{if}\ (T,\vt{X}) \in \mathcal{R}\eqend{,}\\
    -\rho e^{-\kappa t} & \textrm{if}\ (T,\vt{X}) \in \mathcal{F}\eqend{,} \end{cases}
\end{equation}
respectively. Hence, for the global positive-frequency solutions we must let $\rho \to e^{\frac{i\pi}{2}}\rho$ and $t\to t - i\beta_\mathrm{H}/4$, as the point $(T,\vt{X})$ traverses the horizon $X-T=0$ from $\mathcal{R}$, where $X-T > 0$, to $\mathcal{F}$, where $X-T < 0$. This implies that
\begin{equation}\label{eq:continuation_rule}
    \phi_{\omega\sigma}(\mathbf{x})e^{\pm i\omega t}\ (\textrm{in}\ \mathcal{R}) \to \tilde{\phi}_{\omega\sigma}(\mathbf{x})e^{\pm \frac{\omega\beta_\mathrm{H}}{4} \pm i\omega t}\ (\textrm{in}\ \mathcal{F})\eqend{,}
\end{equation}
where $\tilde{\phi}_{\omega\sigma}(\mathbf{x})$ is obtained from $\phi_{\omega\sigma}(\mathbf{x})$ by replacing $\rho$ by $e^{\frac{i\pi}{2}}\rho$.

The coefficient function multiplying creation operators in Eq.~\eqref{eq:state-expansion-R} must be continued as a global negative-frequency mode, for which $\rho\to e^{-\frac{i\pi}{2}}\rho$ and $t \to t + i\beta_\mathrm{H}/4$.  Thus, the free field in the future region $\mathcal{F}$ is obtained from Eq.~\eqref{eq:state-expansion-R} as follows:
\begin{splitequation} \label{eq:state-expansion-Future}
 \Phi_\mathrm{I}^{(\mathrm{F})}(t,\mathbf{x}) 
 & =  \int_0^\infty \frac{d\omega\,e^{-\frac{\omega\beta}{4}}}{\sqrt{2\omega(1-e^{-\omega\beta})}}\sum_\sigma \\
 & \phantom{=} \times \left\{\tilde{\phi}_{\sigma\omega}(\mathbf{x})\left[ A^{(\mathrm{R})}_\sigma(\omega) e^{-i\omega t} + A^{(\mathrm{L})}_\sigma(\omega)e^{i\omega t}\right]\right. \\
 & \phantom{=} \left. +\overline{\tilde{\phi}_{\omega\sigma}(\mathbf{x})}\left[ A^{(\mathrm{R})\dagger}_\sigma (\omega)e^{i\omega t} + A^{(\mathrm{L})\dagger}_\sigma(\omega) e^{- i\omega t}  \right]\right\}\eqend{.} \\
\end{splitequation}
Equations~\eqref{eq:first-eq-for-appB} and \eqref{eq:second-eq-for-appB} readily follow from Eqs.~\eqref{eq:state-expansion-R} and \eqref{eq:state-expansion-Future}.

%%%%%%%%%%%%%%%%%%%%%%%%%%%%%%%%%%%%%%%%%%%%%%%%%%%%%%%%%%%%%%%%%%%%%%%%%%%%%%%%%%%%%%%%%%%%%%%%%%%%%%%%%%%%%%%%%%%%%%%%%%%%%%%%%%%%%%%%%%%%%%%%%%%%%
\section{ANALYTIC CONTINUATION OF THE CORRELATION FUNCTIONS}                                                                                        %
\label{appendix:detail-analytic}                                                                                                                    %
%%%%%%%%%%%%%%%%%%%%%%%%%%%%%%%%%%%%%%%%%%%%%%%%%%%%%%%%%%%%%%%%%%%%%%%%%%%%%%%%%%%%%%%%%%%%%%%%%%%%%%%%%%%%%%%%%%%%%%%%%%%%%%%%%%%%%%%%%%%%%%%%%%%%%

In this Appendix we show that the $N$-point function defined by Eq.~\eqref{eq:expression-of-N-pt-fun2} is analytically continued by changing the real part of the time variables while keeping their imaginary part unchanged, under the assumptions made about the integration over $\Sigma^\mathrm{R}$.

Let $F(x_1,x_2,\ldots,x_N;y_1,y_2,\ldots,y_M)$, with $x_i = (t_i,\mathrm{x}_i)$ and $y_j = (\tau_j,\mathbf{y}_j)$, $\mathbf{x}_i,\mathbf{y}_j\in \Sigma^\mathrm{R}$, denote a product of the free-theory two-point functions $G^{(0)}(x,x')$ with the points from the set of external points $\{x_i\}_{i=1,2,\ldots,N}$ and the set of internal points $\{y_j\}_{j=1,2,\ldots,M}$.  We let each of the external points $x_i$ appear only once as an argument of a two-point function. An example is 
\begin{splitequation}
& F(x_1,x_2,x_3,x_4;y_1,y_2) \\
& = G^{(0)}(x_1,y_1)G^{(0)}(x_2,y_1)\left[G^{(0)}(y_1,y_2)\right]^2 \\
&\phantom{=} \times G^{(0)}(y_2,x_3)G^{(0)}(y_2,x_4)\eqend{,}
\end{splitequation}
which arises in the $\Phi^4$-theory. At each order in perturbation theory, the $N$-point function~\eqref{eq:expression-of-N-pt-fun2} is a finite sum of functions of the form
\begin{splitequation}
& \left(\prod_{j=1}^M \int_C d\tau_j \int_{\Sigma^\mathrm{R}} d^{n-1}\mathbf{y}_j\,\sqrt{-g(\mathbf{y}_j)}\right)\\
& \times F(x_1,x_2,\ldots,x_N;y_1,y_2,\ldots,y_M)\eqend{,}
\end{splitequation}
where $C$ is a contour in $\mathcal{C}_{\beta_\mathrm{H}}$ defined by Eq.~\eqref{eq:definition-of-C_beta} with the points $x_1,x_2,\ldots,x_N$ also on $C$.

Under the assumption we made about the integrals over $\Sigma^\mathrm{R}$---in effect we assume that these integrals are cut off in the ultraviolet and infrared---these integrals do not affect the analytic property with respect to the time variables $t_i$ and $\tau_j$. That is, if the integrand has a certain analytic property, then so does the result of the integration over $\Sigma^\mathrm{R}$.  Thus, we are led to consider
\begin{splitequation}
& I_C(t_1,t_2,\ldots,t_N) \\
&\quad \equiv  \left(\prod_{j=1}^M \int_C d\tau_j\right)
F(x_1,x_2,\ldots,x_N;y_1,y_2,\ldots,y_M)\eqend{.}
\end{splitequation}
What we need to show is that, if $\|\vt{y}_i - \vt{y}_j\|, \|\vt{y}_j - \vt{x}_i\| > \epsilon$ for all $i$ and $j$ for some $\epsilon > 0$, this function is analytically continued by changing the real part of $t_i \in C$ with the contour $C$ with monotonically decreasing imaginary part deformed so that $t_i$ are always on $C$.

Let us define an equivalence relation for $I_C$ as follows: $I_C\sim I_{C'}$ if $I_C(t_1,t_2,\ldots,t_N)$ and $I_{C'}(t_1',t_2',\ldots,t_N')$ are analytic continuations of each other, with the analytic continuation performed by changing the real part of $t_i$ but keeping its imaginary part fixed, i.e., $\textrm{Im}(t_i) = \textrm{Im}(t_i')$ for all $i$.  We now show that $I_C \sim I_{C'}$ for all $C$ and $C'$.
We define the horizontal distance (i.e., the distance along the real axis) between $C$ and $C'$ by
\begin{equation}
|C-C'| \equiv \textrm{max}\left\{|t - t'|: t\in C, t'\in C', \textrm{Im}(t)=\textrm{Im}(t')\right\}\eqend{.}
\end{equation}
Suppose that for some $d>0$ we have $I_C \sim I_{C'}$ for all $C$ and $C'$ satisfying $|C-C'|< d$. Then, since $\sim$ defined here is an equivalence relation, and thus transitive, we have $I_C \sim I_{C'}$ for all $C$ and $C'$ satisfying $|C-C'|< nd$ for any $n\in \mathbb{N}$.  This implies that $I_C \sim I_{C'}$ for all $C$ and $C'$.  Hence, all we need to show is that there is a number $d>0$ such that $I_{C} \sim I_{C'}$ if $|C-C'| < d$.  

In general the two-point function $G^{(0)}(x,x')$ with $x=(t,\mathbf{x})$ and $x'=(t',\mathbf{x})$ for the free scalar field is singular only if $\textrm{Im}(t) = \textrm{Im}(t')$ and the points $(\textrm{Re}(t),\vt{x})$ and $(\textrm{Re}(t'),\vt{x}')$ can be connected by a null geodesic~\cite{fulling_ruijsenaars_pr_1987}.  This implies that there exists a positive number $d$ such that the two-point function $G^{(0)}(t,\mathbf{x};t',\mathbf{x}')$ with the points satisfying $\|\mathbf{x}-\mathbf{x}'\|> \epsilon$ is an analytic function of $t$ and $t'$ in a open neighborhood without holes containing $C$ and $C'$ if $|C-C'| < d$.

Now, assume that $C$ and $C'$ satisfy $|C-C'|<d$ and define $I_{C',C}(t_1',t_2',\ldots,t_N')$ to be the function obtained by shifting each point $t_i$ in the real direction to $t_i'$, which is on $C'$.  Then, the function $I_{C',C}$ is an analytic continuation of $I_C$.  That is, $I_{C',C} \sim I_C$.  Now, the function $I_{C',C}$ is unchanged if we replace the contour of integration for $\tau_j$ from $C$ to $C'$ for any $j$. If we make this change of the contour for all $j$, then the resulting function is $I_{C'}$ by definition. That is, $I_{C'}=I_{C',C}$.  Hence we have $I_{C}\sim I_{C'}$ for all $C$ and $C'$.

%%%%%%%%%%%%%%%%%%%%%%%%%%%%%%%%%%%%%%%%%%%%%%%%%%%%%%%%%%%%%%%%%%%%%%%%%%%%%%%%%%%%%%%%%%%%%%%%%%%%%%%%%%%%%%%%%%%%%%%%%%%%%%%%%%%%%%%%%%%%%%%%%%%%%
\section{INTERACTION PICTURE WITH A TIME-DEPENDENT HAMILTONIAN}                                                                                     %
\label{appendix:int_with_t_dep_hamiltonian}                                                                                                         %
%%%%%%%%%%%%%%%%%%%%%%%%%%%%%%%%%%%%%%%%%%%%%%%%%%%%%%%%%%%%%%%%%%%%%%%%%%%%%%%%%%%%%%%%%%%%%%%%%%%%%%%%%%%%%%%%%%%%%%%%%%%%%%%%%%%%%%%%%%%%%%%%%%%%%

Let the Hamiltonian be given in the Schr\"odinger picture as
\begin{equation}
    H_\mathrm{S}(t) = H_{0,\mathrm{S}}(t) + H_\textrm{I,S}(t)\eqend{,}
\end{equation}
i.e., it is given in terms of the canonical operators $\Phi(t_\mathrm{i},\vt{x})$ and their canonical conjugate momenta with $t_\mathrm{i}$ fixed.  The Hamiltonian $H_{0,\mathrm{S}}(t)$ describes the free-field theory and $H_\textrm{I,S}(t)$ is the nonderivative interaction term. The explicitly time-dependence of $H_\mathrm{S}(t)$ arises from time-dependent functions multiplying these canonical operators.

The time evolution of a state $\Psi$ in the Schr\"odinger picture is given by
\begin{equation}
  i\frac{d\ }{dt}\Psi(t) = H_\mathrm{S}(t)\Psi(t)\eqend{.}
\end{equation}
This can be solved as
\begin{equation}
  \Psi(t) = U(t,t_\mathrm{i})\Psi(t_\mathrm{i})\eqend{,}
\end{equation}
where we have defined the time-evolution operator
\begin{equation}\label{eq:time-evolution-operator}
U(t,t_\mathrm{i}) \equiv \mathcal{P}\exp\left(-i\int_{t_\mathrm{i}}^t H_\mathrm{S}(\tau)d\tau\right)\eqend{,}
\end{equation}
with $\mathcal{P}$ indicating the path-ordering.  That is, products of the operators $H_\mathrm{S}(\tau)$ is time-ordered if $t > t_\mathrm{i}$ and anti-time-ordered if $t < t_\mathrm{i}$. The Heisenberg operator $\Phi(t,\vt{x})$ is given by
\begin{equation}\label{eq:Heisenberg_operator}
 \Phi(t,\vt{x}) = U(t_\mathrm{i},t)\Phi(t_\mathrm{i},\vt{x})U(t,t_\mathrm{i})\eqend{,}
\end{equation}
where $\Phi(t_\mathrm{i},\vt{x})$ is the field operator in the Schr\"odinger picture at any time $t$. The operator $\Phi(t,\vt{x})$ satisfies Heisenberg's equation of motion:
\begin{equation}
    i\frac{d\ }{dt}\Phi(t,\vt{x}) = \left[ \Phi(t,\vt{x}),H(t)\right]\eqend{,}
\end{equation}
where $H(t)$ is the Hamiltonian in the Heisenberg picture:
\begin{equation}\label{eq:Heisenberg_Hamiltonian}
     H(t) = U(t_\mathrm{i},t)H_\mathrm{S}(t)U(t,t_\mathrm{i})\eqend{.}
\end{equation}

The field operator in the interaction picture is defined as 
\begin{equation}
\Phi_\mathrm{I}(t,\vt{x}) = U_0(t_\mathrm{i},t)\Phi(t_\mathrm{i},\vt{x})U_0(t,t_\mathrm{i})\eqend{,}
\end{equation}
where $U_0(t,t_\mathrm{i})$ is the time-evolution operator of the free system, i.e., the operator defined in Eq.~\eqref{eq:time-evolution-operator} but with $H_\mathrm{S}(\tau)$ replaced by $H_{0,\mathrm{S}}(\tau)$.  Defined this way, the operator $\Phi_\mathrm{I}(t,\vt{x})$ satisfies the free-field equation:
\begin{equation}
    i\frac{d\ }{dt}\Phi_\mathrm{I}(t,\vt{x}) = \left[ \Phi_\mathrm{I}(t,\vt{x}),H_{0,\mathrm{I}}(t)\right]\eqend{,}
\end{equation}
where $H_{0,\mathrm{I}}(t)$ is the free-field Hamiltonian in the interaction picture. This operator is defined as 
\begin{equation}
 H_{0,\mathrm{I}}(t) = U_0(t_\mathrm{i},t)H_{0,\mathrm{S}}(t)U_0(t,t_\mathrm{i})\eqend{.}
\end{equation}

From Eq.~\eqref{eq:Heisenberg_operator} and the corresponding expression for $\Phi_\mathrm{I}(t,\vt{x})$ we find
\begin{equation}
    \Phi(t,\vt{x}) = U_\mathrm{I}(t_\mathrm{i},t)\Phi_\mathrm{I}(t,\vt{x})U_\mathrm{I}(t,t_\mathrm{i})\eqend{,}
\end{equation}
where
\begin{equation}
    U_\mathrm{I}(t,t_\mathrm{i}) \equiv U_0(t_\mathrm{i},t)U(t,t_\mathrm{i})\eqend{.}
\end{equation}
Then we find
\begin{equation}\label{eq:app-u-diff-eq}
\frac{d\ }{dt}U_\mathrm{I}(t,t_\mathrm{i}) = -iH_\mathrm{I}(t)U_\mathrm{I}(t,t_\mathrm{i})\eqend{,}
\end{equation}
where $H_\mathrm{I}(t)$ is the interaction term in the Hamiltonian in the interaction picture:
\begin{equation}
H_\mathrm{I}(t) = U_0(t_\mathrm{i},t)H_\textrm{I,S}(t)U_0(t,t_\mathrm{i})\eqend{.}
\end{equation}
From Eq.~\eqref{eq:app-u-diff-eq} we obtain
\begin{equation}
    U_\mathrm{I}(t,t_\mathrm{i}) = \mathcal{P}\exp\left( - i\int_{t_\mathrm{i}}^t H_\mathrm{I}(\tau)d\tau\right)\eqend{.} 
\end{equation}
Then, by defining $U_\mathrm{I}(t,t')$ for general arguments $t$ and $t'$ as
\begin{equation}
    U_\mathrm{I}(t,t') = U_\mathrm{I}(t,t_\mathrm{i})U_\mathrm{I}(t,t_\mathrm{i})^\dagger\eqend{,}
\end{equation}
we find that the Heisenberg operator $\Phi(t,\vt{x})$ is expressed in terms of the interaction-picture operators as in Eq.~\eqref{eq:interaction_pic_field} also for a time-dependent Hamiltonian.

%%%%%%%%%%%%%%%%%%%%%%%%%%%%%%%%%%%%%%%%%%%%%%%%%%%%%%%%%%%%%%%%%%%%%%
%%%%%%%%%%%%%%%%%%%%%%%%%% BIBLIOGRAPHY %%%%%%%%%%%%%%%%%%%%%%%%%%%%%%
%%%%%%%%%%%%%%%%%%%%%%%%%%%%%%%%%%%%%%%%%%%%%%%%%%%%%%%%%%%%%%%%%%%%%%
\bibliography{references}

\begin{thebibliography}{57}%
\makeatletter
\providecommand \@ifxundefined [1]{%
 \@ifx{#1\undefined}
}%
\providecommand \@ifnum [1]{%
 \ifnum #1\expandafter \@firstoftwo
 \else \expandafter \@secondoftwo
 \fi
}%
\providecommand \@ifx [1]{%
 \ifx #1\expandafter \@firstoftwo
 \else \expandafter \@secondoftwo
 \fi
}%
\providecommand \natexlab [1]{#1}%
\providecommand \enquote  [1]{``#1''}%
\providecommand \bibnamefont  [1]{#1}%
\providecommand \bibfnamefont [1]{#1}%
\providecommand \citenamefont [1]{#1}%
\providecommand \href@noop [0]{\@secondoftwo}%
\providecommand \href [0]{\begingroup \@sanitize@url \@href}%
\providecommand \@href[1]{\@@startlink{#1}\@@href}%
\providecommand \@@href[1]{\endgroup#1\@@endlink}%
\providecommand \@sanitize@url [0]{\catcode `\\12\catcode `\$12\catcode
  `\&12\catcode `\#12\catcode `\^12\catcode `\_12\catcode `\%12\relax}%
\providecommand \@@startlink[1]{}%
\providecommand \@@endlink[0]{}%
\providecommand \url  [0]{\begingroup\@sanitize@url \@url }%
\providecommand \@url [1]{\endgroup\@href {#1}{\urlprefix }}%
\providecommand \urlprefix  [0]{URL }%
\providecommand \Eprint [0]{\href }%
\providecommand \doibase [0]{https://doi.org/}%
\providecommand \selectlanguage [0]{\@gobble}%
\providecommand \bibinfo  [0]{\@secondoftwo}%
\providecommand \bibfield  [0]{\@secondoftwo}%
\providecommand \translation [1]{[#1]}%
\providecommand \BibitemOpen [0]{}%
\providecommand \bibitemStop [0]{}%
\providecommand \bibitemNoStop [0]{.\EOS\space}%
\providecommand \EOS [0]{\spacefactor3000\relax}%
\providecommand \BibitemShut  [1]{\csname bibitem#1\endcsname}%
\let\auto@bib@innerbib\@empty
%</preamble>
\bibitem [{\citenamefont {Hartle}\ and\ \citenamefont
  {Hawking}(1976)}]{hartle_hawking_prd_1976}%
  \BibitemOpen
  \bibfield  {author} {\bibinfo {author} {\bibfnamefont {J.~B.}\ \bibnamefont
  {Hartle}}\ and\ \bibinfo {author} {\bibfnamefont {S.~W.}\ \bibnamefont
  {Hawking}},\ }\bibfield  {title} {\bibinfo {title} {{Path-integral derivation
  of black-hole radiance}},\ }\href {https://doi.org/10.1103/PhysRevD.13.2188}
  {\bibfield  {journal} {\bibinfo  {journal} {Phys.~Rev.~D}\ }\textbf {\bibinfo
  {volume} {13}},\ \bibinfo {pages} {2188} (\bibinfo {year}
  {1976})}\BibitemShut {NoStop}%
\bibitem [{\citenamefont {Dowker}(1978)}]{dowker_prd_1978}%
  \BibitemOpen
  \bibfield  {author} {\bibinfo {author} {\bibfnamefont {J.~S.}\ \bibnamefont
  {Dowker}},\ }\bibfield  {title} {\bibinfo {title} {{Thermal properties of
  Green's functions in Rindler, de~Sitter, and Schwarzschild spaces}},\ }\href
  {https://doi.org/10.1103/PhysRevD.18.1856} {\bibfield  {journal} {\bibinfo
  {journal} {Phys.~Rev.~D}\ }\textbf {\bibinfo {volume} {18}},\ \bibinfo
  {pages} {1856} (\bibinfo {year} {1978})}\BibitemShut {NoStop}%
\bibitem [{\citenamefont {Unruh}(1976)}]{unruh_prd_1976}%
  \BibitemOpen
  \bibfield  {author} {\bibinfo {author} {\bibfnamefont {W.~G.}\ \bibnamefont
  {Unruh}},\ }\bibfield  {title} {\bibinfo {title} {{Notes on black hole
  evaporation}},\ }\href {https://doi.org/10.1103/PhysRevD.14.870} {\bibfield
  {journal} {\bibinfo  {journal} {Phys.~Rev.~D}\ }\textbf {\bibinfo {volume}
  {14}},\ \bibinfo {pages} {870} (\bibinfo {year} {1976})}\BibitemShut
  {NoStop}%
%%CITATION = PHRVA,D14,870;%%
\bibitem [{\citenamefont {Chernikov}\ and\ \citenamefont
  {Tagirov}(1968)}]{chernikov_tagirov_aihp_1968}%
  \BibitemOpen
  \bibfield  {author} {\bibinfo {author} {\bibfnamefont {N.~A.}\ \bibnamefont
  {Chernikov}}\ and\ \bibinfo {author} {\bibfnamefont {E.~A.}\ \bibnamefont
  {Tagirov}},\ }\bibfield  {title} {\bibinfo {title} {{Quantum theory of scalar
  fields in de Sitter space-time}},\ }\href
  {http://www.numdam.org/item/AIHPA_1968__9_2_109_0} {\bibfield  {journal}
  {\bibinfo  {journal} {Ann.~Inst.~Henri~Poincar{\'e}~Phys.~Theor.}\ }\textbf
  {\bibinfo {volume} {9}},\ \bibinfo {pages} {109} (\bibinfo {year}
  {1968})}\BibitemShut {NoStop}%
\bibitem [{\citenamefont {Schomblond}\ and\ \citenamefont
  {Spindel}(1976)}]{schomblond_spindel_aihp_1976}%
  \BibitemOpen
  \bibfield  {author} {\bibinfo {author} {\bibfnamefont {C.}~\bibnamefont
  {Schomblond}}\ and\ \bibinfo {author} {\bibfnamefont {P.}~\bibnamefont
  {Spindel}},\ }\bibfield  {title} {\bibinfo {title} {{Conditions d'unicit{\'e}
  pour le propagateur $\Delta^1(x,y)$ du champ scalaire dans l'univers de de
  Sitter}},\ }\href {http://www.numdam.org/item/AIHPA_1976__25_1_67_0}
  {\bibfield  {journal} {\bibinfo  {journal}
  {Ann.~Inst.~Henri~Poincar{\'e}~Phys.~Theor.}\ }\textbf {\bibinfo {volume}
  {25}},\ \bibinfo {pages} {67} (\bibinfo {year} {1976})}\BibitemShut {NoStop}%
\bibitem [{\citenamefont {Bunch}\ and\ \citenamefont
  {Davies}(1978)}]{bunch_davies_prsl_1978}%
  \BibitemOpen
  \bibfield  {author} {\bibinfo {author} {\bibfnamefont {T.~S.}\ \bibnamefont
  {Bunch}}\ and\ \bibinfo {author} {\bibfnamefont {P.~C.~W.}\ \bibnamefont
  {Davies}},\ }\bibfield  {title} {\bibinfo {title} {{Quantum Field Theory in
  de Sitter Space: Renormalization by Point Splitting}},\ }\href
  {https://doi.org/10.1098/rspa.1978.0060} {\bibfield  {journal} {\bibinfo
  {journal} {Proc.~R.~Soc.~A}\ }\textbf {\bibinfo {volume} {360}},\ \bibinfo
  {pages} {117} (\bibinfo {year} {1978})}\BibitemShut {NoStop}%
\bibitem [{\citenamefont {Gibbons}\ and\ \citenamefont
  {Hawking}(1977)}]{gibbons_hawking_prd_1977}%
  \BibitemOpen
  \bibfield  {author} {\bibinfo {author} {\bibfnamefont {G.~W.}\ \bibnamefont
  {Gibbons}}\ and\ \bibinfo {author} {\bibfnamefont {S.~W.}\ \bibnamefont
  {Hawking}},\ }\bibfield  {title} {\bibinfo {title} {{Cosmological event
  horizons, thermodynamics, and particle Creation}},\ }\href
  {https://doi.org/10.1103/PhysRevD.15.2738} {\bibfield  {journal} {\bibinfo
  {journal} {Phys.~Rev.~D}\ }\textbf {\bibinfo {volume} {15}},\ \bibinfo
  {pages} {2738} (\bibinfo {year} {1977})}\BibitemShut {NoStop}%
\bibitem [{\citenamefont {Higuchi}\ \emph {et~al.}(2011)\citenamefont
  {Higuchi}, \citenamefont {Marolf},\ and\ \citenamefont
  {Morrison}}]{higuchi_marolf_morrison_prd_2011}%
  \BibitemOpen
  \bibfield  {author} {\bibinfo {author} {\bibfnamefont {A.}~\bibnamefont
  {Higuchi}}, \bibinfo {author} {\bibfnamefont {D.}~\bibnamefont {Marolf}},\
  and\ \bibinfo {author} {\bibfnamefont {I.~A.}\ \bibnamefont {Morrison}},\
  }\bibfield  {title} {\bibinfo {title} {{Equivalence between Euclidean and
  in-in formalisms in de~Sitter QFT}},\ }\href
  {https://doi.org/10.1103/PhysRevD.83.084029} {\bibfield  {journal} {\bibinfo
  {journal} {Phys.~Rev.~D}\ }\textbf {\bibinfo {volume} {83}},\ \bibinfo
  {pages} {084029} (\bibinfo {year} {2011})},\ \Eprint
  {https://arxiv.org/abs/1012.3415} {arXiv:1012.3415 [gr-qc]} \BibitemShut
  {NoStop}%
%%CITATION = ARXIV:1012.3415;%%
\bibitem [{\citenamefont {Korai}\ and\ \citenamefont
  {Tanaka}(2013)}]{korai_tanaka_2013}%
  \BibitemOpen
  \bibfield  {author} {\bibinfo {author} {\bibfnamefont {Y.}~\bibnamefont
  {Korai}}\ and\ \bibinfo {author} {\bibfnamefont {T.}~\bibnamefont {Tanaka}},\
  }\bibfield  {title} {\bibinfo {title} {{Quantum field theory in the flat
  chart of de~Sitter space}},\ }\href
  {https://doi.org/10.1103/PhysRevD.87.024013} {\bibfield  {journal} {\bibinfo
  {journal} {Phys. Rev. D}\ }\textbf {\bibinfo {volume} {87}},\ \bibinfo
  {pages} {024013} (\bibinfo {year} {2013})},\ \Eprint
  {https://arxiv.org/abs/1210.6544} {arXiv:1210.6544 [gr-qc]} \BibitemShut
  {NoStop}%
\bibitem [{\citenamefont {Marolf}\ and\ \citenamefont
  {Morrison}(2010)}]{marolf_morrison_prd_2010}%
  \BibitemOpen
  \bibfield  {author} {\bibinfo {author} {\bibfnamefont {D.}~\bibnamefont
  {Marolf}}\ and\ \bibinfo {author} {\bibfnamefont {I.~A.}\ \bibnamefont
  {Morrison}},\ }\bibfield  {title} {\bibinfo {title} {{The IR stability of
  de~Sitter: Loop corrections to scalar propagators}},\ }\href
  {https://doi.org/10.1103/PhysRevD.82.105032} {\bibfield  {journal} {\bibinfo
  {journal} {Phys.~Rev.~D}\ }\textbf {\bibinfo {volume} {82}},\ \bibinfo
  {pages} {105032} (\bibinfo {year} {2010})},\ \Eprint
  {https://arxiv.org/abs/1006.0035} {arXiv:1006.0035 [gr-qc]} \BibitemShut
  {NoStop}%
\bibitem [{\citenamefont {Marolf}\ and\ \citenamefont
  {Morrison}(2011)}]{marolf_morrison_prd_2011}%
  \BibitemOpen
  \bibfield  {author} {\bibinfo {author} {\bibfnamefont {D.}~\bibnamefont
  {Marolf}}\ and\ \bibinfo {author} {\bibfnamefont {I.~A.}\ \bibnamefont
  {Morrison}},\ }\bibfield  {title} {\bibinfo {title} {{The IR stability of
  de~Sitter QFT: results at all orders}},\ }\href
  {https://doi.org/10.1103/PhysRevD.84.044040} {\bibfield  {journal} {\bibinfo
  {journal} {Phys.~Rev.~D}\ }\textbf {\bibinfo {volume} {84}},\ \bibinfo
  {pages} {044040} (\bibinfo {year} {2011})},\ \Eprint
  {https://arxiv.org/abs/1010.5327} {arXiv:1010.5327 [gr-qc]} \BibitemShut
  {NoStop}%
\bibitem [{\citenamefont {Hollands}(2013)}]{hollands_cmp_2013}%
  \BibitemOpen
  \bibfield  {author} {\bibinfo {author} {\bibfnamefont {S.}~\bibnamefont
  {Hollands}},\ }\bibfield  {title} {\bibinfo {title} {{Correlators, Feynman
  diagrams, and quantum no-hair in de~Sitter spacetime}},\ }\href
  {https://doi.org/10.1007/s00220-012-1653-2} {\bibfield  {journal} {\bibinfo
  {journal} {Commun.~Math.~Phys.}\ }\textbf {\bibinfo {volume} {319}},\
  \bibinfo {pages} {1} (\bibinfo {year} {2013})},\ \Eprint
  {https://arxiv.org/abs/1010.5367} {arXiv:1010.5367 [gr-qc]} \BibitemShut
  {NoStop}%
\bibitem [{\citenamefont {Rajaraman}(2010)}]{rajaraman_prd_2010}%
  \BibitemOpen
  \bibfield  {author} {\bibinfo {author} {\bibfnamefont {A.}~\bibnamefont
  {Rajaraman}},\ }\bibfield  {title} {\bibinfo {title} {{Proper treatment of
  massless fields in Euclidean de~Sitter space}},\ }\href
  {https://doi.org/10.1103/PhysRevD.82.123522} {\bibfield  {journal} {\bibinfo
  {journal} {Phys.~Rev.~D}\ }\textbf {\bibinfo {volume} {82}},\ \bibinfo
  {pages} {123522} (\bibinfo {year} {2010})},\ \Eprint
  {https://arxiv.org/abs/1008.1271} {arXiv:1008.1271 [hep-th]} \BibitemShut
  {NoStop}%
\bibitem [{\citenamefont {Kay}(1985{\natexlab{a}})}]{kay_cmp_1985}%
  \BibitemOpen
  \bibfield  {author} {\bibinfo {author} {\bibfnamefont {B.~S.}\ \bibnamefont
  {Kay}},\ }\bibfield  {title} {\bibinfo {title} {{The double wedge algebra for
  quantum fields on Schwarzschild and Minkowski spacetimes}},\ }\href
  {https://doi.org/10.1007/BF01212687} {\bibfield  {journal} {\bibinfo
  {journal} {Commun.~Math.~Phys.}\ }\textbf {\bibinfo {volume} {100}},\
  \bibinfo {pages} {57} (\bibinfo {year} {1985}{\natexlab{a}})}\BibitemShut
  {NoStop}%
\bibitem [{\citenamefont {Kruskal}(1960)}]{kruskal_pr_1960}%
  \BibitemOpen
  \bibfield  {author} {\bibinfo {author} {\bibfnamefont {M.~D.}\ \bibnamefont
  {Kruskal}},\ }\bibfield  {title} {\bibinfo {title} {{Maximal extension of
  Schwarzschild metric}},\ }\href {https://doi.org/10.1103/PhysRev.119.1743}
  {\bibfield  {journal} {\bibinfo  {journal} {Phys.~Rev.}\ }\textbf {\bibinfo
  {volume} {119}},\ \bibinfo {pages} {1743} (\bibinfo {year}
  {1960})}\BibitemShut {NoStop}%
\bibitem [{\citenamefont {Szekeres}(2002)}]{szekeres_grg_2002}%
  \BibitemOpen
  \bibfield  {author} {\bibinfo {author} {\bibfnamefont {G.}~\bibnamefont
  {Szekeres}},\ }\bibfield  {title} {\bibinfo {title} {{Golden oldie: On the
  singularities of a Riemannian manifold}},\ }\href
  {https://doi.org/10.1023/A:1020744914721} {\bibfield  {journal} {\bibinfo
  {journal} {Gen.~Relativ.~Gravit.}\ }\textbf {\bibinfo {volume} {34}},\
  \bibinfo {pages} {2001} (\bibinfo {year} {2002})},\ \bibinfo {note}
  {[Publ.~Math.~Debrecen, {\bf 7}, 285 (1960)]}\BibitemShut {NoStop}%
\bibitem [{\citenamefont {Israel}(1976)}]{israel_pla_1976}%
  \BibitemOpen
  \bibfield  {author} {\bibinfo {author} {\bibfnamefont {W.}~\bibnamefont
  {Israel}},\ }\bibfield  {title} {\bibinfo {title} {{Thermo-field dynamics of
  black holes}},\ }\href {https://doi.org/10.1016/0375-9601(76)90178-X}
  {\bibfield  {journal} {\bibinfo  {journal} {Phys.~Lett.}\ }\textbf {\bibinfo
  {volume} {57A}},\ \bibinfo {pages} {107} (\bibinfo {year}
  {1976})}\BibitemShut {NoStop}%
\bibitem [{\citenamefont {Kay}\ and\ \citenamefont
  {Wald}(1991)}]{kay_wald_pr_1991}%
  \BibitemOpen
  \bibfield  {author} {\bibinfo {author} {\bibfnamefont {B.~S.}\ \bibnamefont
  {Kay}}\ and\ \bibinfo {author} {\bibfnamefont {R.~M.}\ \bibnamefont {Wald}},\
  }\bibfield  {title} {\bibinfo {title} {{Theorems on the uniqueness and
  thermal properties of stationary, nonsingular, quasifree states on
  space-times with a bifurcate Killing horizon}},\ }\href
  {https://doi.org/10.1016/0370-1573(91)90015-E} {\bibfield  {journal}
  {\bibinfo  {journal} {Phys.~Rep.}\ }\textbf {\bibinfo {volume} {207}},\
  \bibinfo {pages} {49} (\bibinfo {year} {1991})}\BibitemShut {NoStop}%
%%CITATION = PRPLC,207,49;%%
\bibitem [{\citenamefont {Gibbons}\ and\ \citenamefont
  {Perry}(1978)}]{gibbons_perry_prsl_1978}%
  \BibitemOpen
  \bibfield  {author} {\bibinfo {author} {\bibfnamefont {G.~W.}\ \bibnamefont
  {Gibbons}}\ and\ \bibinfo {author} {\bibfnamefont {M.~J.}\ \bibnamefont
  {Perry}},\ }\bibfield  {title} {\bibinfo {title} {{Black holes and thermal
  Green's functions}},\ }\href {https://doi.org/10.1098/rspa.1978.0022}
  {\bibfield  {journal} {\bibinfo  {journal} {Proc.~R.~Soc.~A}\ }\textbf
  {\bibinfo {volume} {358}},\ \bibinfo {pages} {467} (\bibinfo {year}
  {1978})}\BibitemShut {NoStop}%
%%CITATION = PRSLA,A358,467;%%
\bibitem [{\citenamefont {Unruh}\ and\ \citenamefont
  {Weiss}(1984)}]{unruh_weiss_prd_1984}%
  \BibitemOpen
  \bibfield  {author} {\bibinfo {author} {\bibfnamefont {W.~G.}\ \bibnamefont
  {Unruh}}\ and\ \bibinfo {author} {\bibfnamefont {N.}~\bibnamefont {Weiss}},\
  }\bibfield  {title} {\bibinfo {title} {{Acceleration radiation in interacting
  field theories}},\ }\href {https://doi.org/10.1103/PhysRevD.29.1656}
  {\bibfield  {journal} {\bibinfo  {journal} {Phys.~Rev.~D}\ }\textbf {\bibinfo
  {volume} {29}},\ \bibinfo {pages} {1656} (\bibinfo {year}
  {1984})}\BibitemShut {NoStop}%
\bibitem [{\citenamefont {Barvinsky}\ \emph {et~al.}(1995)\citenamefont
  {Barvinsky}, \citenamefont {Frolov},\ and\ \citenamefont
  {Zelnikov}}]{barvinsky_frolov_zelnikov_prd_1994}%
  \BibitemOpen
  \bibfield  {author} {\bibinfo {author} {\bibfnamefont {A.~O.}\ \bibnamefont
  {Barvinsky}}, \bibinfo {author} {\bibfnamefont {V.~P.}\ \bibnamefont
  {Frolov}},\ and\ \bibinfo {author} {\bibfnamefont {A.~I.}\ \bibnamefont
  {Zelnikov}},\ }\bibfield  {title} {\bibinfo {title} {{Wavefunction of a black
  hole and the dynamical origin of entropy}},\ }\href
  {https://doi.org/10.1103/PhysRevD.51.1741} {\bibfield  {journal} {\bibinfo
  {journal} {Phys.~Rev.~D}\ }\textbf {\bibinfo {volume} {51}},\ \bibinfo
  {pages} {1741} (\bibinfo {year} {1995})},\ \Eprint
  {https://arxiv.org/abs/gr-qc/9404036} {arXiv:gr-qc/9404036} \BibitemShut
  {NoStop}%
\bibitem [{\citenamefont {Jacobson}(1994)}]{jacobson_prd_1994}%
  \BibitemOpen
  \bibfield  {author} {\bibinfo {author} {\bibfnamefont {T.}~\bibnamefont
  {Jacobson}},\ }\bibfield  {title} {\bibinfo {title} {{A note on
  Hartle-Hawking vacua}},\ }\href {https://doi.org/10.1103/PhysRevD.50.R6031}
  {\bibfield  {journal} {\bibinfo  {journal} {Phys.~Rev.~D}\ }\textbf {\bibinfo
  {volume} {50}},\ \bibinfo {pages} {6031} (\bibinfo {year} {1994})},\ \Eprint
  {https://arxiv.org/abs/gr-qc/9407022} {arXiv:gr-qc/9407022} \BibitemShut
  {NoStop}%
\bibitem [{\citenamefont {Sanders}(2015)}]{sanders_lmp_2015}%
  \BibitemOpen
  \bibfield  {author} {\bibinfo {author} {\bibfnamefont {K.}~\bibnamefont
  {Sanders}},\ }\bibfield  {title} {\bibinfo {title} {{On the construction of
  Hartle-Hawking-Israel states across a static bifurcate Killing horizon}},\
  }\href {https://doi.org/10.1007/s11005-015-0745-2} {\bibfield  {journal}
  {\bibinfo  {journal} {Lett.~Math.~Phys.}\ }\textbf {\bibinfo {volume}
  {105}},\ \bibinfo {pages} {575} (\bibinfo {year} {2015})},\ \Eprint
  {https://arxiv.org/abs/1310.5537} {arXiv:1310.5537 [gr-qc]} \BibitemShut
  {NoStop}%
\bibitem [{\citenamefont {Schwinger}(1961)}]{schwinger_jmp_1961}%
  \BibitemOpen
  \bibfield  {author} {\bibinfo {author} {\bibfnamefont {J.~S.}\ \bibnamefont
  {Schwinger}},\ }\bibfield  {title} {\bibinfo {title} {{Brownian motion of a
  quantum oscillator}},\ }\href {https://doi.org/10.1063/1.1703727} {\bibfield
  {journal} {\bibinfo  {journal} {J.~Math.~Phys.~(N.Y.)}\ }\textbf {\bibinfo
  {volume} {2}},\ \bibinfo {pages} {407} (\bibinfo {year} {1961})}\BibitemShut
  {NoStop}%
%%CITATION = JMAPA,2,407;%%
\bibitem [{\citenamefont {Keldysh}(1964)}]{keldysh_zetf_1964}%
  \BibitemOpen
  \bibfield  {author} {\bibinfo {author} {\bibfnamefont {L.~V.}\ \bibnamefont
  {Keldysh}},\ }\bibfield  {title} {\bibinfo {title} {{Diagram technique for
  nonequilibrium processes}},\ }\href@noop {} {\bibfield  {journal} {\bibinfo
  {journal} {Zh.~Eksp.~Teor.~Fiz.}\ }\textbf {\bibinfo {volume} {47}},\
  \bibinfo {pages} {1515} (\bibinfo {year} {1964})},\ \bibinfo {note}
  {[\href{http://jetp.ras.ru/cgi-bin/dn/e_020_04_1018.pdf}{Sov.~Phys.~JETP {\bf
  20}, 1018 (1965)}]}\BibitemShut {NoStop}%
%%CITATION = ZETFA,47,1515;%%
\bibitem [{\citenamefont {Calzetta}\ and\ \citenamefont
  {Hu}(2008)}]{calzetta_hu_book}%
  \BibitemOpen
  \bibfield  {author} {\bibinfo {author} {\bibfnamefont {E.~A.}\ \bibnamefont
  {Calzetta}}\ and\ \bibinfo {author} {\bibfnamefont {B.-L.~B.}\ \bibnamefont
  {Hu}},\ }\href {https://doi.org/10.1017/CBO9780511535123} {\emph {\bibinfo
  {title} {{Nonequilibrium Quantum Field Theory}}}},\ Cambridge Monographs on
  Mathematical Physics\ (\bibinfo  {publisher} {Cambridge University Press},\
  \bibinfo {address} {Cambridge, England},\ \bibinfo {year} {2008})\BibitemShut
  {NoStop}%
\bibitem [{\citenamefont {Bisognano}\ and\ \citenamefont
  {Wichmann}(1975)}]{bisognano_wichmann_jmp_1975}%
  \BibitemOpen
  \bibfield  {author} {\bibinfo {author} {\bibfnamefont {J.~J.}\ \bibnamefont
  {Bisognano}}\ and\ \bibinfo {author} {\bibfnamefont {E.~H.}\ \bibnamefont
  {Wichmann}},\ }\bibfield  {title} {\bibinfo {title} {{On the duality
  condition for a Hermitian scalar field}},\ }\href
  {https://doi.org/10.1063/1.522605} {\bibfield  {journal} {\bibinfo  {journal}
  {J.~Math.~Phys.~(N.Y.)}\ }\textbf {\bibinfo {volume} {16}},\ \bibinfo {pages}
  {985} (\bibinfo {year} {1975})}\BibitemShut {NoStop}%
%%CITATION = JMAPA,16,985;%%
\bibitem [{\citenamefont {Bisognano}\ and\ \citenamefont
  {Wichmann}(1976)}]{bisognano_wichmann_jmp_1976}%
  \BibitemOpen
  \bibfield  {author} {\bibinfo {author} {\bibfnamefont {J.~J.}\ \bibnamefont
  {Bisognano}}\ and\ \bibinfo {author} {\bibfnamefont {E.~H.}\ \bibnamefont
  {Wichmann}},\ }\bibfield  {title} {\bibinfo {title} {{On the duality
  condition for quantum fields}},\ }\href {https://doi.org/10.1063/1.522898}
  {\bibfield  {journal} {\bibinfo  {journal} {J.~Math.~Phys.~(N.Y.)}\ }\textbf
  {\bibinfo {volume} {17}},\ \bibinfo {pages} {303} (\bibinfo {year}
  {1976})}\BibitemShut {NoStop}%
%%CITATION = JMAPA,17,303;%%
\bibitem [{\citenamefont {Sewell}(1982)}]{sewell_ap_1982}%
  \BibitemOpen
  \bibfield  {author} {\bibinfo {author} {\bibfnamefont {G.~L.}\ \bibnamefont
  {Sewell}},\ }\bibfield  {title} {\bibinfo {title} {{Quantum fields on
  manifolds: PCT and gravitationally induced thermal states}},\ }\href
  {https://doi.org/10.1016/0003-4916(82)90285-8} {\bibfield  {journal}
  {\bibinfo  {journal} {Ann.~Phys.~(N.Y.)}\ }\textbf {\bibinfo {volume}
  {141}},\ \bibinfo {pages} {201} (\bibinfo {year} {1982})}\BibitemShut
  {NoStop}%
%%CITATION = APNYA,141,201;%%
\bibitem [{\citenamefont {Wald}(1984)}]{wald_gr_book}%
  \BibitemOpen
  \bibfield  {author} {\bibinfo {author} {\bibfnamefont {R.~M.}\ \bibnamefont
  {Wald}},\ }\href {https://doi.org/10.7208/chicago/9780226870373.001.0001}
  {\emph {\bibinfo {title} {{General Relativity}}}}\ (\bibinfo  {publisher}
  {The University of Chicago Press},\ \bibinfo {address} {Chicago, USA},\
  \bibinfo {year} {1984})\BibitemShut {NoStop}%
\bibitem [{\citenamefont {Boyer}(1969)}]{boyer_prsla_1969}%
  \BibitemOpen
  \bibfield  {author} {\bibinfo {author} {\bibfnamefont {R.~H.}\ \bibnamefont
  {Boyer}},\ }\bibfield  {title} {\bibinfo {title} {{Geodesic Killing orbits
  and bifurcate Killing horizons}},\ }\href
  {https://doi.org/10.1098/rspa.1969.0116} {\bibfield  {journal} {\bibinfo
  {journal} {Proc.~R.~Soc.~A}\ }\textbf {\bibinfo {volume} {311}},\ \bibinfo
  {pages} {245} (\bibinfo {year} {1969})}\BibitemShut {NoStop}%
\bibitem [{\citenamefont {Takahashi}\ and\ \citenamefont
  {Umezawa}(1996)}]{takahashi_umezawa_ijmpb_1996}%
  \BibitemOpen
  \bibfield  {author} {\bibinfo {author} {\bibfnamefont {Y.}~\bibnamefont
  {Takahashi}}\ and\ \bibinfo {author} {\bibfnamefont {H.}~\bibnamefont
  {Umezawa}},\ }\bibfield  {title} {\bibinfo {title} {{Thermo field
  dynamics}},\ }\href {https://doi.org/10.1142/S0217979296000817} {\bibfield
  {journal} {\bibinfo  {journal} {Int.~J.~Mod.~Phys.~B}\ }\textbf {\bibinfo
  {volume} {10}},\ \bibinfo {pages} {1755} (\bibinfo {year} {1996})},\ \bibinfo
  {note} {[Collective Phenomena {\bf 2}, 55 (1975)]}\BibitemShut {NoStop}%
\bibitem [{\citenamefont {Kubo}(1957)}]{kubo_jpsj_1957}%
  \BibitemOpen
  \bibfield  {author} {\bibinfo {author} {\bibfnamefont {R.}~\bibnamefont
  {Kubo}},\ }\bibfield  {title} {\bibinfo {title} {{Statistical mechanical
  theory of irreversible processes. I. General theory and simple applications
  in magnetic and conduction problems}},\ }\href
  {https://doi.org/10.1143/JPSJ.12.570} {\bibfield  {journal} {\bibinfo
  {journal} {J.~Phys.~Soc.~Jpn.}\ }\textbf {\bibinfo {volume} {12}},\ \bibinfo
  {pages} {570} (\bibinfo {year} {1957})}\BibitemShut {NoStop}%
%%CITATION = JUPSA,12,570;%%
\bibitem [{\citenamefont {Martin}\ and\ \citenamefont
  {Schwinger}(1959)}]{martin_schwinger_pr_1959}%
  \BibitemOpen
  \bibfield  {author} {\bibinfo {author} {\bibfnamefont {P.~C.}\ \bibnamefont
  {Martin}}\ and\ \bibinfo {author} {\bibfnamefont {J.~S.}\ \bibnamefont
  {Schwinger}},\ }\bibfield  {title} {\bibinfo {title} {{Theory of many
  particle systems. I}},\ }\href {https://doi.org/10.1103/PhysRev.115.1342}
  {\bibfield  {journal} {\bibinfo  {journal} {Phys. Rev.}\ }\textbf {\bibinfo
  {volume} {115}},\ \bibinfo {pages} {1342} (\bibinfo {year}
  {1959})}\BibitemShut {NoStop}%
%%CITATION = PHRVA,115,1342;%%
\bibitem [{\citenamefont {Haag}\ \emph {et~al.}(1967)\citenamefont {Haag},
  \citenamefont {Hugenholtz},\ and\ \citenamefont
  {Winnink}}]{haag_hugenholtz_Winnink_cmp_1967}%
  \BibitemOpen
  \bibfield  {author} {\bibinfo {author} {\bibfnamefont {R.}~\bibnamefont
  {Haag}}, \bibinfo {author} {\bibfnamefont {N.~M.}\ \bibnamefont
  {Hugenholtz}},\ and\ \bibinfo {author} {\bibfnamefont {M.}~\bibnamefont
  {Winnink}},\ }\bibfield  {title} {\bibinfo {title} {{On the equilibrium
  states in quantum statistical mechanics}},\ }\href
  {https://doi.org/10.1007/BF01646342} {\bibfield  {journal} {\bibinfo
  {journal} {Commun.~Math.~Phys.}\ }\textbf {\bibinfo {volume} {5}},\ \bibinfo
  {pages} {215} (\bibinfo {year} {1967})}\BibitemShut {NoStop}%
%%CITATION = CMPHA,5,215;%%
\bibitem [{\citenamefont {Kay}(1985{\natexlab{b}})}]{kay_hpa_1985}%
  \BibitemOpen
  \bibfield  {author} {\bibinfo {author} {\bibfnamefont {B.~S.}\ \bibnamefont
  {Kay}},\ }\bibfield  {title} {\bibinfo {title} {{Purification of KMS
  states}},\ }\href {https://doi.org/10.5169/seals-115635} {\bibfield
  {journal} {\bibinfo  {journal} {Helv.~Phys.~Acta}\ }\textbf {\bibinfo
  {volume} {58}},\ \bibinfo {pages} {1030} (\bibinfo {year}
  {1985}{\natexlab{b}})}\BibitemShut {NoStop}%
\bibitem [{\citenamefont {Fulling}\ and\ \citenamefont
  {Ruijsenaars}(1987)}]{fulling_ruijsenaars_pr_1987}%
  \BibitemOpen
  \bibfield  {author} {\bibinfo {author} {\bibfnamefont {S.~A.}\ \bibnamefont
  {Fulling}}\ and\ \bibinfo {author} {\bibfnamefont {S.~N.~M.}\ \bibnamefont
  {Ruijsenaars}},\ }\bibfield  {title} {\bibinfo {title} {Temperature,
  periodicity and horizons},\ }\href
  {https://doi.org/https://doi.org/10.1016/0370-1573(87)90136-0} {\bibfield
  {journal} {\bibinfo  {journal} {Phys.~Rep.}\ }\textbf {\bibinfo {volume}
  {152}},\ \bibinfo {pages} {135} (\bibinfo {year} {1987})}\BibitemShut
  {NoStop}%
\bibitem [{\citenamefont {Kay}(1985{\natexlab{c}})}]{kay_hpa_1985second}%
  \BibitemOpen
  \bibfield  {author} {\bibinfo {author} {\bibfnamefont {B.~S.}\ \bibnamefont
  {Kay}},\ }\bibfield  {title} {\bibinfo {title} {{A uniqueness result for
  quasi-free KMS states}},\ }\href {https://doi.org/10.5169/seals-115634}
  {\bibfield  {journal} {\bibinfo  {journal} {Helv.~Phys.~Acta}\ }\textbf
  {\bibinfo {volume} {58}},\ \bibinfo {pages} {1017} (\bibinfo {year}
  {1985}{\natexlab{c}})}\BibitemShut {NoStop}%
\bibitem [{\citenamefont {Higuchi}\ \emph {et~al.}(2017)\citenamefont
  {Higuchi}, \citenamefont {Iso}, \citenamefont {Ueda},\ and\ \citenamefont
  {Yamamoto}}]{yamamoto_prd_minkowski}%
  \BibitemOpen
  \bibfield  {author} {\bibinfo {author} {\bibfnamefont {A.}~\bibnamefont
  {Higuchi}}, \bibinfo {author} {\bibfnamefont {S.}~\bibnamefont {Iso}},
  \bibinfo {author} {\bibfnamefont {K.}~\bibnamefont {Ueda}},\ and\ \bibinfo
  {author} {\bibfnamefont {K.}~\bibnamefont {Yamamoto}},\ }\bibfield  {title}
  {\bibinfo {title} {{Entanglement of the vacuum between left, right, future,
  and past: The origin of entanglement-induced quantum radiation}},\ }\href
  {https://doi.org/10.1103/PhysRevD.96.083531} {\bibfield  {journal} {\bibinfo
  {journal} {Phys. Rev. D}\ }\textbf {\bibinfo {volume} {96}},\ \bibinfo
  {pages} {083531} (\bibinfo {year} {2017})},\ \Eprint
  {https://arxiv.org/abs/1709.05757} {arXiv:1709.05757 [hep-th]} \BibitemShut
  {NoStop}%
\bibitem [{\citenamefont {Higuchi}\ and\ \citenamefont
  {Yamamoto}(2018)}]{yamamoto_prd_de_sitter}%
  \BibitemOpen
  \bibfield  {author} {\bibinfo {author} {\bibfnamefont {A.}~\bibnamefont
  {Higuchi}}\ and\ \bibinfo {author} {\bibfnamefont {K.}~\bibnamefont
  {Yamamoto}},\ }\bibfield  {title} {\bibinfo {title} {{Vacuum state in de
  Sitter spacetime with static charts}},\ }\href
  {https://doi.org/10.1103/PhysRevD.98.065014} {\bibfield  {journal} {\bibinfo
  {journal} {Phys. Rev. D}\ }\textbf {\bibinfo {volume} {98}},\ \bibinfo
  {pages} {065014} (\bibinfo {year} {2018})},\ \Eprint
  {https://arxiv.org/abs/1808.02147} {arXiv:1808.02147 [gr-qc]} \BibitemShut
  {NoStop}%
\bibitem [{\citenamefont {Dyson}(1949)}]{dyson_pr_1949a}%
  \BibitemOpen
  \bibfield  {author} {\bibinfo {author} {\bibfnamefont {F.~J.}\ \bibnamefont
  {Dyson}},\ }\bibfield  {title} {\bibinfo {title} {{The radiation theories of
  Tomonaga, Schwinger, and Feynman}},\ }\href
  {https://doi.org/10.1103/PhysRev.75.486} {\bibfield  {journal} {\bibinfo
  {journal} {Phys.~Rev.}\ }\textbf {\bibinfo {volume} {75}},\ \bibinfo {pages}
  {486} (\bibinfo {year} {1949})}\BibitemShut {NoStop}%
%%CITATION = PHRVA,75,486;%%
\bibitem [{\citenamefont {Matsubara}(1955)}]{matsubara_ptp_1955}%
  \BibitemOpen
  \bibfield  {author} {\bibinfo {author} {\bibfnamefont {T.}~\bibnamefont
  {Matsubara}},\ }\bibfield  {title} {\bibinfo {title} {{A new approach to
  quantum statistical mechanics}},\ }\href {https://doi.org/10.1143/PTP.14.351}
  {\bibfield  {journal} {\bibinfo  {journal} {Prog.~Theor.~Phys.}\ }\textbf
  {\bibinfo {volume} {14}},\ \bibinfo {pages} {351} (\bibinfo {year}
  {1955})}\BibitemShut {NoStop}%
%%CITATION = PTPKA,14,351;%%
\bibitem [{\citenamefont {Higuchi}\ and\ \citenamefont
  {Lima}(2020)}]{higuchi_lima_prd_2020}%
  \BibitemOpen
  \bibfield  {author} {\bibinfo {author} {\bibfnamefont {A.}~\bibnamefont
  {Higuchi}}\ and\ \bibinfo {author} {\bibfnamefont {W.~C.~C.}\ \bibnamefont
  {Lima}},\ }\bibfield  {title} {\bibinfo {title} {{Equivalence between the
  in-in perturbation theories for quantum fields in Minkowski spacetime and in
  the Rindler wedge}},\ }\href {https://doi.org/10.1103/PhysRevD.101.065009}
  {\bibfield  {journal} {\bibinfo  {journal} {Phys.~Rev.~D}\ }\textbf {\bibinfo
  {volume} {101}},\ \bibinfo {pages} {065009} (\bibinfo {year} {2020})},\
  \Eprint {https://arxiv.org/abs/2001.05500} {arXiv:2001.05500 [gr-qc]}
  \BibitemShut {NoStop}%
\bibitem [{\citenamefont {Landsman}\ and\ \citenamefont {van
  Weert}(1987)}]{landsman_van_weert_pr_1987}%
  \BibitemOpen
  \bibfield  {author} {\bibinfo {author} {\bibfnamefont {N.~P.}\ \bibnamefont
  {Landsman}}\ and\ \bibinfo {author} {\bibfnamefont {C.~G.}\ \bibnamefont {van
  Weert}},\ }\bibfield  {title} {\bibinfo {title} {{Real- and imaginary-time
  field theory at finite temperature and density}},\ }\href
  {https://doi.org/10.1016/0370-1573(87)90121-9} {\bibfield  {journal}
  {\bibinfo  {journal} {Phys.~Rep.}\ }\textbf {\bibinfo {volume} {145}},\
  \bibinfo {pages} {141} (\bibinfo {year} {1987})}\BibitemShut {NoStop}%
%%CITATION = PRPLC,145,141;%%
\bibitem [{\citenamefont {Miao}\ \emph {et~al.}(2014)\citenamefont {Miao},
  \citenamefont {Mora}, \citenamefont {Tsamis},\ and\ \citenamefont
  {Woodard}}]{miao_et_al_prd_2014}%
  \BibitemOpen
  \bibfield  {author} {\bibinfo {author} {\bibfnamefont {S.~P.}\ \bibnamefont
  {Miao}}, \bibinfo {author} {\bibfnamefont {P.~J.}\ \bibnamefont {Mora}},
  \bibinfo {author} {\bibfnamefont {N.~C.}\ \bibnamefont {Tsamis}},\ and\
  \bibinfo {author} {\bibfnamefont {R.~P.}\ \bibnamefont {Woodard}},\
  }\bibfield  {title} {\bibinfo {title} {{Perils of analytic continuation}},\
  }\href {https://doi.org/10.1103/PhysRevD.89.104004} {\bibfield  {journal}
  {\bibinfo  {journal} {Phys.~Rev.~D}\ }\textbf {\bibinfo {volume} {89}},\
  \bibinfo {pages} {104004} (\bibinfo {year} {2014})},\ \Eprint
  {https://arxiv.org/abs/1306.5410} {arXiv:1306.5410 [gr-qc]} \BibitemShut
  {NoStop}%
\bibitem [{\citenamefont {Woodard}(2014)}]{woodard_ijmpd_2014}%
  \BibitemOpen
  \bibfield  {author} {\bibinfo {author} {\bibfnamefont {R.~P.}\ \bibnamefont
  {Woodard}},\ }\bibfield  {title} {\bibinfo {title} {{Perturbative quantum
  gravity comes of age}},\ }\href {https://doi.org/10.1142/S0218271814300201}
  {\bibfield  {journal} {\bibinfo  {journal} {Int.~J.~Mod.~Phys.~D}\ }\textbf
  {\bibinfo {volume} {23}},\ \bibinfo {pages} {1430020} (\bibinfo {year}
  {2014})},\ \Eprint {https://arxiv.org/abs/1407.4748} {arXiv:1407.4748
  [gr-qc]} \BibitemShut {NoStop}%
\bibitem [{\citenamefont {Allen}\ and\ \citenamefont
  {Turyn}(1987)}]{allen_turyn}%
  \BibitemOpen
  \bibfield  {author} {\bibinfo {author} {\bibfnamefont {B.}~\bibnamefont
  {Allen}}\ and\ \bibinfo {author} {\bibfnamefont {M.}~\bibnamefont {Turyn}},\
  }\bibfield  {title} {\bibinfo {title} {{An evaluation of the graviton
  propagator in de Sitter space}},\ }\href
  {https://doi.org/10.1016/0550-3213(87)90672-9} {\bibfield  {journal}
  {\bibinfo  {journal} {Nucl. Phys.}\ }\textbf {\bibinfo {volume} {B292}},\
  \bibinfo {pages} {813} (\bibinfo {year} {1987})}\BibitemShut {NoStop}%
\bibitem [{\citenamefont {Higuchi}\ and\ \citenamefont
  {Kouris}(2001)}]{higuchi_kouris}%
  \BibitemOpen
  \bibfield  {author} {\bibinfo {author} {\bibfnamefont {A.}~\bibnamefont
  {Higuchi}}\ and\ \bibinfo {author} {\bibfnamefont {S.~S.}\ \bibnamefont
  {Kouris}},\ }\bibfield  {title} {\bibinfo {title} {{The covariant graviton
  propagator in de Sitter space-time}},\ }\href
  {https://doi.org/10.1088/0264-9381/18/20/311} {\bibfield  {journal} {\bibinfo
   {journal} {Classical~Quantum~Gravity}\ }\textbf {\bibinfo {volume} {18}},\
  \bibinfo {pages} {4317} (\bibinfo {year} {2001})},\ \Eprint
  {https://arxiv.org/abs/gr-qc/0107036} {arXiv:gr-qc/0107036} \BibitemShut
  {NoStop}%
\bibitem [{\citenamefont {Faizal}\ and\ \citenamefont
  {Higuchi}(2012)}]{faizal_higuchi_2011}%
  \BibitemOpen
  \bibfield  {author} {\bibinfo {author} {\bibfnamefont {M.}~\bibnamefont
  {Faizal}}\ and\ \bibinfo {author} {\bibfnamefont {A.}~\bibnamefont
  {Higuchi}},\ }\bibfield  {title} {\bibinfo {title} {{Physical equivalence
  between the covariant and physical graviton two-point functions in de Sitter
  spacetime}},\ }\href {https://doi.org/10.1103/PhysRevD.85.124021} {\bibfield
  {journal} {\bibinfo  {journal} {Phys. Rev. D}\ }\textbf {\bibinfo {volume}
  {85}},\ \bibinfo {pages} {124021} (\bibinfo {year} {2012})},\ \Eprint
  {https://arxiv.org/abs/1107.0395} {arXiv:1107.0395 [gr-qc]} \BibitemShut
  {NoStop}%
\bibitem [{\citenamefont {Allen}(1986)}]{allen_prd_1986}%
  \BibitemOpen
  \bibfield  {author} {\bibinfo {author} {\bibfnamefont {B.}~\bibnamefont
  {Allen}},\ }\bibfield  {title} {\bibinfo {title} {{The graviton propagator in
  de~Sitter space}},\ }\href {https://doi.org/10.1103/PhysRevD.34.3670}
  {\bibfield  {journal} {\bibinfo  {journal} {Phys.~Rev.~D}\ }\textbf {\bibinfo
  {volume} {34}},\ \bibinfo {pages} {3670} (\bibinfo {year}
  {1986})}\BibitemShut {NoStop}%
\bibitem [{\citenamefont {Polchinski}(1989)}]{polchinski_plb_1989}%
  \BibitemOpen
  \bibfield  {author} {\bibinfo {author} {\bibfnamefont {J.}~\bibnamefont
  {Polchinski}},\ }\bibfield  {title} {\bibinfo {title} {{The phase of the sum
  over spheres}},\ }\href {https://doi.org/10.1016/0370-2693(89)90387-0}
  {\bibfield  {journal} {\bibinfo  {journal} {Phys.~Lett.~B}\ }\textbf
  {\bibinfo {volume} {219}},\ \bibinfo {pages} {251} (\bibinfo {year}
  {1989})}\BibitemShut {NoStop}%
\bibitem [{\citenamefont {Taylor}\ and\ \citenamefont
  {Veneziano}(1990)}]{taylor_veneziano_npb_1990}%
  \BibitemOpen
  \bibfield  {author} {\bibinfo {author} {\bibfnamefont {T.~R.}\ \bibnamefont
  {Taylor}}\ and\ \bibinfo {author} {\bibfnamefont {G.}~\bibnamefont
  {Veneziano}},\ }\bibfield  {title} {\bibinfo {title} {{Quantum gravity at
  large distances and the cosmological constant}},\ }\href
  {https://doi.org/10.1016/0550-3213(90)90615-K} {\bibfield  {journal}
  {\bibinfo  {journal} {Nucl.~Phys.}\ }\textbf {\bibinfo {volume} {B345}},\
  \bibinfo {pages} {210} (\bibinfo {year} {1990})}\BibitemShut {NoStop}%
\bibitem [{\citenamefont {Faizal}\ and\ \citenamefont
  {Higuchi}(2008)}]{faizal_higuchi_2008}%
  \BibitemOpen
  \bibfield  {author} {\bibinfo {author} {\bibfnamefont {M.}~\bibnamefont
  {Faizal}}\ and\ \bibinfo {author} {\bibfnamefont {A.}~\bibnamefont
  {Higuchi}},\ }\bibfield  {title} {\bibinfo {title} {{On the FP-ghost
  propagators for Yang-Mills theories and perturbative quantum gravity in the
  covariant gauge in de Sitter spacetime}},\ }\href
  {https://doi.org/10.1103/PhysRevD.78.067502} {\bibfield  {journal} {\bibinfo
  {journal} {Phys. Rev. D}\ }\textbf {\bibinfo {volume} {78}},\ \bibinfo
  {pages} {067502} (\bibinfo {year} {2008})},\ \Eprint
  {https://arxiv.org/abs/0806.3735} {arXiv:0806.3735 [gr-qc]} \BibitemShut
  {NoStop}%
\bibitem [{\citenamefont {Gibbons}\ \emph {et~al.}(2021)\citenamefont
  {Gibbons}, \citenamefont {Higuchi},\ and\ \citenamefont
  {Lima}}]{gibbons_higuchi_lima_prd_2021}%
  \BibitemOpen
  \bibfield  {author} {\bibinfo {author} {\bibfnamefont {J.}~\bibnamefont
  {Gibbons}}, \bibinfo {author} {\bibfnamefont {A.}~\bibnamefont {Higuchi}},\
  and\ \bibinfo {author} {\bibfnamefont {W.~C.~C.}\ \bibnamefont {Lima}},\
  }\bibfield  {title} {\bibinfo {title} {{Infrared problem in the
  Faddeev-Popov-ghost propagator in perturbative quantum gravity in de~Sitter
  spacetime}},\ }\href {https://doi.org/10.1103/PhysRevD.103.065016} {\bibfield
   {journal} {\bibinfo  {journal} {Phys.~Rev.~D}\ }\textbf {\bibinfo {volume}
  {103}},\ \bibinfo {pages} {065016} (\bibinfo {year} {2021})},\ \Eprint
  {https://arxiv.org/abs/2101.07268} {arXiv:2101.07268 [gr-qc]} \BibitemShut
  {NoStop}%
\bibitem [{\citenamefont {Gibbons}\ \emph {et~al.}(1978)\citenamefont
  {Gibbons}, \citenamefont {Hawking},\ and\ \citenamefont
  {Perry}}]{gibbons_hawking_perry_npb_1978}%
  \BibitemOpen
  \bibfield  {author} {\bibinfo {author} {\bibfnamefont {G.~W.}\ \bibnamefont
  {Gibbons}}, \bibinfo {author} {\bibfnamefont {S.~W.}\ \bibnamefont
  {Hawking}},\ and\ \bibinfo {author} {\bibfnamefont {M.~J.}\ \bibnamefont
  {Perry}},\ }\bibfield  {title} {\bibinfo {title} {{Path integrals and the
  indefiniteness of the gravitational action}},\ }\href
  {https://doi.org/10.1016/0550-3213(78)90161-X} {\bibfield  {journal}
  {\bibinfo  {journal} {Nucl.~Phys.}\ }\textbf {\bibinfo {volume} {B138}},\
  \bibinfo {pages} {141} (\bibinfo {year} {1978})}\BibitemShut {NoStop}%
\bibitem [{\citenamefont {Schleich}(1987)}]{schleich_prd_1987}%
  \BibitemOpen
  \bibfield  {author} {\bibinfo {author} {\bibfnamefont {K.}~\bibnamefont
  {Schleich}},\ }\bibfield  {title} {\bibinfo {title} {{Conformal rotation in
  perturbative gravity}},\ }\href {https://doi.org/10.1103/PhysRevD.36.2342}
  {\bibfield  {journal} {\bibinfo  {journal} {Phys.~Rev.~D}\ }\textbf {\bibinfo
  {volume} {36}},\ \bibinfo {pages} {2342} (\bibinfo {year}
  {1987})}\BibitemShut {NoStop}%
\bibitem [{\citenamefont {Mazur}\ and\ \citenamefont
  {Mottola}(1990)}]{mazur_mottola_npb_1990}%
  \BibitemOpen
  \bibfield  {author} {\bibinfo {author} {\bibfnamefont {P.~O.}\ \bibnamefont
  {Mazur}}\ and\ \bibinfo {author} {\bibfnamefont {E.}~\bibnamefont
  {Mottola}},\ }\bibfield  {title} {\bibinfo {title} {{The gravitational
  measure, solution of the conformal factor problem and stability of the ground
  state of quantum gravity}},\ }\href
  {https://doi.org/10.1016/0550-3213(90)90268-I} {\bibfield  {journal}
  {\bibinfo  {journal} {Nucl.~Phys.}\ }\textbf {\bibinfo {volume} {B341}},\
  \bibinfo {pages} {187} (\bibinfo {year} {1990})}\BibitemShut {NoStop}%
\end{thebibliography}%

\end{document}